# Implementing the Availability Model of a Software-Defined Backbone Network in Möbius

*(Technical Report)*

Gianfranco Nencioni, Bjarne E. Helvik and Poul E. Heegaard

Department of Information Security and Communication Technology,
NTNU – Norwegian University of Science and Technology, Trondheim, Norway
{gianfranco.nencioni, bjarne.e.helvik, poul.heegaard}@ntnu.no

*Abstract*—**Software-defined networking (SDN) promises to improve the programmability and flexibility of networks, but it may bring also new challenges that need to be explored. One open issue is the quantitative assessment of the properties of SDN backbone networks to determine whether they can provide similar availability to the traditional IP backbone networks. To achieve this goal, a two-level availability model that is able to capture the global network connectivity without neglecting the essential details and which includes a failure correlation assessment should be considered. The two-level availability model is composed by a structural model and the dynamic models of the principal minimal-cut sets of the network. The purpose of this technical report is to extensively present the implementation on Möbius of the Stochastic Activity Network (SAN) availability model of the network elements and the principal minimal-cut sets of a SDN backbone network and the corresponding traditional backbone network.**

## I. INTRODUCTION

During the recent years, the SDN has emerged as a new network paradigm, which mainly consists of a programmable network approach where the forwarding plane is decoupled from the control plane [1], [2]. Despite programmable networks having been studied for decades, SDN is experiencing a growing success because it is expected that the ease of changing protocols and provide support for adding new services and applications will foster future network innovation, which is limited and expensive in todays legacy systems.

A simplified sketch of the SDN architecture from IRFT RFC 7426 [1] without the management plane is depicted in Figure 1. The control plane and data plane are separated. Here the control plane is logically centralised in a software-based controller ("network brain"), while the data plane is composed of the network devices ("network arms") that conduct the packet forwarding.

The control plane has a northbound and a southbound interface. The northbound interface provides an network abstraction to the network applications (e.g. routing protocol, firewall, load balancer, anomaly detection, etc...), while the southbound interface (e.g. OpenFlow) standardises the information exchange between control and data planes.

In [3], the following set of potential advantages of SDN were pointed out:

- centralised control;
- simplified algorithms;

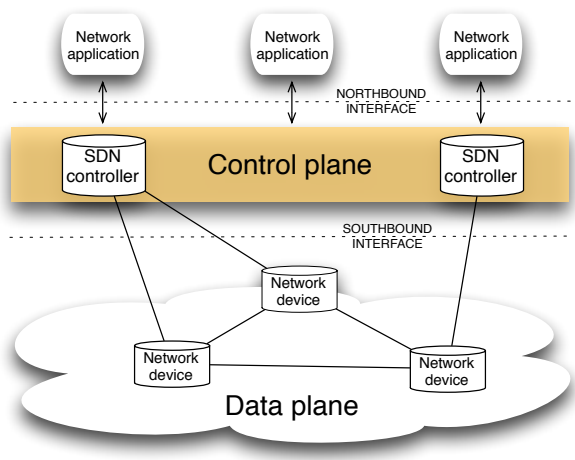

Fig. 1: SDN architecture (exclusive the management plane)

- commoditising network hardware;
- eliminating middle-boxes;
- enabling the design and deployment of third-party applications.

However, from a dependability perspective, the SDN poses a set of new vulnerabilities and challenges compared with traditional networking, as discussed in [4]:

- consistency of network information (user plane state information) and controller decisions;
- consistency between the distributed SDN controllers in the control plane;
- increased failure intensities of (commodity) network elements;
- compatibility and interoperability between general purpose, non-standard network elements
- interdependency between path setup in network elements and monitoring of the data plane in the control plane;
- load sharing (to avoid performance bottleneck) and fault tolerance in the control plane have conflicting requirements;

In [5], a two-level availability model has been proposed in order to capture the global network connectivity without neglecting the essential details and which includes a failure correlation assessment. The two-level availability model is composed by a structural model and the dynamic models of the principal minimal-cut sets of both the SDN backbone network and the corresponding traditional backbone network.

The purpose of this technical report is the detailed presentation of the implementation on Möbius [14] of the Stochastic Activity Network (SAN) availability model of both the network elements and the principal minimal-cut sets. These models have been used in [5]

In Section II, we introduce the nation-wide backbone network that has been used for computing the principal minimal-cut sets. The SAN models of the network elements and the principal minimal-cut sets are presented in Section III and Section IV, respectively. Finally, the conclusions are summarized in Section V.

## II. MODEL CASE STUDY

In this technical report and in [5], we consider a nation-wide backbone network that consists of 10 nodes across 4 cities, and two dual-homed SDN controllers. See Figure 2 for an illustration of the topology. The nodes are located in the four major cities in Norway, Bergen (BRG), Trondheim (TRD), Stavanger (STV), and Oslo (OSL). Each town has duplicated nodes, except Oslo which has four nodes (OSL1 and OSL2). The duplicated nodes are labelled, $X_1$ and $X_2$, where $X$=OSL1, OSL2, BRG, STV, and TRD. In addition to the forwarding nodes, there are two dual-homed SDN controllers (SC$_1$ and SC$_2$), which are connected to TRD and OSL1.

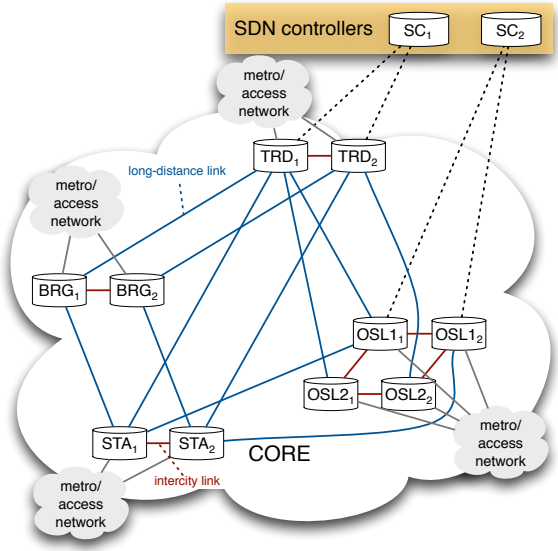

Fig. 2: Nation-wide backbone network

Given this network, for computing and comparing the network availability of SDN with a traditional IP network we need to calculate the availability of the single network elements [12] or of the principal minimal-cut sets [5].

## III. SAN MODEL OF THE NETWORK ELEMENTS

In the following, we present the SAN models of the network elements: links (which are the same in both SDN and traditional network), traditional IP routers, SDN switches, and SDN controllers.

### A. Link

The model of a link is assumed to be dominated by physical link failures. Therefore, a simple two-state Markov model is used. Figure 3 shows the SAN representation. The links are either up or down due to hardware failure. We use the same model for both traditional network and SDN. Given failure rate $\lambda_L$ and repair rate $\mu_L$, the availability of a link is $A_L = \frac{\mu_L}{\lambda_L + \mu_L}$. This model is assumed for each of the link components in the structural model. We don't know the geographical location of the nodes and therefor the distance between them either, which implies that the length of the links connecting the nodes in the network can't be determined. Hence, in our case studies we have to assume that the link failure rate is not dependent of the link length. Note that in general the failure rate is expected to be proportional to the length of the link.

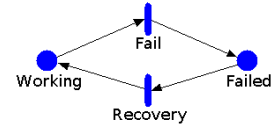

Fig. 3: SAN model of a link

### B. Traditional IP router

The SAN model of a traditional router is depicted in Figure 4. In the model we focus on the router functionalities and the related failure sources, each component of the router has not been considered because it would be dependent on a particular router architecture. In any case, we assume 1+1 redundancy of the controller hardware, which is a common best practice in any architecture. Multiple failures are not included in the model since they are assumed to be less frequent and will probably not have significant impact on the expected accuracy of the approach.

The SAN model of the traditional router is composed of eight places:

- *Working* represents the state when the system is fully working and it is initialized with one token;
- *failed_MAN* is equal to 1 when there is a failure of the Operation and Management (O&M), 0 otherwise;
- *spare_CHW* represents the state when one of the two redundant control hardware is failed but the other one is correctly working;
- *sys_down* is a coverage state and is equal to 1 if there is an unsuccessful activation of the stand-by hardware after a failure (manual recovery).
- *failed_CHW* represents the state when both controllers has an hardware failure;

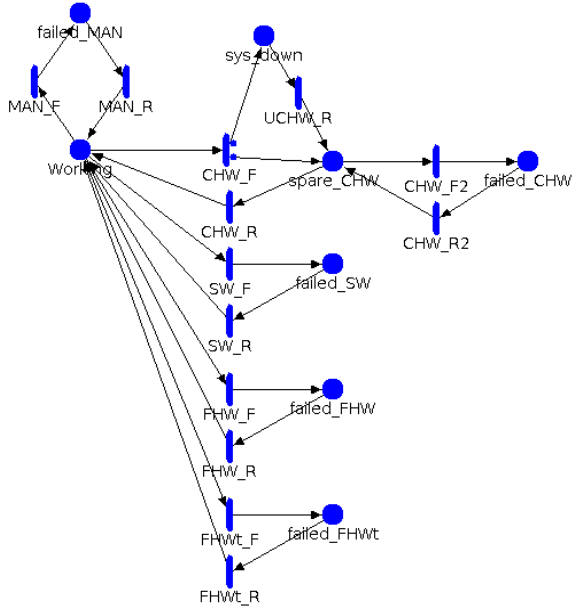

Fig. 4: SAN model of a traditional IP router

TABLE I: Model parameters for the IP network with numerical values used in the case studies

| intensity | [time] | description |
|---|---|---|
| $1/\lambda_L = 4$ | [months] | expected time to next link failure |
| $1/\mu_L = 15$ | [minutes] | expected time to link repair |
| $1/\lambda_{dF} = 6$ | [months] | expected time to next permanent forwarding hardware failure |
| $1/\mu_{dF} = 12$ | [hours] | expected time to repair permanent forwarding hardware |
| $1/\lambda_{dFt} = 1$ | [week] | expected time to next transient forwarding hardware failure |
| $1/\mu_{dFt} = 3$ | [minutes] | expected time to repair transient forwarding hardware |
| $1/\lambda_{dC} = 6$ | [months] | expected time to next control hardware failure |
| $1/\mu_{dC} = 12$ | [hours] | expected time to repair control hardware |
| $1/\lambda_{dS} = 1$ | [week] | expected time to next software failure |
| $1/\mu_{dS} = 3$ | [minutes] | expected time to software repair |
| $1/\lambda_{dO} = 1$ | [month] | expected time to next O&M failure |
| $1/\mu_{dO} = 3$ | [hours] | expected time to O&M repair |
| $1/\mu_{dUC} = 8$ | [hours] | expected time to recover from uncovered control hardware failure |
| $C_{dC} = 0.97$ | | coverage factor |

TABLE II: Model parameters for the SDN switch

| intensity | description |
|---|---|
| $\lambda_F = \lambda_{dF}$ | intensity of permanent hardware failures |
| $\mu_F = \mu_{dF}$ | repair intensity of permanent hardware failures |
| $\lambda_{Ft} = \lambda_{dFt}$ | intensity of transient hardware failures |
| $\mu_{Ft} = \mu_{dFt}$ | restoration intensity after transient hardware failures |
| $\lambda_{sS} = 0$ | intensity of software failure |

- *failed_SW* is equal to 1 when there is a software failure, 0 otherwise;
- *failed_FHW* represents the state when there is a permanent hardware failure in forwarding plane
- *failed_FHWt* represents the presence of a transient hardware failure in forwarding plane;

The router is failed when the token is not in *Working* or *spare_CHW*.

The places are connected by mean of the following timed activities with exponential time distribution:

- *MAN_F* and *MAN_R* represent the failure and the recovery events of the O&M with a rate of $\lambda_{dO}$ and $\mu_{dO}$, respectively;
- *CHW_F* represents the failure event of the control hardware with a rate of 2 $\lambda_{dC}$ and there are two cases, with probability $C_{dC}$ a token is put into *spare_CHW*, otherwise (with probability $1 - C_{dC}$) the system is not able to manage the control hardware failure and the system goes down;
- *CHW_F2* represents the failure event of the spare control with a rate of $\lambda_{dC}$;
- *CHW_R* and *CHW_R2* represent the recovery of the control hardware with a rate of $\mu_{dC}$;
- *UCHW_R* represents the recovery after an unsuccessful activation of the stand-by hardware with a rate of $\mu_{dUC}$;
- *SW_F* and *SW_R* represent the failure and the recovery events of the software with a rate of $\lambda_{dS}$ and $\mu_{dS}$, respectively;
- *FHW_F* and *FHW_R* represent the permanent failure and the recovery events of the forwarding hardware with a rate of $\lambda_{dF}$ and $\mu_{dF}$, respectively;

- *FHWt_F* and *FHWt_R* represent the transient failure and the recovery events of the forwarding hardware with a rate of $\lambda_{dFt}$ and $\mu_{dFt}$, respectively;

All the model parameters are defined in Table I. Note that for sake of simplicity we have assumed homogeneous equipment. The table includes the numerical values used in the case studies and that are inspired by and taken from several studies [9], [10], [11].

### C. SDN switch

Figure 5 shows the model of the switch in an SDN, which is significantly simpler than the router in a traditional network. The states related to the control hardware failures are not contained in this model, since all the control logic is located in the controller. O&M associated with the SDN switch has been also omitted because we assume that the complexity of the O&M operations done on a single switch is likely to be small relative to a router and globally in the controller. The software is still present but its failure rate will be very low since the functionality is much simpler.

Table II describes the parameters for modelling the SDN switch.

All SDN parameters are expressed relative to the parameters for the traditional network (Table I). In an SDN switch, the failure/repair intensities of (permanent/transient) hardware failures are the same because failures with the same cause, have the same intensities in both models. However, we assume that the software on an SDN switch will be much less complicated than on a traditional IP router because the control logic has been moved to the controllers, and we have set the failure rate to zero, for the sake of simplicity.

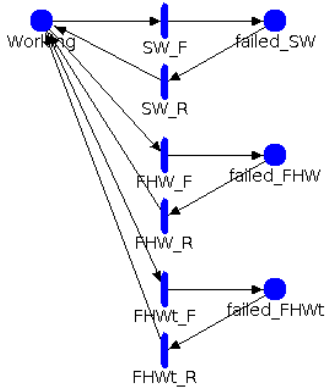

Fig. 5: SAN model of a SDN switch

### D. SDN controller

The SDN controller has been modelled with the SAN model depicted in Figure 6. We have assumed that the SDN controller is a cluster of $M$ processors and the system is working, i.e., possesses sufficient capacity if $K$ out of the $M$ processors are active, which means that both software and hardware are working. The other main assumptions of the model are:

- single repairman for a hardware failure;
- load dependency of software failure when the system is working, $\lambda_S(N_a) = \lambda_S/N_a$, where $N_a$ is the number of active processors;
- when the entire system fails, only processors failed due to hardware failures will be down until the system recovers;
- load independence of software failure when the system has failed, $\lambda_S(N_a) = \lambda_S$, since the remaining unfailed processors are working at the full capacity.

The SAN model of the SDN controller is composed of six places:

- *Active_proc* represents the number of active processors and it is initialized to the total number of processors;
- *failed_MAN* is equal to 1 when there is a failure of the O&M, 0 otherwise;
- *failed_SW* represents the number of processors where the software has failed;
- *failed_HW* represents the number of processors where the hardware has failed;
- *sys_down* is a coverage state and is equal to 1 if the hardware failure in one processor forces all the system to be down;
- *sw_sys_down* is a coverage state and is equal to 1 if the software failure in one processor causes the crash of all the processors.

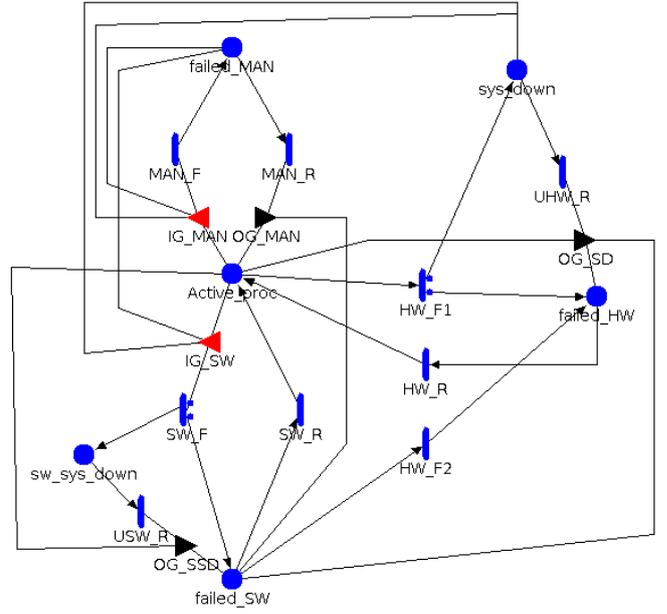

Fig. 6: SAN model of SDN controller

The places are connected by mean of the following timed activities with exponential time distribution:

- *MAN_F* and *MAN_R* represent the failure and the recovery of the O&M with a rate of $\lambda_O$ and $\mu_O$, respectively;
- *SW_F* represents the failure of the software with a rate of $\lambda_S$, if the number of active processors is at least $K$, or $N_a$ $\lambda_S$, otherwise; there are two cases, with probability $C_S$ a token is put into *failed_SW* (if there are enough working processors, the system is still working), otherwise (with probability $1 - C_S$) the system is not able to manage the software failure and the system goes down;
- *SW_R* represents the recovery of the software with a rate of $\mu_S$;
- *USW_R* represents the recovery of the software crash with a rate of $\mu_{US}$;
- *HW_F1* represents the failure of the hardware of the active processors with a rate of $N_a$ $\lambda_H$ and there are two cases, with probability $C_C$ a token is put into *failed_HW* (the hardware is failed but if there are enough working processors, the system is working), otherwise (with probability $1-C_C$) the system is not able to manage the hardware failure and the system goes down (note that if there is already a token in *failed_MAN* or *sys_down*, the token is forced to be put in *failed_HW*);
- *HW_F2* represents the failure of the hardware of the processors with a failed software with a rate of $N_s$ $\lambda_H$, where $N_s$ is the number of token in *failed_SW*;
- *HW_R* represents the recovery of the hardware with a rate of $\mu_H$;
- *UHW_R* represent the recovery after an unsuccessful activation of the stand-by hardware with a rate of $\mu_{UH}$;

TABLE III: Model parameters for the SDN controller

| intensity | description |
|---|---|
| $\lambda_H = \alpha_H \; \lambda_{dC} \; N/K$ | intensity of hardware failures |
| $\mu_H = \mu_{dC}$ | hardware repair intensity |
| $1/\mu_{UH} = 0.5h$ | restoration time after uncovered hardware failure |
| $\lambda_S = \alpha_S \; \lambda_{dS} \; N$ | intensity of software failures |
| $\mu_S = \mu_{dS}$ | restoration intensity after software failure |
| $1/\mu_{US} = 0.5h$ | restoration time after uncovered software failure |
| $\lambda_O = \alpha_O \; \lambda_{dO} \; N$ | intensity of O&M failures |
| $\mu_O = \mu_{dO}$ | rectification intensity after O&M failures |
| $C_H = C_{dC}$ | hardware failure coverage factor |
| $C_S = 0.9$ | software failure coverage factor |

Furthermore, the following input and output gates are included:

- *IG_MAN* enables the O&M failure activity only if there are no tokens in *failed_MAN*, *sys_down*, and *sw_sys_down*;
- *IG_SW* enables the software failure activity only if there are no tokens in *failed_MAN*, *sys_down*, *sw_sys_down*, and there are active processors and implies the decrease of the number of active processors;
- *OG_MAN* and *OG_SSD* resets the number of software failures and sets the number of active processors to the total number of processors minus the number of processors with failed hardware;
- *OG_SD* increases the number of failed hardware, resets the number of software failure, and sets the number of active processors to the total number of processors minus the number of processors with failed hardware.

In the proposed model the system is down where the number of tokens in *Active_proc* is lower than $K$ or there is a token in *failed_MAN*, in *sys_down*, or in *sw_sys_down*.

The parameters the SDN controller model are listed in Table III.

In an SDN controller, all failure rates are $N$-times larger than in the traditional network, where $N$ is the number of network nodes (10 in the addressed nation-wide backbone network). This is because we assume that the SDN needs roughly the same processing capacity and amount of hardware than in the traditional network. Therefore, the failure intensity is assumed to be proportional to $N$, and of the same order of magnitude as the total failure intensity of the traditional distributed IP router system. For the hardware failures the total failure intensity is divided by the number of needed processors $K = \lfloor 0.8 \cdot M \rfloor$, where $M = N$ is the total number of processors. Moreover, we set the proportionality factors $\alpha_H$, $\alpha_S$, and $\alpha_O$ as follows by basing on previous work [12]: $\alpha_H = 1$, $\alpha_S = 1$, $\alpha_O = 0.2$, and $\alpha_C = 1$.

## IV. SAN MODEL OF THE PRINCIPAL MINIMAL-CUT SETS

In [5], we have determined the minimal-cut sets for the different networks (TN: traditional network, F-SDN: forwarding part of SDN, C-SDN: control part of SDN), then we have identifies the principal minimal-cut sets, i.e. the ones with lower cardinality, (see Table IV).

Successively, we have evaluated which are the the failure correlation sources among the elements composing the principal minimal-cut set. Table V maps the failure correlation sources to the elements composing the 12 kinds of minimal-cut sets (4 for the traditional network, 8 for the SDN). The considered failure correlation sources are the following: Geographical Proximity (GEO), Physical Proximity (PHY), Common O&M (COM), Misconfiguration (MIS), Compatibility Issue (CIS), Homogeneous Equipment (HEQ), Traffic Migration (TMI).

TABLE V: Type of minimal-cut sets for the different networks vs failure correlation source

| type | network | GEO | PHY | COM | MIS | CIS | TMI | HEQ |
|---|---|---|---|---|---|---|---|---|
| {n,n} | TN | ✓ | | ✓ | | | ✓ | |
| | F-SDN | ✓ | | | ✓ | | ✓ | |
| | C-SDN | | | | ✓ | | ✓ | |
| {n,n,n} | TN | | | | | | | ✓ |
| | F-SDN | | | | | | | ✓ |
| | C-SDN | ✓ | | | | ✓ | | |
| {n,n,l} | TN | ✓ | | | | | | ✓ |
| | F-SDN | | | | | | | ✓ |
| | C-SDN | ✓ | | | | ✓ | | |
| {n,l,l} | TN | ✓ | ✓ | | | | | |
| | F-SDN | ✓ | ✓ | | | | | |
| | C-SDN | ✓ | ✓ | | | | | |

For modelling the availability of the minimal-cuts sets, in [5] we have used a modular and systematic approach to compose the SAN model of the network elements. In the composition, for considering the failure correlation among the network elements we have "added", "modified", or "merged" dependency models. In particular, we have added for GEO, PHY, MIS, and CIS, modified for TMI and HEQ, and merged for COM.

Table VI shows the parameters related to the failure correlation. In [13], the authors discovered that around of the 10% failures are actually multiple simultaneous failures. Based on this consideration we have consider an intensity of the correlated failures that is ten times lower than the "original" one. In particular, the "original" intensity of the GEO, PHY, MIS, and CIS are related to the permanent forwarding hardware or link (depending on the correlated elements), link, O&M, and SDN controller software, respectively. Since the COM failure is a merge failure correlation, we have considered a failure intensity equal to the intensity of distributed O&M failure. For the GEO and CIS recovery, we have considered a rate three times lower than the "original" rate since they need more time for restoring from the failure source (e.g. blackout) or to discover the origin of the failure. Instead, for the PHY, MIS and COM recovery, the rate for restoring the single element as been considered. Moreover, for conducting our sensitivity analysis we use the multiplicative factors $\alpha_{GEO}$, $\alpha_{PHY}$, $\alpha_{MIS}$, $\alpha_{COM}$, and $\alpha_{CIS}$ and the addends $\beta_{TMI}$ and $\beta_{HEQ}$. In particular we have considered $\alpha_{GEO,PHY,MIS,COM,CIS} \in \{10^i\}$ $i = 0, \pm1, \pm2$, $\beta_{TMI} \in \{\pm0.05, \pm0.02, 0\}$, and $\beta_{HEQ} \in \{\pm0.01, 0\}$.

TABLE IV: Principal minimal-cut sets (2 and 3 cardinality) for the different networks

| cardinality | type | TN & F-SDN | C-SDN |
|---|---|---|---|
| 2 | {n,n} | $\{n_{BRG_1}, n_{BRG_2}\}$ <br> $\{n_{STV_1}, n_{STV_2}\}$ <br> $\{n_{TRD_1}, n_{TRD_2}\}$ | $\{n_{SC_1}, n_{SC_2}\}$ |
| 3 | {n,n,n} | $\{n_{BRG_1}, n_{STV_2}, n_{TRD_2}\}$ <br> $\{n_{BRG_2}, n_{STV_1}, n_{TRD_1}\}$ | $\{n_{OSL1_1}, n_{OSL1_2}, n_{SC_1}\}$ |
| | {n,n,l} | $\{n_{BRG_1}, n_{STV_2}, l_{TRD_2-BRG_2}\}$ <br> $\{n_{BRG_1}, n_{TRD_2}, l_{STV_2-BRG_2}\}$ <br> $\{n_{BRG_2}, n_{STV_1}, l_{TRD_1-BRG_1}\}$ <br> $\{n_{BRG_2}, n_{TRD_1}, l_{STV_1-BRG_1}\}$ | $\{n_{OSL1_1}, n_{SC_1}, l_{OSL1_2-SC_2}\}$ <br> $\{n_{OSL1_2}, n_{SC_1}, l_{OSL1_1-SC_2}\}$ <br> $\{n_{SC_2}, n_{TRD_2}, l_{TRD_2-SC_1}\}$ <br> $\{n_{SC_2}, n_{TRD_2}, l_{TRD_1-SC_1}\}$ |
| | {n,l,l} | $\{n_{BRG_1}, l_{STV_2-BRG_2}, l_{TRD_2-BRG_2}\}$ <br> $\{n_{BRG_2}, l_{STV_1-BRG_1}, l_{TRD_1-BRG_1}\}$ | $\{n_{SC_1}, l_{OSL1_1-SC_2}, l_{OSL1_2-SC_2}\}$ <br> $\{n_{SC_2}, l_{TRD_1-SC_1}, l_{TRD_2-SC_1}\}$ |

TABLE VI: Model parameters for failure correlation sources

| intensity | description |
|---|---|
| $\lambda_{GEO} = \frac{\alpha_{GEO}\ \lambda_{FHW}}{10}$ | intensity of geographical-spread failure |
| $\mu_{GEO} = \mu_{FHW}/3$ | repair rate after a geographical-spread failure |
| $\lambda_{PHY} = \alpha_{PHY}\ \lambda_L/10$ | intensity of physical-spread failure |
| $\mu_{PHY} = \mu_L$ | repair rate after a physical-spread failure |
| $\lambda_{COM} = \alpha_{COM}\ \lambda_{dO}$ | failure intensity caused by a shared O&M |
| $\mu_{COM} = \mu_{dO}$ | recovery rate from a shared-O&M failure |
| $\lambda_{MIS} = \alpha_{MIS}\ \lambda_O/10$ | misconfiguration failure intensity |
| $\mu_{MIS} = \mu_O$ | intensity to recover from a misconfiguration failure |
| $\lambda_{CIS} = \alpha_{CIS}\ \lambda_S/10$ | failure intensity caused by a compatibility issue among different elements |
| $\mu_{CIS} = \mu_S/3$ | recovery rate from a incompatibility failure |
| $C_{TMI} = 0.95 + \beta_{TMI}$ | coverage factor for considering failures induced by traffic migration |
| $C_{HEQ} = 0.99 + \beta_{HEQ}$ | coverage factor for taking into account failures due to homogeneous equipment |

In the remainder of the section, we briefly describe the SAN model of the principal minimal-cut sets, for further details the reader can find the Möbius documentation in the appendix.

### A. $\{n, n\}$ in TN

Figure 7 depicts the SAN model of $\{n, n\}$ in TN, where the two routers are in the same city (*GEO*), share the O&M (*COM*), and if one fails all the traffic is managed by the other one (*TMI*). The SAN model is composed of the SAN of the two routers (*_S1* and *_S2*), where the single O&M failure places have been deleted and the following places are added:

- *GEO* is equal to 1 when there is a GEO failure, 0 otherwise;
- *failed_MAN* is equal to 1 when there is a COM failure, 0 otherwise.

The places are connected by mean of the following timed activities with exponential time distribution:

- *MAN_F* and *MAN_R* represent the failure and the recovery of the common O&M with a rate of $\lambda_{COM}$ and $\mu_{COM}$, respectively;

- *GEO_F* and *GEO_R* represent the failure and the recovery from GEO failure with a rate of $\lambda_{GEO}$ and $\mu_{GEO}$, respectively.

For considering the *TMI* failure, the *SW_F*, *FHW_F*, *FHWt_F*, *CHW_F2* time activities of the two routers are modified by creating two cases: if both routers are working, with probability $C_{TMI}$ only one router is failing and instead with probability $1 - C_{TMI}$ both the routers are failing, otherwise only one router is failing.

Furthermore, the following input and output gates are included:

- *IG_GF* and *IG_MF* enable the GEO and COM failure activities only if there in a token in both *Working_S1* and *Working_S2*, i.e. both routers are working, and reset the token in both *Working_S1* and *Working_S2*;
- *OG_GF* and *OG_MF* set the token in both *Working_S1* and *Working_S2* again;
- *OG_SW*, *OG_FHW*, *OG_FHWt*, and *OG_CHW* reset the token in both *Working_S1* and *Working_S2* and set *failed_SW*, *failed_FHW*, *failed_FHWt*, and *failed_CHW*, respectively, of both routers.

The minimal-cut set is unavailable when there are not token in *Working_S1*, *Working_S2*, *spare_CHW_S1*, and *spare_CHW_S2*.

Further details on the implementation in Möbius of the SAN model and the related simulation can be found in the Appendix A6 and B8, respectively.

### B. $\{n, n\}$ in F-SDN

Figure 8 shows the SAN model of $\{n, n\}$ in F-SDN, where the two SDN switches (*_S1* and *_S2*) are in the same city (*GEO*), if one fails all the traffic is managed by the other one (*TMI*), and share a common configuration (*MIS*). The SAN model is similar to the one for the two routers (see Figure 7): there is not the part related to the control hardware and there is the MIS failure instead of the COM failure.

The minimal-cut set is unavailable when there are not token in *Working_S1* and *Working_S2*.

Further details on the implementation in Möbius of the SAN model and the related simulation can be found in the Appendix A10 and B12, respectively.

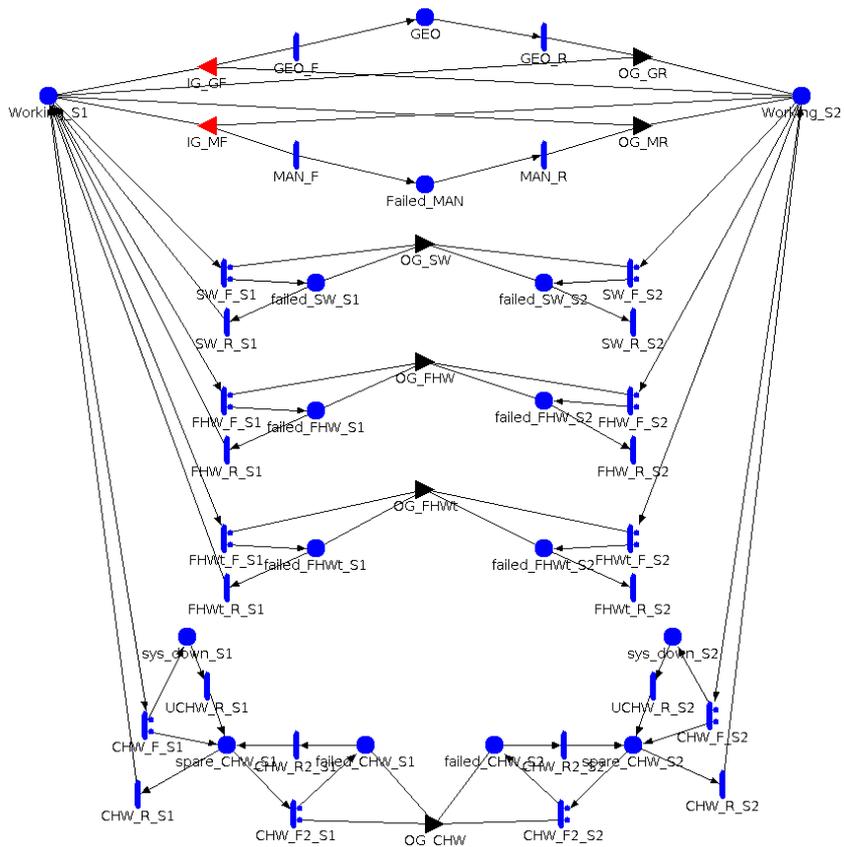

Fig. 7: SAN model of $\{n, n\}$ in TN

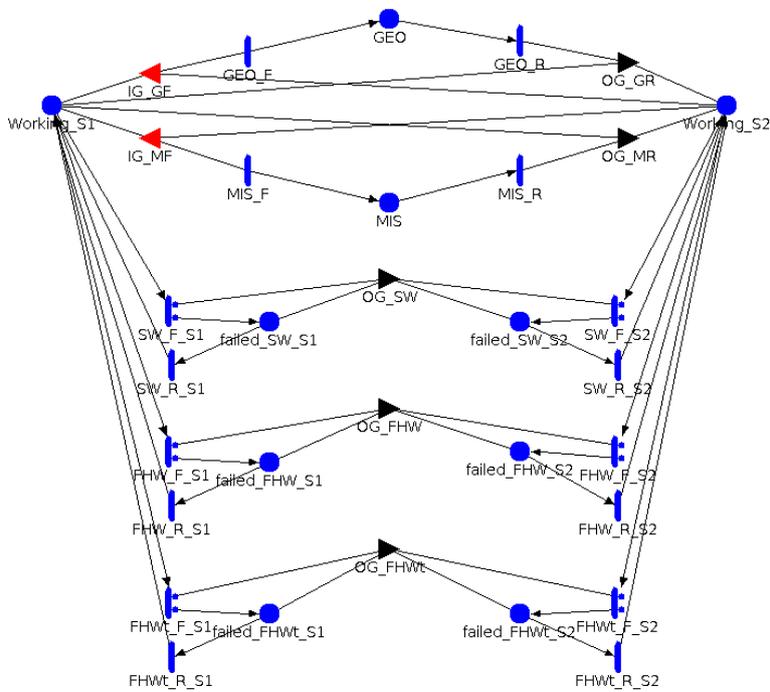

Fig. 8: SAN model of $\{n, n\}$ in F-SDN

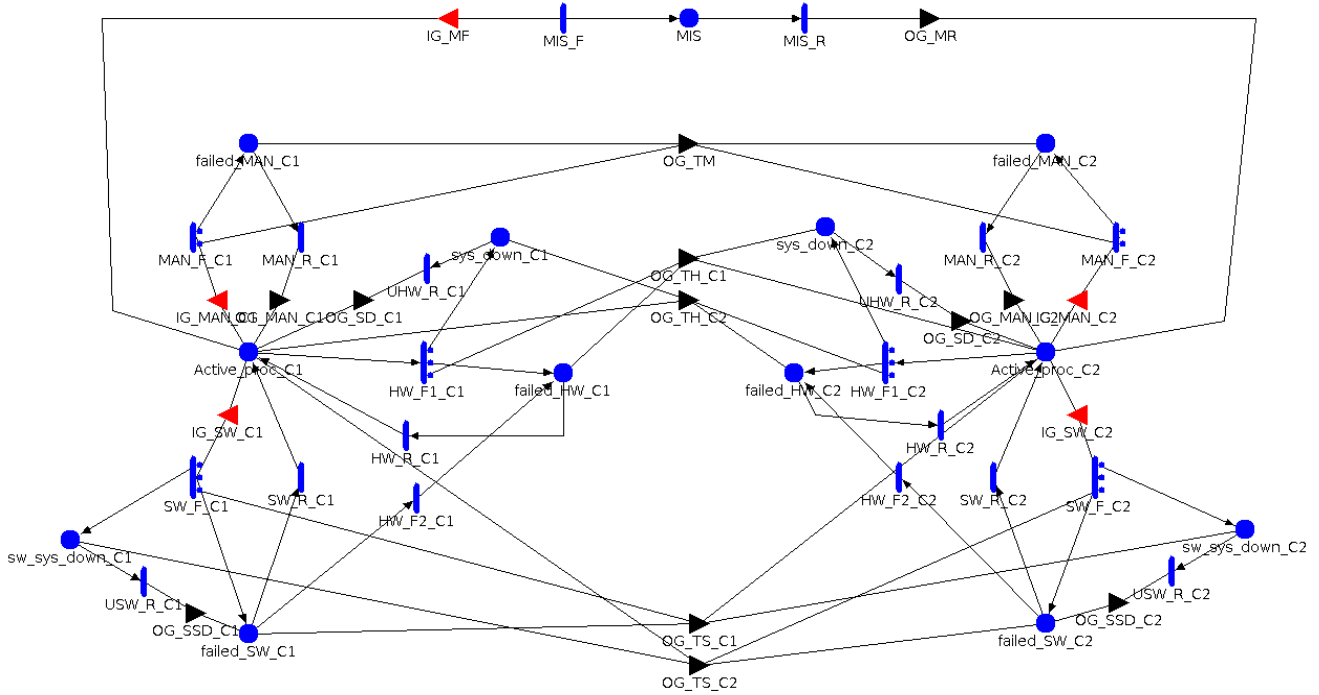

Fig. 9: SAN model of $\{n, n\}$ in C-SDN

### C. $\{n, n\}$ in C-SDN

Figure 9 depicts the SAN model of $\{n, n\}$ in C-SDN, where the two SDN controllers share a common configuration (*MIS*) and if one fails the other one takes over the control (*TMI*). The SAN model is composed of the SAN of the two SDN controllers (*_C1* and *_C2*), where the *MIS* place has been added and it is by mean of *MIS_F* and *MIS_R* timed activities, which represent the failure and the recovery of the common O&M with a rate of $\lambda_{MIS}$ and $\mu_{MIS}$, respectively.

Moreover, similarly to the two routers and two switches cases, several time activities of the two controllers are modified for considering the *TMI* failure. In particular, the *MAN_F*, *HW_F1*, *FHWt_F*, *CHW_F2* time activities of the two controllers are modified by adding one case: if *sys_down*, *sw_sys_down*, and *failed_MAN* of the other router have zero token and if *Active_proc* of the addressed router is equal to $K$, i.e. the addressed router is not able to fulfil the demand and the other router is working, then with probability $C_{TMI}$ only one router is failing and instead with probability $1 - C_{TMI}$ both the routers are failing.

Furthermore, the following output gates are included:

- *OG_TM* sets *failed_MAN* of both controllers;
- *OG_TH_C1* (and *OG_TH_C2*) sets *sys_down_C1* (or *sys_down_C2*), decreases the tokens in *Active_proc_C2* (or *Active_proc_C1*) and increases the tokens in *failed_HW_C1* (or *failed_HW_C2*);
- *OG_TS_C1* (and *OG_TS_C2*) similarly sets *sys_down_C1* (or *sys_down_C2*), decreases the tokens in

*Active_proc_C2* (or *Active_proc_C1*) and increases the tokens in *failed_SW_C1* (or *failed_SW_C2*).

The minimal-cut set is unavailable when both the SDN controllers are "singularly" failed or there is a token in *MIS* place. A SDN controller is "singularly" failed when $Active\_proc < K$ or there is a token in one of these places: *failed_MAN*, *sys_down*, *sw_sys_down*.

Further details on the implementation in Möbius of the SAN model and the related simulation can be found in the Appendix A1 and B1, respectively.

### D. $\{n, n, n\}$ in TN

Figure 10 shows the SAN model of $\{n, n, n\}$ in TN, where the three routers have both HW and SW homogeneous equipment (*HEQ*). Similarly as for the TMI in the two routers case, time activities are modified and output gates are added for considering the HEQ failure.

The minimal-cut set is unavailable when there are not token in *Working_S1*, *Working_S2*, *Working_S3*, *spare_CHW_S1*, *spare_CHW_S2*, and *spare_CHW_S3*.

Further details on the implementation in Möbius of the SAN model and the related simulation can be found in the Appendix A8 and B7, respectively.

### E. $\{n, n, n\}$ in F-SDN

Figure 11 depicts the SAN model of $\{n, n, n\}$ in F-SDN where the three SDN switches have mainly HW homogeneous equipment (*HEQ*). The SAN model is similar to the one for the three routers (see Figure 10): there is not the part related

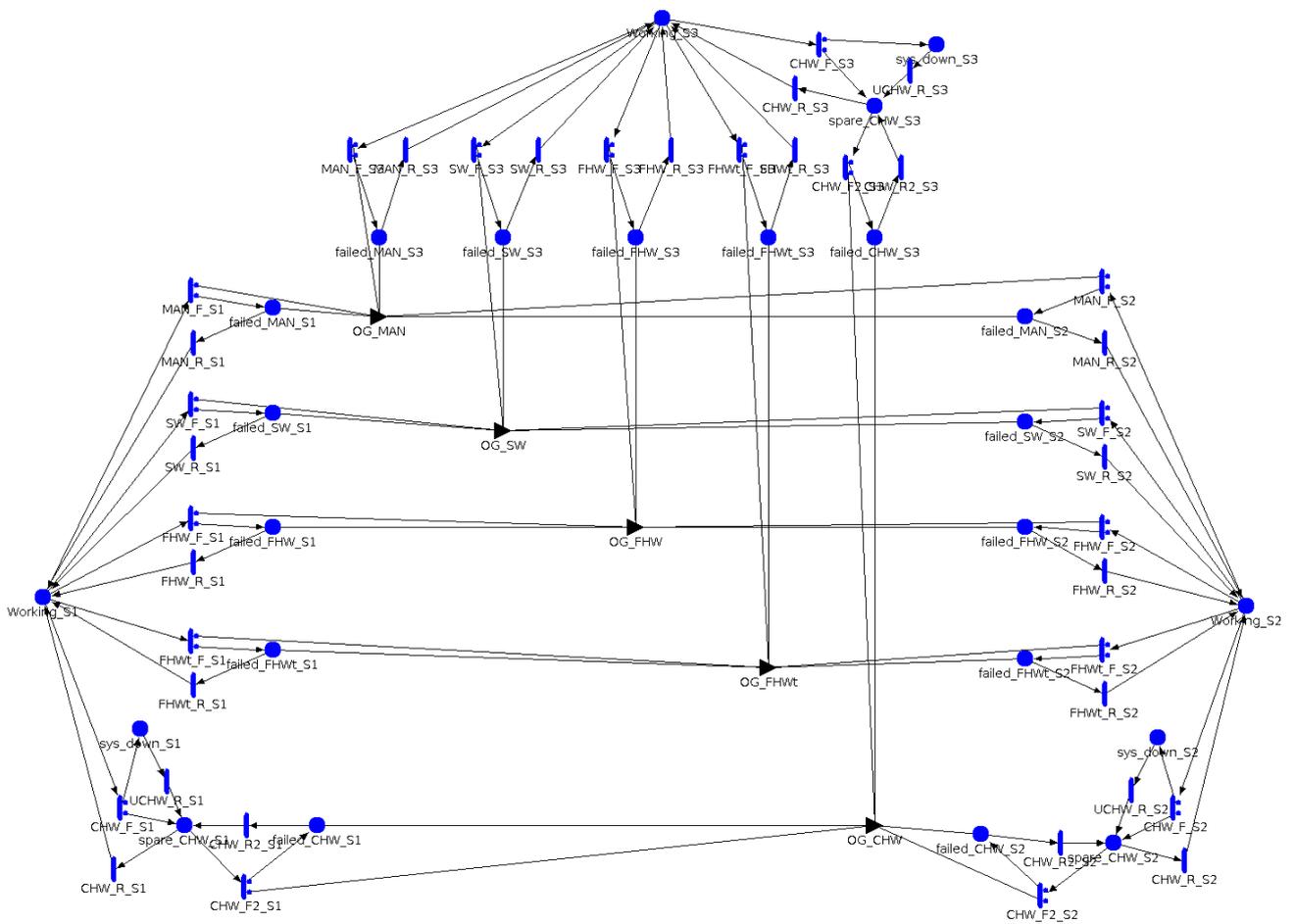

Fig. 10: SAN model of $\{n, n, n\}$ in TN

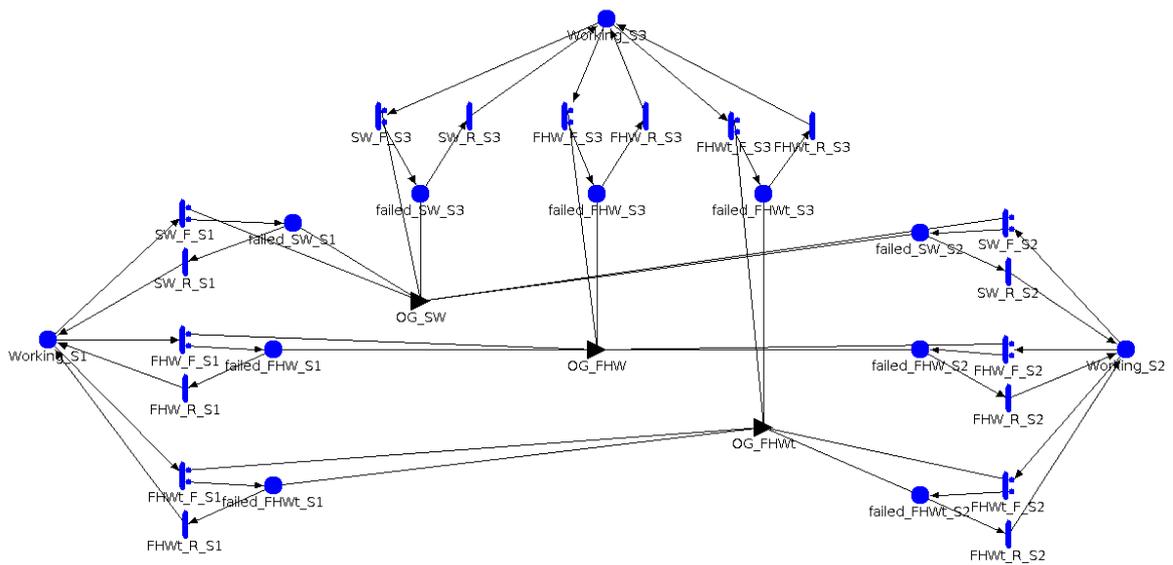

Fig. 11: SAN model of $\{n, n, n\}$ in F-SDN

to the control hardware and there is the MIS failure instead of the O&M failure.

The minimal-cut set is unavailable when there are not token in *Working_S1*, *Working_S2*, and *Working_S3*.

Further details on the implementation in Möbius of the SAN model and the related simulation can be found in the Appendix A12 and B11, respectively.

### F. {n, n, n} in C-SDN

Figure 12 depicts the SAN model of {n, n, n} in C-SDN, where the SDN switches are in the same city (*GEO*), instead the controller and the switches can have compatibility issues (CIS). The GEO failure is included as in the two router case (see Figure 8). For the CIS failure, the following places (with the related timed activities and output gates) are added:

- *CIS* that assesses the CIS between the SDN controller and both the switches;
- *CIS_S1* and *CIS_S2* consider the CIS between the SDN controller and the single switch (S1 and S2, respectively).

The minimal-cut set is unavailable when both SDN switches are failed and the SDN controller is "singularly" failed or there is a token in the CIS places.

Further details on the implementation in Möbius of the SAN model and the related simulation can be found in the Appendix A3 and B3, respectively.

### G. {n, n, l} in TN

Figure 13 shows the SAN model of SAN model of {n, n, l} in TN, where one router and the link are in the same city (*GEO*) and the two routers have homogeneous equipment (*HEQ*). The HEQ failure is added as in the case of only two routers (see Figure 7), nut note that in this case there is the O&M failure places are not merged because there is not COM failure here. The GEO place (with the related timed activities and output gates) is added between the working places of the link and one of the SDN switches.

The minimal-cut set is unavailable when there are not token in *Working_S1*, *Working_S2*, *Working_L*, *spare_CHW_S1*, and *spare_CHW_S2*.

Further details on the implementation in Möbius of the SAN model and the related simulation can be found in the Appendix A7 and B6, respectively.

### H. {n, n, l} in F-SDN

Figure 14 depicts the SAN model of {n, n, l} in F-SDN, where one SDN switch and the link are in the same city (*GEO*) and the two SDN switches have homogeneous equipment (*HEQ*). The SAN model is similar to the one for the three routers: there is not the part related to the control hardware and the O&M failures.

The minimal-cut set is unavailable when there are not token in *Working_S1*, *Working_S2*, and *Working_L*.

Further details on the implementation in Möbius of the SAN model and the related simulation can be found in the Appendix A11 and B10, respectively.

### I. {n, n, l} in C-SDN

Figure 15 depicts the SAN model of {n, n, l} in C-SDN, where the SDN switch and the link are in the same city (*GEO*), instead the controller and the switch can have compatibility issues (CIS). The GEO failure is similar to the one of the two switches and the link (see Figure 14).The CIS failure is similar to the one of the two switches and the controller (see Figure 12).

The minimal-cut set is unavailable when both SDN switch and link are failed and the SDN controller is "singularly" failed or there is a token the CIS place.

Further details on the implementation in Möbius of the SAN model and the related simulation can be found in the Appendix A2 and B2, respectively.

### J. {n, l, l} in TN

Figure 16 shows the SAN model of {n, l, l} in TN, where the two links are connected to the same router (*PHY*) and the router and the two links are in the same city (*GEO*). The PHY place (with the related timed activities and output gates) is added between the working place of the links. Instead, the GEO place is connected to the working places of each network element, i.e.the links and the SDN switch.

The minimal-cut set is unavailable when there are not token in *Working_L1*, *Working_L2*, *Working_R*, and *spare_CHW*.

Further details on the implementation in Möbius of the SAN model and the related simulation can be found in the Appendix A5 and B5, respectively.

### K. {n, l, l} in F-SDN

Figure 17 shows the SAN model of {n, l, l} in F-SDN, where the two links are connected to the same SDN switch (*PHY*) and the SDN switch and the two links are in the same city (*GEO*). As in the previous cases, the SAN model is similar to the one for the router and the two links: there is not the part related to the control hardware and the O&M failures.

The minimal-cut set is unavailable when there are not token in *Working_L1*, *Working_L2*, and *Working_S*.

Further details on the implementation in Möbius of the SAN model and the related simulation can be found in the Appendix A9 and B9, respectively.

### L. {n, l, l} in C-SDN

Figure 18 shows the SAN model of {l, l} in C-SDN, where the two links are connected to the same SDN switch (*GEO*, *PHY*). The SDN controller is independent.

The two links are unavailable when there are not token in *Working_L1* and *Working_L2*. The unavailability of the minimal-cut set is the multiplication of the unavailability of the two links and the unavailability of the SDN controller.

Further details on the implementation in Möbius of the SAN models and the related simulations can be found in the Appendix A4 and B4, respectively.

Fig. 12: SAN model of $\{n, n, n\}$ in C-SDN

## V. Conclusion

The technical report has detailed presented of the implementation on Möbius of the SAN availability model of both the network elements and the principal minimal-cut sets. The models of principal minimal-cut sets have been have been used in [5].

## Appendix
## Möbius Documentation

In the following appendix, the Möbius documentation of the SAN model and the simulation for the principal minimal-cut sets is introduced by indicating the pages of the attached document.

### A. Documentation on SAN models

Firstly, we introduce the documentation on the implementation in Möbius of the SAN model of the principal minimal-cut sets.

*1)* $\{n, n\}$ *in C-SDN:* Form page A-1 to page A-5.

*2)* $\{n, n, l\}$ *in C-SDN:* Form page A-5 to page A-9.

*3)* $\{n, n, n\}$ *in C-SDN:* Form page A-9 to page A-13.

*4)* $\{n, l, l\}$ *in C-SDN:* Form page A-13 to page A-14 the model of the two links and form page A-55 to page A-57 the model of the SDN controller.

*5)* $\{n, l, l\}$ *in TN:* Form page A-14 to page A-17.

*6)* $\{n, n\}$ *in TN:* Form page A-17 to page A-22.

*7)* $\{n, n, l\}$ *in TN:* Form page A-22 to page A-28.

*8)* $\{n, n, n\}$ *in TN:* Form page A-28 to page A-35.

*9)* $\{n, l, l\}$ *in F-SDN:* Form page A-35 to page A-37.

*10)* $\{n, n\}$ *in F-SDN:* Form page A-37 to page A-40.

*11)* $\{n, n, l\}$ *in F-SDN:* Form page A-40 to page A-43.

*12)* $\{n, n, n\}$ *in F-SDN:* Form page A-43 to page A-47.

### B. Documentation on simulations

Secondly, we introduce the documentation on the simulation (reward and study) in Möbius of the SAN model of the principal minimal-cut sets.

*1)* $\{n, n\}$ *in C-SDN:* In page A-48.

*2)* $\{n, n, l\}$ *in C-SDN:* Form page A-48 to page A-49.

*3)* $\{n, n, n\}$ *in C-SDN:* Form page A-49 to page A-50.

*4)* $\{n, l, l\}$ *in C-SDN:* Form page A-50 to page A-51 the model of the SDN controller and form page A-51 to page A-52 the model of the two links.

*5)* $\{n, l, l\}$ *in TN:* Form page A-52 to page A-53.

*6)* $\{n, n, l\}$ *in TN:* In page A-53.

*7)* $\{n, n, n\}$ *in TN:* In page A-54.

*8)* $\{n, n\}$ *in TN:* Form page A-54 to page A-55.

*9)* $\{n, l, l\}$ *in F-SDN:* Form page A-57 to page A-58.

*10)* $\{n, n, l\}$ *in F-SDN:* Form page A-58 to page A-59.

*11)* $\{n, n, n\}$ *in F-SDN:* In page A-59.

*12)* $\{n, n\}$ *in F-SDN:* Form page A-59 to page A-60.

## References

[1] E. Haleplidis, K. Pentikousis, S. Denazis, J. H. Salim, D. Meyer, and O. Koufopavlou, "Software-defined networking (SDN): Layers and architecture terminology," Internet Research Task Force (IRTF), Request for Comments RFC 7426, January 2015.

[2] D. Kreutz, F. M. V. Ramos, P. J. E. Veríssimo, C. E. Rothenberg, S. Azodolmolky, and S. Uhlig, "Software-defined networking: A comprehensive survey," *Proceedings of the IEEE*, vol. 103, no. 1, pp. 14–76, 2015.

[3] B. Nunes, M. Mendonca, X.-N. Nguyen, K. Obraczka, and T. Turletti, "A survey of software-defined networking: Past, present, and future of programmable networks," *Communications Surveys Tutorials, IEEE*, vol. 16, no. 3, pp. 1617–1634, Third 2014.

[4] P. E. Heegaard, V. B. Mendiratta, and B. E. Helvik., "Achieving dependability in software-defined networking - a perspective," in *7th International Workshop on Reliable Networks Design and Modeling (RNDM)*, Munich, Germany, October 2015.

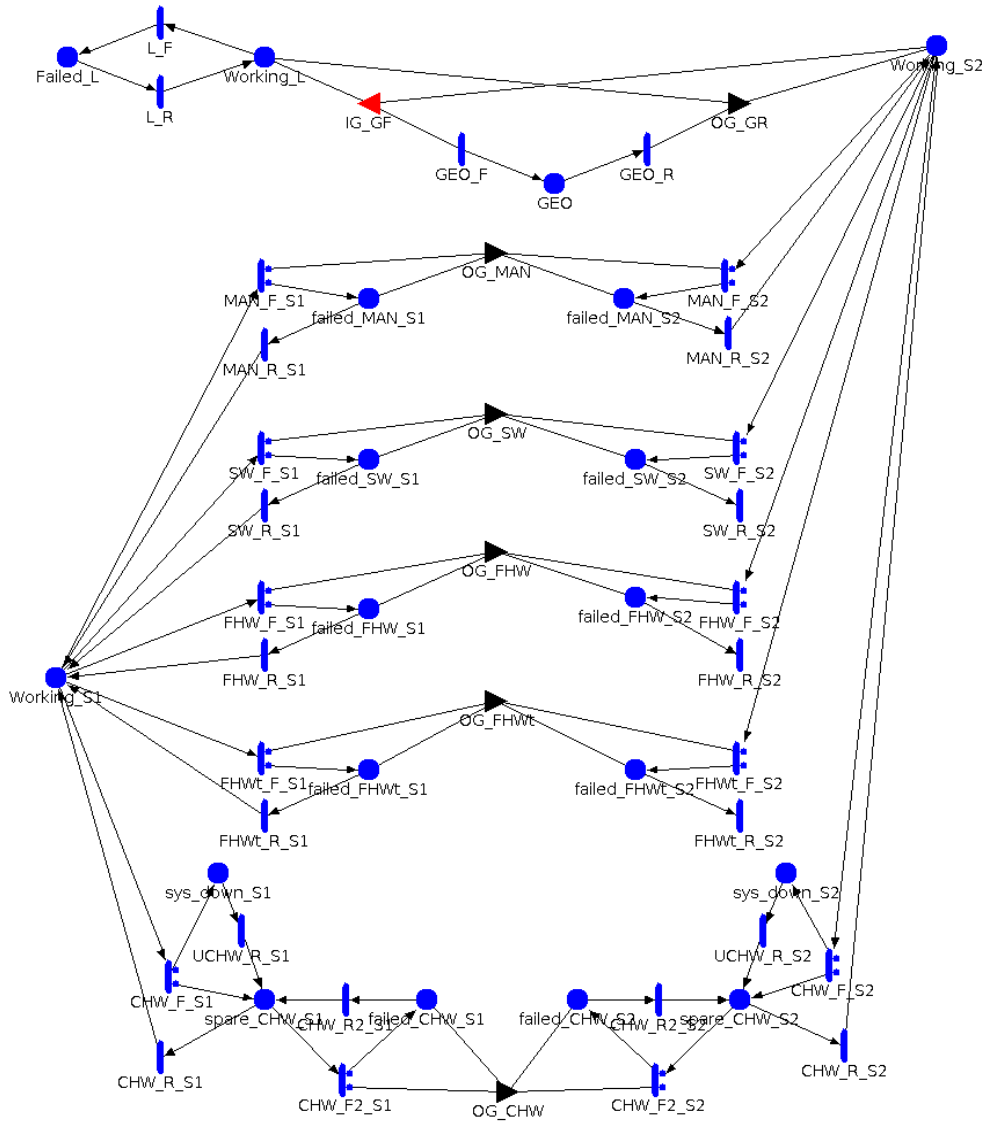

Fig. 13: SAN model of $\{n, n, l\}$ in TN

[5] G. Nencioni, B. E. Helvik, and P. E. Heegaard, "Including Failure Correlation in Availability Modelling of a Software-Defined Backbone Network," *submitted to IEEE TNSM Special Issue on "Advances in Management of Softwarized Networks"*.

[6] G. Ciardo and K. S. Trivedi, "A decomposition approach for stochastic reward net models," *Perf. Eval.*, vol. 18, pp. 37–59, 1993.

[7] R. E. Barlow and F. Proschan, *Statistical Theory of Reliability and Life Testing: Probability Models*. To BEGIN WITH, 1975.

[8] K. Kanoun, M. Borrel, T. Morteveille, and A. Peytavin, "Availability of CAUTRA, a Subset of the French Air Traffic Control System." *IEEE Trans. Computers*, no. 5, pp. 528–535.

[9] A. J. Gonzalez and B. E. Helvik, "Characterization of router and link failure processes in UNINETT's IP backbone network," *International Journal of Space-Based and Situated Computing*, 2012.

[10] P. Kuusela and I. Norros, "On/off process modeling of ip network failures," in *Dependable Systems and Networks (DSN), 2010 IEEE/IFIP International Conference on*, June 2010, pp. 585–594.

[11] S. Verbrugge, D. Colle, P. Demeester, R. Huelsermann, and M. Jaeger, "General availability model for multilayer transport networks," in *Proceedings.5th International Workshop on Design of Reliable Communi-*

*cation Networks, 2005. (DRCN 2005)*. IEEE, October 16-19 2005, pp. 85 – 92.

[12] G. Nencioni, B. E. Helvik, A. J. Gonzalez, P. E. Heegaard, and A. Kamisinski, "Availability modelling of software-defined backbone networks," in *2016 46th Annual IEEE/IFIP International Conference on Dependable Systems and Networks Workshop (DSN-W)*, June 2016, pp. 105–112.

[13] A. J. Gonzalez, B. E. Helvik, J. K. Hellan, and P. Kuusela, "Analysis of dependencies between failures in the uninett ip backbone network," in *2010 IEEE 16th Pacific Rim International Symposium on Dependable Computing*, Dec 2010, pp. 149–156.

[14] "Möbius: Model-based environment for validation of system reliability, availability, security, and performance," https://www.mobius.illinois.edu/, accessed: 2017-03-02.

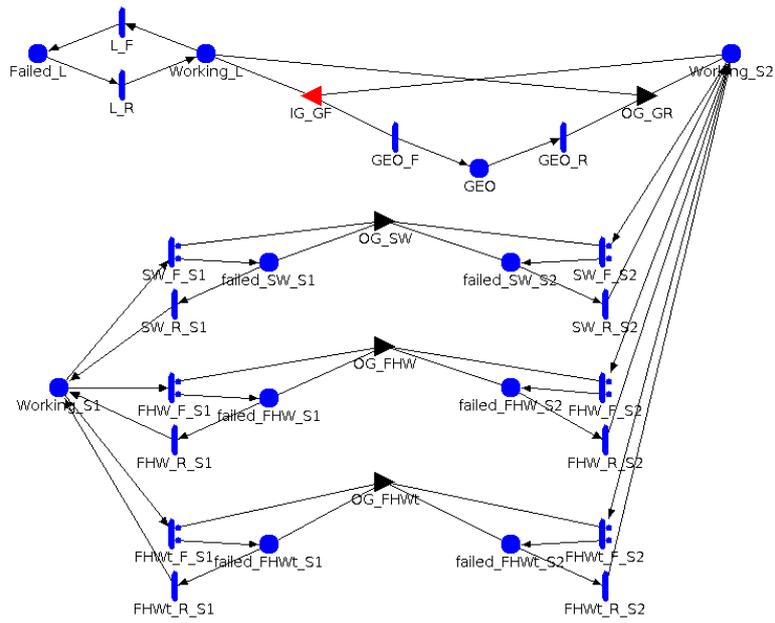

Fig. 14: SAN model of $\{n, n, l\}$ in F-SDN

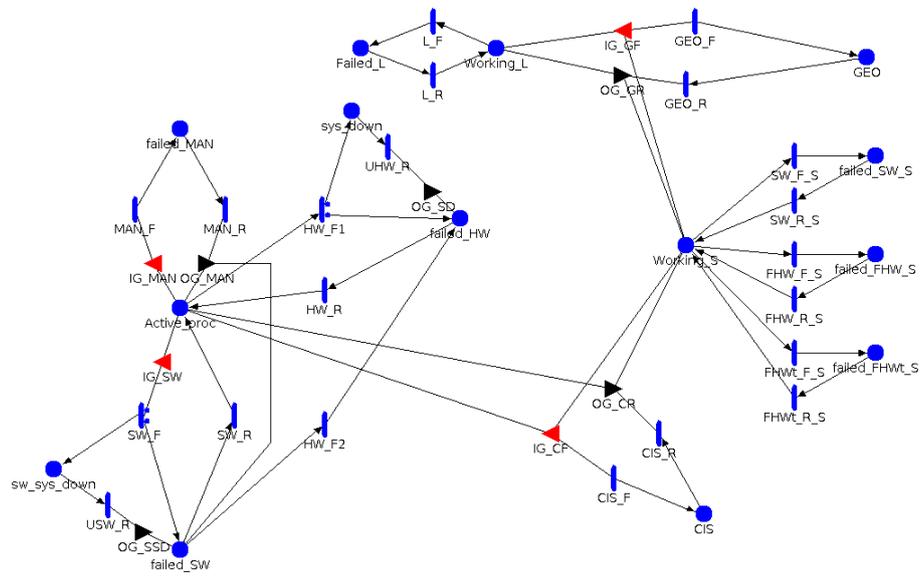

Fig. 15: SAN model of $\{n, n, l\}$ in C-SDN

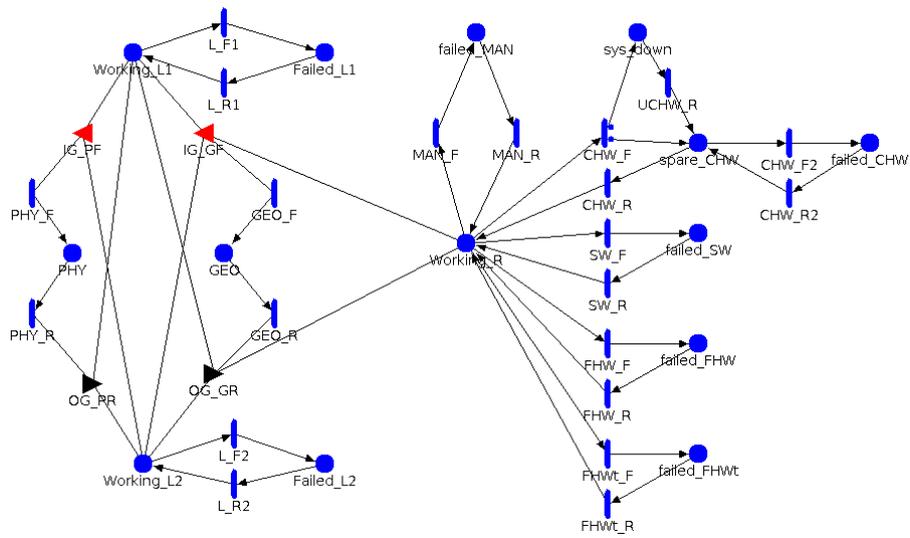

Fig. 16: SAN model of $\{n, l, l\}$ in TN

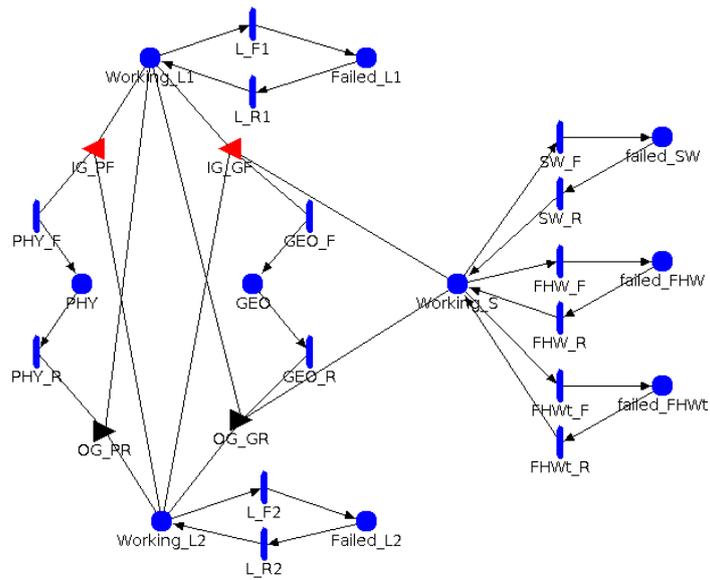

Fig. 17: SAN model of $\{n, l, l\}$ in F-SDN

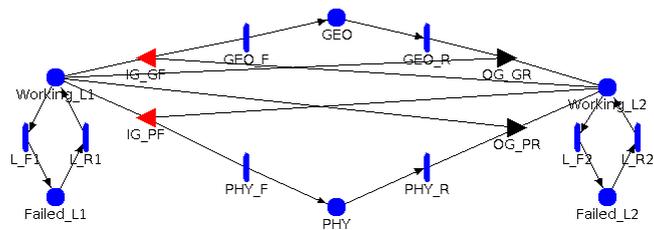

Fig. 18: SAN model of $\{l, l\}$ in C-SDN

**Model: cc**

**Place Attributes**:

| Place Names | Initial Markings |
|---|---|
| Active_proc_C1 | N_proc |
| Active_proc_C2 | N_proc |
| MIS | 0 |
| failed_HW_C1 | 0 |
| failed_HW_C2 | 0 |
| failed_MAN_C1 | 0 |
| failed_MAN_C2 | 0 |
| failed_SW_C1 | 0 |
| failed_SW_C2 | 0 |
| sw_sys_down_C1 | 0 |
| sw_sys_down_C2 | 0 |
| sys_down_C1 | 0 |
| sys_down_C2 | 0 |

| Timed Activity: | **HW_F1_C1** |
|---|---|
| **Distribution Parameters** | Rate<br><br>Active_proc_C1->Mark() * hw_fail_rate |
| **Activation Predicate** | (none) |
| **Reactivation Predicate** | (none) |
| **Case Distributions** | case 1<br><br>if (MIS->Mark() == 0 && sys_down_C1->Mark() == 0 && sw_sys_down_C1->Mark() == 0 && failed_MAN_C1->Mark() == 0)<br>        return(1-hw_cvg);<br>else<br>        return(0);<br><br>case 2<br><br>if (MIS->Mark() == 0 && sys_down_C1->Mark() == 0 && sw_sys_down_C1->Mark() == 0 && failed_MAN_C1->Mark() == 0)<br>{<br>        if (sys_down_C2->Mark()==0 && sw_sys_down_C2->Mark()==0 && failed_MAN_C2->Mark()==0 && Active_proc_C1->Mark()==K_th)<br>                return(hw_cvg*tmi_cvg);<br>        else<br>                return(hw_cvg);<br>}<br>else<br>        return(1);<br><br>case 3<br><br>if (MIS->Mark() == 0 && sys_down_C1->Mark() == 0 && sw_sys_down_C1->Mark() == 0 && failed_MAN_C1->Mark() == 0)<br>{<br>        if (sys_down_C2->Mark()==0 && sw_sys_down_C2->Mark()==0 && failed_MAN_C2->Mark()==0 && Active_proc_C1->Mark()==K_th)<br>                return(hw_cvg*(1-tmi_cvg));<br>        else<br>                return(0);<br>}<br>else<br>        return(0); |

| Timed Activity: | **HW_F1_C2** |
|---|---|
| **Distribution Parameters** | Rate<br><br>Active_proc_C2->Mark() * hw_fail_rate |
| **Activation Predicate** | (none) |
| **Reactivation Predicate** | (none) |
| **Case Distributions** | case 1<br><br>if (MIS->Mark() == 0 && sys_down_C2->Mark() == 0 && sw_sys_down_C2->Mark() == 0 && failed_MAN_C2->Mark() == 0)<br>        return(0);<br>else<br>        return(1-hw_cvg);<br><br>case 2<br><br>if (MIS->Mark() == 0 && sys_down_C2->Mark() == 0 && sw_sys_down_C2->Mark() == 0 && failed_MAN_C2->Mark() == 0)<br>{<br>        if (sys_down_C2->Mark()==0 && sw_sys_down_C1->Mark()==0 && failed_MAN_C1->Mark()==0 && Active_proc_C2->Mark()==K_th)<br>                return(hw_cvg*tmi_cvg);<br>        else<br>                return(hw_cvg);<br>}<br>else<br>        return(1); |

```
case 3

if (MIS->Mark() == 0 && sys_down_C2->Mark() == 0 && sw_sys_down_C2->Mark() == 0 && failed_MAN_C2->Mark() == 0)
{
        if (sys_down_C1->Mark()==0 && sw_sys_down_C1->Mark()==0 && failed_MAN_C1->Mark()==0 && Active_proc_C2->Mark()==K_th)
                return(hw_cvg*(1-tmi_cvg));
        else
                return(0);
}
else
        return(0);
```

| Timed Activity: | HW_F2_C1 |
|---|---|
| Distribution Parameters | **Rate**<br><br>hw_fail_rate * failed_SW_C1->Mark() |
| Activation Predicate | (none) |
| Reactivation Predicate | (none) |

| Timed Activity: | HW_F2_C2 |
|---|---|
| Distribution Parameters | **Rate**<br><br>hw_fail_rate * failed_SW_C2->Mark() |
| Activation Predicate | (none) |
| Reactivation Predicate | (none) |

| Timed Activity: | HW_R_C1 |
|---|---|
| Distribution Parameters | **Rate**<br><br>hw_rcv_rate |
| Activation Predicate | (none) |
| Reactivation Predicate | (none) |

| Timed Activity: | HW_R_C2 |
|---|---|
| Distribution Parameters | **Rate**<br><br>hw_rcv_rate |
| Activation Predicate | (none) |
| Reactivation Predicate | (none) |

| Timed Activity: | MAN_F_C1 |
|---|---|
| Distribution Parameters | **Rate**<br><br>man_fail_rate |
| Activation Predicate | (none) |
| Reactivation Predicate | (none) |
| Case Distributions | **case 1**<br><br>if(failed_MAN_C2->Mark() == 0 && sys_down_C2->Mark() == 0 && sw_sys_down_C2->Mark() == 0)<br>    return(tmi_cvg);<br>else<br>    return(1);<br>**case 2**<br><br>if(failed_MAN_C2->Mark() == 0 && sys_down_C2->Mark() == 0 && sw_sys_down_C2->Mark() == 0)<br>    return(1-tmi_cvg);<br>else<br>    return(0); |

| Timed Activity: | MAN_F_C2 |
|---|---|
| Distribution Parameters | **Rate**<br><br>man_fail_rate |
| Activation Predicate | (none) |
| Reactivation Predicate | (none) |
| Case Distributions | **case 1**<br><br>if(failed_MAN_C1->Mark() == 0 && sys_down_C1->Mark() == 0 && sw_sys_down_C1->Mark() == 0)<br>    return(tmi_cvg);<br>else<br>    return(1);<br>**case 2**<br><br>if(failed_MAN_C1->Mark() == 0 && sys_down_C1->Mark() == 0 && sw_sys_down_C1->Mark() == 0)<br>    return(1-tmi_cvg);<br>else<br>    return(0); |

| Timed Activity: | MAN_R_C1 |
|---|---|
| Distribution Parameters | **Rate**<br><br>man_rcv_rate |
| Activation Predicate | (none) |
| Reactivation Predicate | (none) |

| Timed Activity: | MAN_R_C2 |
|---|---|
| Distribution Parameters | **Rate**<br><br>man_rcv_rate |
| Activation Predicate | (none) |
| Reactivation Predicate | (none) |

| Timed Activity: | MIS_F |
|---|---|
| Distribution Parameters | **Rate**<br><br>mis_fail_rate |
| Activation Predicate | (none) |
| Reactivation Predicate | (none) |

| Timed Activity: | MIS_R |
|---|---|
| Distribution Parameters | **Rate**<br><br>mis_rcv_rate |
| Activation Predicate | (none) |
| Reactivation Predicate | (none) |

| Timed Activity: | SW_F_C1 |
|---|---|
| Distribution Parameters | **Rate**<br><br>`if(Active_proc_C1->Mark() >= K_th)`<br>`        return(sw_fail_rate);`<br>`else`<br>`        return(sw_fail_rate * Active_proc_C1->Mark());` |
| Activation Predicate | (none) |
| Reactivation Predicate | (none) |
| Case Distributions | **case 1**<br><br>`1-sw_cvg`<br><br>**case 2**<br><br>`if (sys_down_C2->Mark()==0 && sw_sys_down_C2->Mark()==0 && failed_MAN_C2->Mark()==0 && Active_proc_C1->Mark()==K_th)`<br>`        return(sw_cvg*tmi_cvg);`<br>`else`<br>`        return(sw_cvg);`<br><br>**case 3**<br><br>`if (sys_down_C2->Mark()==0 && sw_sys_down_C2->Mark()==0 && failed_MAN_C2->Mark()==0 && Active_proc_C1->Mark()==K_th)`<br>`        return(sw_cvg*(1-tmi_cvg));`<br>`else`<br>`        return(0);` |

| Timed Activity: | SW_F_C2 |
|---|---|
| Distribution Parameters | **Rate**<br><br>`if(Active_proc_C2->Mark() >= K_th)`<br>`        return(sw_fail_rate);`<br>`else`<br>`        return(sw_fail_rate * Active_proc_C2->Mark());` |
| Activation Predicate | (none) |
| Reactivation Predicate | (none) |
| Case Distributions | **case 1**<br><br>`1-sw_cvg`<br><br>**case 2**<br><br>`if (sys_down_C1->Mark()==0 && sw_sys_down_C1->Mark()==0 && failed_MAN_C1->Mark()==0 && Active_proc_C2->Mark()==K_th)`<br>`        return(sw_cvg*tmi_cvg);`<br>`else`<br>`        return(sw_cvg);`<br><br>**case 3**<br><br>`if (sys_down_C1->Mark()==0 && sw_sys_down_C1->Mark()==0 && failed_MAN_C1->Mark()==0 && Active_proc_C2->Mark()==K_th)`<br>`        return(sw_cvg*(1-tmi_cvg));` |

```
else
        return(0);
```

| Timed Activity: | SW_R_C1 |
|---|---|
| **Distribution Parameters** | **Rate**<br><br>sw_rcv_rate |
| **Activation Predicate** | (none) |
| **Reactivation Predicate** | (none) |

| Timed Activity: | SW_R_C2 |
|---|---|
| **Distribution Parameters** | **Rate**<br><br>sw_rcv_rate |
| **Activation Predicate** | (none) |
| **Reactivation Predicate** | (none) |

| Timed Activity: | UHW_R_C1 |
|---|---|
| **Distribution Parameters** | **Rate**<br><br>uhw_rcv_rate |
| **Activation Predicate** | (none) |
| **Reactivation Predicate** | (none) |

| Timed Activity: | UHW_R_C2 |
|---|---|
| **Distribution Parameters** | **Rate**<br><br>uhw_rcv_rate |
| **Activation Predicate** | (none) |
| **Reactivation Predicate** | (none) |

| Timed Activity: | USW_R_C1 |
|---|---|
| **Distribution Parameters** | **Rate**<br><br>usw_rcv_rate |
| **Activation Predicate** | (none) |
| **Reactivation Predicate** | (none) |

| Timed Activity: | USW_R_C2 |
|---|---|
| **Distribution Parameters** | **Rate**<br><br>usw_rcv_rate |
| **Activation Predicate** | (none) |
| **Reactivation Predicate** | (none) |

| Input Gate: | IG_MAN_C1 |
|---|---|
| **Predicate** | (MIS->Mark() == 0 && failed_MAN_C1->Mark() == 0 && sys_down_C1->Mark() == 0 && sw_sys_down_C1->Mark() == 0) |
| **Function** | ; |

| Input Gate: | IG_MAN_C2 |
|---|---|
| **Predicate** | (MIS->Mark()==0 && failed_MAN_C2->Mark() == 0 && sys_down_C2->Mark() == 0 && sw_sys_down_C2->Mark() == 0) |
| **Function** | ; |

| Input Gate: | IG_MF |
|---|---|
| **Predicate** | (MIS->Mark() == 0 &&<br>failed_MAN_C1->Mark() == 0 && sys_down_C1->Mark() == 0 && sw_sys_down_C1->Mark() == 0 &&<br>failed_MAN_C2->Mark() == 0 && sys_down_C2->Mark() == 0 && sw_sys_down_C2->Mark() == 0) |
| **Function** | ; |

| Input Gate: | IG_SW_C1 |
|---|---|
| **Predicate** | (MIS->Mark() ==0 && failed_MAN_C1->Mark() ==0 && sys_down_C1->Mark() ==0 && sw_sys_down_C1->Mark() == 0 && Active_proc_C1->Mark() > 0) |
| **Function** | Active_proc_C1->Mark()--; |

| Input Gate: | IG_SW_C2 |
|---|---|
| **Predicate** | (failed_MAN_C2->Mark() ==0 && sys_down_C2->Mark() ==0 && sw_sys_down_C2->Mark() == 0 && Active_proc_C2->Mark() > 0) |

| Function | `Active_proc_C2->Mark()--;` |
|---|---|

| Output Gate: | **OG_MAN_C1** |
|---|---|
| Function | `Active_proc_C1->Mark() = N_proc - failed_HW_C1->Mark();`<br>`failed_SW_C1->Mark()=0;` |

| Output Gate: | **OG_MAN_C2** |
|---|---|
| Function | `Active_proc_C2->Mark() = N_proc - failed_HW_C2->Mark();`<br>`failed_SW_C2->Mark()=0;` |

| Output Gate: | **OG_MR** |
|---|---|
| Function | `;` |

| Output Gate: | **OG_SD_C1** |
|---|---|
| Function | `failed_HW_C1->Mark()++;`<br>`Active_proc_C1->Mark() = N_proc - failed_HW_C1->Mark();`<br>`failed_SW_C1->Mark()=0;` |

| Output Gate: | **OG_SD_C2** |
|---|---|
| Function | `failed_HW_C2->Mark()++;`<br>`Active_proc_C2->Mark() = N_proc - failed_HW_C2->Mark();`<br>`failed_SW_C2->Mark()=0;` |

| Output Gate: | **OG_SSD_C1** |
|---|---|
| Function | `Active_proc_C1->Mark() = N_proc - failed_HW_C1->Mark();`<br>`failed_SW_C1->Mark()=0;` |

| Output Gate: | **OG_SSD_C2** |
|---|---|
| Function | `Active_proc_C2->Mark() = N_proc - failed_HW_C2->Mark();`<br>`failed_SW_C2->Mark()=0;` |

| Output Gate: | **OG_TH_C1** |
|---|---|
| Function | `sys_down_C2->Mark()=1;`<br>`Active_proc_C2->Mark()--;`<br>`failed_HW_C1->Mark()++;` |

| Output Gate: | **OG_TH_C2** |
|---|---|
| Function | `sys_down_C1->Mark()=1;`<br>`Active_proc_C1->Mark()--;`<br>`failed_HW_C2->Mark()++;` |

| Output Gate: | **OG_TM** |
|---|---|
| Function | `failed_MAN_C1->Mark()=1;`<br>`failed_MAN_C2->Mark()=1;` |

| Output Gate: | **OG_TS_C1** |
|---|---|
| Function | `sw_sys_down_C2->Mark()=1;`<br>`Active_proc_C2->Mark()--;`<br>`failed_SW_C1->Mark()++;` |

| Output Gate: | **OG_TS_C2** |
|---|---|
| Function | `sw_sys_down_C1->Mark()=1;`<br>`Active_proc_C1->Mark()--;`<br>`failed_SW_C2->Mark()++;` |

## Model: csl

**Place Attributes**:

| Place Names | Initial Markings |
|---|---|
| Active_proc | N_proc |
| CIS | 0 |
| Failed_L | 0 |
| GEO | 0 |
| Working_L | 1 |
| Working_S | 1 |
| failed_FHW_S | 0 |
| failed_FHWt_S | 0 |
| failed_HW | 0 |
| failed_MAN | 0 |

| failed_SW | 0 |
|---|---|
| failed_SW_S | 0 |
| sw_sys_down | 0 |
| sys_down | 0 |

| Timed Activity: | CIS_F |
|---|---|
| Distribution Parameters | **Rate**<br><br>cis_fail_rate |
| Activation Predicate | (none) |
| Reactivation Predicate | (none) |

| Timed Activity: | CIS_R |
|---|---|
| Distribution Parameters | **Rate**<br><br>cis_rcv_rate |
| Activation Predicate | (none) |
| Reactivation Predicate | (none) |

| Timed Activity: | FHW_F_S |
|---|---|
| Distribution Parameters | **Rate**<br><br>fhw_fail_rate |
| Activation Predicate | (none) |
| Reactivation Predicate | (none) |

| Timed Activity: | FHW_R_S |
|---|---|
| Distribution Parameters | **Rate**<br><br>fhw_rcv_rate |
| Activation Predicate | (none) |
| Reactivation Predicate | (none) |

| Timed Activity: | FHWt_F_S |
|---|---|
| Distribution Parameters | **Rate**<br><br>fhwt_fail_rate |
| Activation Predicate | (none) |
| Reactivation Predicate | (none) |

| Timed Activity: | FHWt_R_S |
|---|---|
| Distribution Parameters | **Rate**<br><br>fhwt_rcv_rate |
| Activation Predicate | (none) |
| Reactivation Predicate | (none) |

| Timed Activity: | GEO_F |
|---|---|
| Distribution Parameters | **Rate**<br><br>geo_fail_rate |
| Activation Predicate | (none) |
| Reactivation Predicate | (none) |

| Timed Activity: | GEO_R |
|---|---|
| Distribution Parameters | **Rate**<br><br>geo_rcv_rate |
| Activation Predicate | (none) |
| Reactivation Predicate | (none) |

| Timed Activity: | HW_F1 |
|---|---|
| Distribution Parameters | **Rate**<br><br>Active_proc->Mark() * hw_fail_rate |
| Activation Predicate | (none) |
| Reactivation Predicate | (none) |
| Case Distributions | case 1 |

```
if (sys_down->Mark() == 0 && sw_sys_down->Mark() == 0 && failed_MAN->Mark() == 0)
        return(1-hw_cvg);
else
        return(0);
case 2

if (sys_down->Mark() == 0 && sw_sys_down->Mark() == 0 && failed_MAN->Mark() == 0)
        return(hw_cvg);
else
        return(1);
```

| Timed Activity: | HW_F2 |
|---|---|
| Distribution Parameters | **Rate**<br><br>hw_fail_rate * failed_SW->Mark() |
| Activation Predicate | (none) |
| Reactivation Predicate | (none) |

| Timed Activity: | HW_R |
|---|---|
| Distribution Parameters | **Rate**<br><br>hw_rcv_rate |
| Activation Predicate | (none) |
| Reactivation Predicate | (none) |

| Timed Activity: | L_F |
|---|---|
| Distribution Parameters | **Rate**<br><br>link_fail_rate |
| Activation Predicate | (none) |
| Reactivation Predicate | (none) |

| Timed Activity: | L_R |
|---|---|
| Distribution Parameters | **Rate**<br><br>link_rcv_rate |
| Activation Predicate | (none) |
| Reactivation Predicate | (none) |

| Timed Activity: | MAN_F |
|---|---|
| Distribution Parameters | **Rate**<br><br>man_fail_rate |
| Activation Predicate | (none) |
| Reactivation Predicate | (none) |

| Timed Activity: | MAN_R |
|---|---|
| Distribution Parameters | **Rate**<br><br>man_rcv_rate |
| Activation Predicate | (none) |
| Reactivation Predicate | (none) |

| Timed Activity: | SW_F |
|---|---|
| Distribution Parameters | **Rate**<br><br>if(Active_proc->Mark() >= K_th)<br>        return(csw_fail_rate);<br>else<br>        return(csw_fail_rate * Active_proc->Mark()); |
| Activation Predicate | (none) |
| Reactivation Predicate | (none) |
| Case Distributions | **case 1**<br><br>1-sw_cvg<br>**case 2**<br><br>sw_cvg |

| Timed Activity: | SW_F_S |
|---|---|
| Distribution Parameters | **Rate** |

| | sw_fail_rate |
|---|---|
| **Activation Predicate** | (none) |
| **Reactivation Predicate** | (none) |

| **Timed Activity:** | **SW_R** |
|---|---|
| **Distribution Parameters** | Rate<br><br>csw_rcv_rate |
| **Activation Predicate** | (none) |
| **Reactivation Predicate** | (none) |

| **Timed Activity:** | **SW_R_S** |
|---|---|
| **Distribution Parameters** | Rate<br><br>sw_rcv_rate |
| **Activation Predicate** | (none) |
| **Reactivation Predicate** | (none) |

| **Timed Activity:** | **UHW_R** |
|---|---|
| **Distribution Parameters** | Rate<br><br>uhw_rcv_rate |
| **Activation Predicate** | (none) |
| **Reactivation Predicate** | (none) |

| **Timed Activity:** | **USW_R** |
|---|---|
| **Distribution Parameters** | Rate<br><br>usw_rcv_rate |
| **Activation Predicate** | (none) |
| **Reactivation Predicate** | (none) |

| **Input Gate:** | **IG_CF** |
|---|---|
| **Predicate** | (CIS->Mark() == 0 && Working_S->Mark()==1 && failed_MAN->Mark() == 0 && sys_down->Mark() == 0 && sw_sys_down->Mark() == 0) |
| **Function** | Working_S->Mark()=0; |

| **Input Gate:** | **IG_GF** |
|---|---|
| **Predicate** | (Working_L->Mark()==1 && Working_S->Mark()==1) |
| **Function** | Working_L->Mark()=0;<br>Working_S->Mark()=0; |

| **Input Gate:** | **IG_MAN** |
|---|---|
| **Predicate** | (failed_MAN->Mark() == 0 && sys_down->Mark() == 0 && sw_sys_down->Mark() == 0) |
| **Function** | ; |

| **Input Gate:** | **IG_SW** |
|---|---|
| **Predicate** | (failed_MAN->Mark() ==0 && sys_down->Mark() ==0 && sw_sys_down->Mark() == 0 && Active_proc->Mark() > 0) |
| **Function** | Active_proc->Mark()--; |

| **Output Gate:** | **OG_CR** |
|---|---|
| **Function** | Working_S->Mark()=1; |

| **Output Gate:** | **OG_GR** |
|---|---|
| **Function** | Working_L->Mark()=1;<br>Working_S->Mark()=1; |

| **Output Gate:** | **OG_MAN** |
|---|---|
| **Function** | Active_proc->Mark() = N_proc - failed_HW->Mark();<br>failed_SW->Mark()=0; |

| **Output Gate:** | **OG_SD** |
|---|---|
| **Function** | failed_HW->Mark()++;<br>Active_proc->Mark() = N_proc - failed_HW->Mark();<br>failed_SW->Mark()=0; |

| Output Gate: | OG_SSD |
|---|---|
| Function | Active_proc->Mark() = N_proc - failed_HW->Mark();<br>failed_SW->Mark()=0; |

## Model: css

**Place Attributes:**

| Place Names | Initial Markings |
|---|---|
| Active_proc | N_proc |
| CIS | 0 |
| CIS_S1 | 0 |
| CIS_S2 | 0 |
| GEO | 0 |
| Working_S1 | 1 |
| Working_S2 | 1 |
| failed_FHW_S1 | 0 |
| failed_FHW_S2 | 0 |
| failed_FHWt_S1 | 0 |
| failed_FHWt_S2 | 0 |
| failed_HW | 0 |
| failed_MAN | 0 |
| failed_SW | 0 |
| failed_SW_S1 | 0 |
| failed_SW_S2 | 0 |
| sw_sys_down | 0 |
| sys_down | 0 |

| Timed Activity: | CIS_F |
|---|---|
| Distribution Parameters | **Rate**<br><br>cis_fail_rate |
| Activation Predicate | (none) |
| Reactivation Predicate | (none) |

| Timed Activity: | CIS_F_S1 |
|---|---|
| Distribution Parameters | **Rate**<br><br>cis_fail_rate |
| Activation Predicate | (none) |
| Reactivation Predicate | (none) |

| Timed Activity: | CIS_F_S2 |
|---|---|
| Distribution Parameters | **Rate**<br><br>cis_fail_rate |
| Activation Predicate | (none) |
| Reactivation Predicate | (none) |

| Timed Activity: | CIS_R |
|---|---|
| Distribution Parameters | **Rate**<br><br>cis_rcv_rate |
| Activation Predicate | (none) |
| Reactivation Predicate | (none) |

| Timed Activity: | CIS_R_S1 |
|---|---|
| Distribution Parameters | **Rate**<br><br>cis_rcv_rate |
| Activation Predicate | (none) |
| Reactivation Predicate | (none) |

| Timed Activity: | CIS_R_S2 |
|---|---|
| Distribution Parameters | **Rate**<br><br>cis_rcv_rate |
| Activation Predicate | (none) |

| Reactivation Predicate | (none) |
|---|---|

| Timed Activity: | FHW_F_S1 |
|---|---|
| **Distribution Parameters** | **Rate**<br><br>fhw_fail_rate |
| Activation Predicate | (none) |
| Reactivation Predicate | (none) |

| Timed Activity: | FHW_F_S2 |
|---|---|
| **Distribution Parameters** | **Rate**<br><br>fhw_fail_rate |
| Activation Predicate | (none) |
| Reactivation Predicate | (none) |

| Timed Activity: | FHW_R_S1 |
|---|---|
| **Distribution Parameters** | **Rate**<br><br>fhw_rcv_rate |
| Activation Predicate | (none) |
| Reactivation Predicate | (none) |

| Timed Activity: | FHW_R_S2 |
|---|---|
| **Distribution Parameters** | **Rate**<br><br>fhw_rcv_rate |
| Activation Predicate | (none) |
| Reactivation Predicate | (none) |

| Timed Activity: | FHWt_F_S1 |
|---|---|
| **Distribution Parameters** | **Rate**<br><br>fhwt_fail_rate |
| Activation Predicate | (none) |
| Reactivation Predicate | (none) |

| Timed Activity: | FHWt_F_S2 |
|---|---|
| **Distribution Parameters** | **Rate**<br><br>fhwt_fail_rate |
| Activation Predicate | (none) |
| Reactivation Predicate | (none) |

| Timed Activity: | FHWt_R_S1 |
|---|---|
| **Distribution Parameters** | **Rate**<br><br>fhwt_rcv_rate |
| Activation Predicate | (none) |
| Reactivation Predicate | (none) |

| Timed Activity: | FHWt_R_S2 |
|---|---|
| **Distribution Parameters** | **Rate**<br><br>fhwt_rcv_rate |
| Activation Predicate | (none) |
| Reactivation Predicate | (none) |

| Timed Activity: | GEO_F |
|---|---|
| **Distribution Parameters** | **Rate**<br><br>geo_fail_rate |
| Activation Predicate | (none) |
| Reactivation Predicate | (none) |

| Timed Activity: | GEO_R |
|---|---|
| **Distribution Parameters** | **Rate** |
| | geo_rcv_rate |
| **Activation Predicate** | (none) |
| **Reactivation Predicate** | (none) |

| Timed Activity: | HW_F1 |
|---|---|
| **Distribution Parameters** | **Rate**<br><br>Active_proc->Mark() * hw_fail_rate |
| **Activation Predicate** | (none) |
| **Reactivation Predicate** | (none) |
| **Case Distributions** | **case 1**<br><br>if (sys_down->Mark() == 0 && sw_sys_down->Mark() == 0 && failed_MAN->Mark() == 0)<br>     return(1-hw_cvg);<br>else<br>     return(0);<br><br>**case 2**<br><br>if (sys_down->Mark() == 0 && sw_sys_down->Mark() == 0 && failed_MAN->Mark() == 0)<br>     return(hw_cvg);<br>else<br>     return(1); |

| Timed Activity: | HW_F2 |
|---|---|
| **Distribution Parameters** | **Rate**<br><br>hw_fail_rate * failed_SW->Mark() |
| **Activation Predicate** | (none) |
| **Reactivation Predicate** | (none) |

| Timed Activity: | HW_R |
|---|---|
| **Distribution Parameters** | **Rate**<br><br>hw_rcv_rate |
| **Activation Predicate** | (none) |
| **Reactivation Predicate** | (none) |

| Timed Activity: | MAN_F |
|---|---|
| **Distribution Parameters** | **Rate**<br><br>man_fail_rate |
| **Activation Predicate** | (none) |
| **Reactivation Predicate** | (none) |

| Timed Activity: | MAN_R |
|---|---|
| **Distribution Parameters** | **Rate**<br><br>man_rcv_rate |
| **Activation Predicate** | (none) |
| **Reactivation Predicate** | (none) |

| Timed Activity: | SW_F |
|---|---|
| **Distribution Parameters** | **Rate**<br><br>if(Active_proc->Mark() >= K_th)<br>     return(csw_fail_rate);<br>else<br>     return(csw_fail_rate * Active_proc->Mark()); |
| **Activation Predicate** | (none) |
| **Reactivation Predicate** | (none) |
| **Case Distributions** | **case 1**<br><br>1-sw_cvg<br>**case 2**<br><br>sw_cvg |

| Timed Activity: | SW_F_S1 |
|---|---|
| **Distribution Parameters** | **Rate**<br><br>sw_fail_rate |

| Activation Predicate | (none) |
|---|---|
| Reactivation Predicate | (none) |

| Timed Activity: | SW_F_S2 |
|---|---|
| Distribution Parameters | **Rate**<br>sw_fail_rate |
| Activation Predicate | (none) |
| Reactivation Predicate | (none) |

| Timed Activity: | SW_R |
|---|---|
| Distribution Parameters | **Rate**<br>csw_rcv_rate |
| Activation Predicate | (none) |
| Reactivation Predicate | (none) |

| Timed Activity: | SW_R_S1 |
|---|---|
| Distribution Parameters | **Rate**<br>sw_rcv_rate |
| Activation Predicate | (none) |
| Reactivation Predicate | (none) |

| Timed Activity: | SW_R_S2 |
|---|---|
| Distribution Parameters | **Rate**<br>sw_rcv_rate |
| Activation Predicate | (none) |
| Reactivation Predicate | (none) |

| Timed Activity: | UHW_R |
|---|---|
| Distribution Parameters | **Rate**<br>uhw_rcv_rate |
| Activation Predicate | (none) |
| Reactivation Predicate | (none) |

| Timed Activity: | USW_R |
|---|---|
| Distribution Parameters | **Rate**<br>usw_rcv_rate |
| Activation Predicate | (none) |
| Reactivation Predicate | (none) |

| Input Gate: | IG_CF |
|---|---|
| Predicate | (Working_S1->Mark()==1 && Working_S2->Mark()==1 && failed_MAN->Mark() == 0 && sys_down->Mark() == 0 && sw_sys_down->Mark() == 0) |
| Function | Working_S1->Mark()=0;<br>Working_S2->Mark()=0; |

| Input Gate: | IG_CF_S1 |
|---|---|
| Predicate | (CIS_S2->Mark() == 0 && Working_S1->Mark()==1 && failed_MAN->Mark() == 0 && sys_down->Mark() == 0 && sw_sys_down->Mark() == 0) |
| Function | Working_S1->Mark()=0; |

| Input Gate: | IG_CF_S2 |
|---|---|
| Predicate | (CIS_S1->Mark()==0 && Working_S2->Mark()==1 && failed_MAN->Mark() == 0 && sys_down->Mark() == 0 && sw_sys_down->Mark() == 0) |
| Function | Working_S2->Mark()=0; |

| Input Gate: | IG_GF |
|---|---|
| Predicate | (Working_S1->Mark()==1 && Working_S2->Mark()==1) |
| Function | Working_S1->Mark()=0;<br>Working_S2->Mark()=0; |

| Input Gate: | IG_MAN |
|---|---|
| Predicate | (failed_MAN->Mark() == 0 && sys_down->Mark() == 0 && sw_sys_down->Mark() == 0) |

| Function | ; |
|---|---|

| Input Gate: | **IG_SW** |
|---|---|
| Predicate | (failed_MAN->Mark() ==0 && sys_down->Mark() ==0 && sw_sys_down->Mark() == 0 && Active_proc->Mark() > 0) |
| Function | Active_proc->Mark()--; |

| Output Gate: | **OG_CR** |
|---|---|
| Function | Working_S1->Mark()=1;<br>Working_S2->Mark()=1; |

| Output Gate: | **OG_CR_S1** |
|---|---|
| Function | Working_S1->Mark()=1; |

| Output Gate: | **OG_CR_S2** |
|---|---|
| Function | Working_S2->Mark()=1; |

| Output Gate: | **OG_GR** |
|---|---|
| Function | Working_S1->Mark()=1;<br>Working_S2->Mark()=1; |

| Output Gate: | **OG_MAN** |
|---|---|
| Function | Active_proc->Mark() = N_proc - failed_HW->Mark();<br>failed_SW->Mark()=0; |

| Output Gate: | **OG_SD** |
|---|---|
| Function | failed_HW->Mark()++;<br>Active_proc->Mark() = N_proc - failed_HW->Mark();<br>failed_SW->Mark()=0; |

| Output Gate: | **OG_SSD** |
|---|---|
| Function | Active_proc->Mark() = N_proc - failed_HW->Mark();<br>failed_SW->Mark()=0; |

## Model: ll

**Place Attributes**:

| Place Names | Initial Markings |
|---|---|
| Failed_L1 | 0 |
| Failed_L2 | 0 |
| GEO | 0 |
| PHY | 0 |
| Working_L1 | 1 |
| Working_L2 | 1 |

| Timed Activity: | **GEO_F** | |
|---|---|---|
| Distribution Parameters | Rate<br><br>geo_fail_rate | |
| Activation Predicate | | (none) |
| Reactivation Predicate | | (none) |

| Timed Activity: | **GEO_R** | |
|---|---|---|
| Distribution Parameters | Rate<br><br>geo_rcv_rate | |
| Activation Predicate | | (none) |
| Reactivation Predicate | | (none) |

| Timed Activity: | **L_F1** | |
|---|---|---|
| Distribution Parameters | Rate<br><br>link_fail_rate | |
| Activation Predicate | | (none) |
| Reactivation Predicate | | (none) |

| Timed Activity: | **L_F2** |
|---|---|

| Distribution Parameters | Rate<br><br>link_fail_rate |
|---|---|
| Activation Predicate | (none) |
| Reactivation Predicate | (none) |

| Timed Activity: | **L_R1** |
|---|---|
| Distribution Parameters | Rate<br><br>link_rcv_rate |
| Activation Predicate | (none) |
| Reactivation Predicate | (none) |

| Timed Activity: | **L_R2** |
|---|---|
| Distribution Parameters | Rate<br><br>link_rcv_rate |
| Activation Predicate | (none) |
| Reactivation Predicate | (none) |

| Timed Activity: | **PHY_F** |
|---|---|
| Distribution Parameters | Rate<br><br>phy_fail_rate |
| Activation Predicate | (none) |
| Reactivation Predicate | (none) |

| Timed Activity: | **PHY_R** |
|---|---|
| Distribution Parameters | Rate<br><br>phy_rcv_rate |
| Activation Predicate | (none) |
| Reactivation Predicate | (none) |

| Input Gate: | **IG_GF** |
|---|---|
| Predicate | (Working_L1->Mark()==1 && Working_L2->Mark()==1) |
| Function | Working_L1->Mark()=0;<br>Working_L2->Mark()=0; |

| Input Gate: | **IG_PF** |
|---|---|
| Predicate | (Working_L1->Mark()==1 && Working_L2->Mark()==1) |
| Function | Working_L1->Mark()=0;<br>Working_L2->Mark()=0; |

| Output Gate: | **OG_GR** |
|---|---|
| Function | Working_L1->Mark()=1;<br>Working_L2->Mark()=1; |

| Output Gate: | **OG_PR** |
|---|---|
| Function | Working_L1->Mark()=1;<br>Working_L2->Mark()=1; |

## Model: rll

**Place Attributes**:

| Place Names | Initial Markings |
|---|---|
| Failed_L1 | 0 |
| Failed_L2 | 0 |
| GEO | 0 |
| PHY | 0 |
| Working_L1 | 1 |
| Working_L2 | 1 |
| Working_R | 1 |
| failed_CHW | 0 |
| failed_FHW | 0 |
| failed_FHWt | 0 |

| failed_MAN | 0 |
|---|---|
| failed_SW | 0 |
| spare_CHW | 0 |
| sys_down | 0 |

| Timed Activity: | CHW_F |
|---|---|
| Distribution Parameters | Rate<br><br>2 * chw_fail_rate |
| Activation Predicate | (none) |
| Reactivation Predicate | (none) |
| Case Distributions | case 1<br><br>1-chw_cvg<br>case 2<br><br>chw_cvg |

| Timed Activity: | CHW_F2 |
|---|---|
| Distribution Parameters | Rate<br><br>chw_fail_rate |
| Activation Predicate | (none) |
| Reactivation Predicate | (none) |

| Timed Activity: | CHW_R |
|---|---|
| Distribution Parameters | Rate<br><br>chw_fail_rate |
| Activation Predicate | (none) |
| Reactivation Predicate | (none) |

| Timed Activity: | CHW_R2 |
|---|---|
| Distribution Parameters | Rate<br><br>chw_rcv_rate |
| Activation Predicate | (none) |
| Reactivation Predicate | (none) |

| Timed Activity: | FHW_F |
|---|---|
| Distribution Parameters | Rate<br><br>fhw_fail_rate |
| Activation Predicate | (none) |
| Reactivation Predicate | (none) |

| Timed Activity: | FHW_R |
|---|---|
| Distribution Parameters | Rate<br><br>fhw_rcv_rate |
| Activation Predicate | (none) |
| Reactivation Predicate | (none) |

| Timed Activity: | FHWt_F |
|---|---|
| Distribution Parameters | Rate<br><br>fhwt_fail_rate |
| Activation Predicate | (none) |
| Reactivation Predicate | (none) |

| Timed Activity: | FHWt_R |
|---|---|
| Distribution Parameters | Rate<br><br>fhwt_rcv_rate |
| Activation Predicate | (none) |
| Reactivation Predicate | (none) |

| Timed Activity: | GEO_F |
|---|---|

| Distribution Parameters | **Rate**<br><br>geo_fail_rate |
|---|---|
| **Activation Predicate** | (none) |
| **Reactivation Predicate** | (none) |

| **Timed Activity:** | **GEO_R** |
|---|---|
| **Distribution Parameters** | **Rate**<br><br>geo_rcv_rate |
| **Activation Predicate** | (none) |
| **Reactivation Predicate** | (none) |

| **Timed Activity:** | **L_F1** |
|---|---|
| **Distribution Parameters** | **Rate**<br><br>link_fail_rate |
| **Activation Predicate** | (none) |
| **Reactivation Predicate** | (none) |

| **Timed Activity:** | **L_F2** |
|---|---|
| **Distribution Parameters** | **Rate**<br><br>link_fail_rate |
| **Activation Predicate** | (none) |
| **Reactivation Predicate** | (none) |

| **Timed Activity:** | **L_R1** |
|---|---|
| **Distribution Parameters** | **Rate**<br><br>link_rcv_rate |
| **Activation Predicate** | (none) |
| **Reactivation Predicate** | (none) |

| **Timed Activity:** | **L_R2** |
|---|---|
| **Distribution Parameters** | **Rate**<br><br>link_rcv_rate |
| **Activation Predicate** | (none) |
| **Reactivation Predicate** | (none) |

| **Timed Activity:** | **MAN_F** |
|---|---|
| **Distribution Parameters** | **Rate**<br><br>man_fail_rate |
| **Activation Predicate** | (none) |
| **Reactivation Predicate** | (none) |

| **Timed Activity:** | **MAN_R** |
|---|---|
| **Distribution Parameters** | **Rate**<br><br>man_rcv_rate |
| **Activation Predicate** | (none) |
| **Reactivation Predicate** | (none) |

| **Timed Activity:** | **PHY_F** |
|---|---|
| **Distribution Parameters** | **Rate**<br><br>phy_fail_rate |
| **Activation Predicate** | (none) |
| **Reactivation Predicate** | (none) |

| **Timed Activity:** | **PHY_R** |
|---|---|
| **Distribution Parameters** | **Rate**<br><br>phy_rcv_rate |
| **Activation Predicate** | (none) |
| **Reactivation Predicate** | (none) |

| Timed Activity: | SW_F |
|---|---|
| **Distribution Parameters** | **Rate**<br><br>sw_fail_rate |
| **Activation Predicate** | (none) |
| **Reactivation Predicate** | (none) |

| Timed Activity: | SW_R |
|---|---|
| **Distribution Parameters** | **Rate**<br><br>sw_rcv_rate |
| **Activation Predicate** | (none) |
| **Reactivation Predicate** | (none) |

| Timed Activity: | UCHW_R |
|---|---|
| **Distribution Parameters** | **Rate**<br><br>uchw_rcv_rate |
| **Activation Predicate** | (none) |
| **Reactivation Predicate** | (none) |

| Input Gate: | IG_GF |
|---|---|
| **Predicate** | (Working_L1->Mark()==1 && Working_L2->Mark()==1 && Working_R->Mark()==1) |
| **Function** | Working_L1->Mark()=0;<br>Working_L2->Mark()=0;<br>Working_R->Mark()=0; |

| Input Gate: | IG_PF |
|---|---|
| **Predicate** | (Working_L1->Mark()==1 && Working_L2->Mark()==1) |
| **Function** | Working_L1->Mark()=0;<br>Working_L2->Mark()=0; |

| Output Gate: | OG_GR |
|---|---|
| **Function** | Working_L1->Mark()=1;<br>Working_L2->Mark()=1;<br>Working_R->Mark()=1; |

| Output Gate: | OG_PR |
|---|---|
| **Function** | Working_L1->Mark()=1;<br>Working_L2->Mark()=1; |

**Model: rr**

**Place Attributes**:

| Place Names | Initial Markings |
|---|---|
| Failed_MAN | 0 |
| GEO | 0 |
| Working_S1 | 1 |
| Working_S2 | 1 |
| failed_CHW_S1 | 0 |
| failed_CHW_S2 | 0 |
| failed_FHW_S1 | 0 |
| failed_FHW_S2 | 0 |
| failed_FHWt_S1 | 0 |
| failed_FHWt_S2 | 0 |
| failed_SW_S1 | 0 |
| failed_SW_S2 | 0 |
| spare_CHW_S1 | 0 |
| spare_CHW_S2 | 0 |
| sys_down_S1 | 0 |
| sys_down_S2 | 0 |

| Timed Activity: | CHW_F2_S1 |
|---|---|
| **Distribution Parameters** | **Rate**<br><br>chw_fail_rate |
| **Activation Predicate** | (none) |

| Reactivation Predicate | (none) |
|---|---|
| Case Distributions | **case 1**<br><br>```if (Working_S2->Mark() == 1)```<br>```        return(tmi_cvg);```<br>```else```<br>```        return(1);```<br>**case 2**<br><br>```if (Working_S2->Mark() == 1)```<br>```        return(1-tmi_cvg);```<br>```else```<br>```        return(0);``` |

| Timed Activity: | CHW_F2_S2 |
|---|---|
| Distribution Parameters | **Rate**<br><br>```chw_fail_rate``` |
| Activation Predicate | (none) |
| Reactivation Predicate | (none) |
| Case Distributions | **case 1**<br><br>```if (Working_S1->Mark() == 1)```<br>```        return(tmi_cvg);```<br>```else```<br>```        return(1);```<br>**case 2**<br><br>```if (Working_S1->Mark() == 1)```<br>```        return(1-tmi_cvg);```<br>```else```<br>```        return(0);``` |

| Timed Activity: | CHW_F_S1 |
|---|---|
| Distribution Parameters | **Rate**<br><br>```2 * chw_fail_rate``` |
| Activation Predicate | (none) |
| Reactivation Predicate | (none) |
| Case Distributions | **case 1**<br><br>```1-chw_cvg```<br>**case 2**<br><br>```chw_cvg``` |

| Timed Activity: | CHW_F_S2 |
|---|---|
| Distribution Parameters | **Rate**<br><br>```2 * chw_fail_rate``` |
| Activation Predicate | (none) |
| Reactivation Predicate | (none) |
| Case Distributions | **case 1**<br><br>```1-chw_cvg```<br>**case 2**<br><br>```chw_cvg``` |

| Timed Activity: | CHW_R2_S1 |
|---|---|
| Distribution Parameters | **Rate**<br><br>```chw_rcv_rate``` |
| Activation Predicate | (none) |
| Reactivation Predicate | (none) |

| Timed Activity: | CHW_R2_S2 |
|---|---|
| Distribution Parameters | **Rate**<br><br>```chw_rcv_rate``` |
| Activation Predicate | (none) |
| Reactivation Predicate | (none) |

| Timed Activity: | CHW_R_S1 |
|---|---|
| Distribution Parameters | **Rate**<br><br>chw_fail_rate |
| Activation Predicate | (none) |
| Reactivation Predicate | (none) |

| Timed Activity: | CHW_R_S2 |
|---|---|
| Distribution Parameters | **Rate**<br><br>chw_fail_rate |
| Activation Predicate | (none) |
| Reactivation Predicate | (none) |

| Timed Activity: | FHW_F_S1 |
|---|---|
| Distribution Parameters | **Rate**<br><br>fhw_fail_rate |
| Activation Predicate | (none) |
| Reactivation Predicate | (none) |
| Case Distributions | **case 1**<br><br>`if(Working_S1->Mark()==1 && Working_S2->Mark()==1)`<br>`        return(1-tmi_cvg);`<br>`else`<br>`        return(0);`<br>**case 2**<br><br>`if(Working_S1->Mark()==1 && Working_S2->Mark()==1)`<br>`        return(tmi_cvg);`<br>`else`<br>`        return(1);` |

| Timed Activity: | FHW_F_S2 |
|---|---|
| Distribution Parameters | **Rate**<br><br>fhw_fail_rate |
| Activation Predicate | (none) |
| Reactivation Predicate | (none) |
| Case Distributions | **case 1**<br><br>`if(Working_S1->Mark()==1 && Working_S2->Mark()==1)`<br>`        return(1-tmi_cvg);`<br>`else`<br>`        return(0);`<br>**case 2**<br><br>`if(Working_S1->Mark()==1 && Working_S2->Mark()==1)`<br>`        return(tmi_cvg);`<br>`else`<br>`        return(1);` |

| Timed Activity: | FHW_R_S1 |
|---|---|
| Distribution Parameters | **Rate**<br><br>fhw_rcv_rate |
| Activation Predicate | (none) |
| Reactivation Predicate | (none) |

| Timed Activity: | FHW_R_S2 |
|---|---|
| Distribution Parameters | **Rate**<br><br>fhw_rcv_rate |
| Activation Predicate | (none) |
| Reactivation Predicate | (none) |

| Timed Activity: | FHWt_F_S1 |
|---|---|
| Distribution Parameters | **Rate**<br><br>fhwt_fail_rate |
| Activation Predicate | (none) |
| Reactivation Predicate | (none) |
| Case Distributions | **case 1** |

```
if(Working_S1->Mark()==1 && Working_S2->Mark()==1)
        return(1-tmi_cvg);
else
        return(0);
case 2

if(Working_S1->Mark()==1 && Working_S2->Mark()==1)
        return(tmi_cvg);
else
        return(1);
```

| Timed Activity: | FHWt_F_S2 |
|---|---|
| **Distribution Parameters** | **Rate**<br><br>fhwt_fail_rate |
| **Activation Predicate** | (none) |
| **Reactivation Predicate** | (none) |
| **Case Distributions** | **case 1**<br><br>if(Working_S1->Mark()==1 && Working_S2->Mark()==1)<br>        return(1-tmi_cvg);<br>else<br>        return(0);<br>**case 2**<br><br>if(Working_S1->Mark()==1 && Working_S2->Mark()==1)<br>        return(tmi_cvg);<br>else<br>        return(1); |

| Timed Activity: | FHWt_R_S1 |
|---|---|
| **Distribution Parameters** | **Rate**<br><br>fhwt_rcv_rate |
| **Activation Predicate** | (none) |
| **Reactivation Predicate** | (none) |

| Timed Activity: | FHWt_R_S2 |
|---|---|
| **Distribution Parameters** | **Rate**<br><br>fhwt_rcv_rate |
| **Activation Predicate** | (none) |
| **Reactivation Predicate** | (none) |

| Timed Activity: | GEO_F |
|---|---|
| **Distribution Parameters** | **Rate**<br><br>geo_fail_rate |
| **Activation Predicate** | (none) |
| **Reactivation Predicate** | (none) |

| Timed Activity: | GEO_R |
|---|---|
| **Distribution Parameters** | **Rate**<br><br>geo_rcv_rate |
| **Activation Predicate** | (none) |
| **Reactivation Predicate** | (none) |

| Timed Activity: | MAN_F |
|---|---|
| **Distribution Parameters** | **Rate**<br><br>man_fail_rate |
| **Activation Predicate** | (none) |
| **Reactivation Predicate** | (none) |

| Timed Activity: | MAN_R |
|---|---|
| **Distribution Parameters** | **Rate**<br><br>man_rcv_rate |
| **Activation Predicate** | (none) |
| **Reactivation Predicate** | (none) |

| Timed Activity: | SW_F_S1 |
|---|---|

| Distribution Parameters | **Rate**<br><br>sw_fail_rate |
|---|---|
| **Activation Predicate** | (none) |
| **Reactivation Predicate** | (none) |
| **Case Distributions** | **case 1**<br><br>`if(Working_S1->Mark()==1 && Working_S2->Mark()==1)`<br>`        return(1-tmi_cvg);`<br>`else`<br>`        return(0);`<br>**case 2**<br><br>`if(Working_S1->Mark()==1 && Working_S2->Mark()==1)`<br>`        return(tmi_cvg);`<br>`else`<br>`        return(1);` |

| Timed Activity: | **SW_F_S2** |
|---|---|
| **Distribution Parameters** | **Rate**<br><br>sw_fail_rate |
| **Activation Predicate** | (none) |
| **Reactivation Predicate** | (none) |
| **Case Distributions** | **case 1**<br><br>`if(Working_S1->Mark()==1 && Working_S2->Mark()==1)`<br>`        return(1-tmi_cvg);`<br>`else`<br>`        return(0);`<br>**case 2**<br><br>`if(Working_S1->Mark()==1 && Working_S2->Mark()==1)`<br>`        return(tmi_cvg);`<br>`else`<br>`        return(1);` |

| Timed Activity: | **SW_R_S1** |
|---|---|
| **Distribution Parameters** | **Rate**<br><br>sw_rcv_rate |
| **Activation Predicate** | (none) |
| **Reactivation Predicate** | (none) |

| Timed Activity: | **SW_R_S2** |
|---|---|
| **Distribution Parameters** | **Rate**<br><br>sw_rcv_rate |
| **Activation Predicate** | (none) |
| **Reactivation Predicate** | (none) |

| Timed Activity: | **UCHW_R_S1** |
|---|---|
| **Distribution Parameters** | **Rate**<br><br>uchw_rcv_rate |
| **Activation Predicate** | (none) |
| **Reactivation Predicate** | (none) |

| Timed Activity: | **UCHW_R_S2** |
|---|---|
| **Distribution Parameters** | **Rate**<br><br>uchw_rcv_rate |
| **Activation Predicate** | (none) |
| **Reactivation Predicate** | (none) |

| Input Gate: | **IG_GF** |
|---|---|
| **Predicate** | `(Working_S1->Mark()==1 && Working_S2->Mark()==1)` |
| **Function** | `Working_S1->Mark()=0;`<br>`Working_S2->Mark()=0;` |

| Input Gate: | **IG_MF** |
|---|---|
| **Predicate** | |

| | (Working_S1->Mark()==1 && Working_S2->Mark()==1) |
|---|---|
| **Function** | Working_S1->Mark()=0;<br>Working_S2->Mark()=0; |

| **Output Gate:** | **OG_CHW** |
|---|---|
| **Function** | Working_S1->Mark()=0;<br>Working_S2->Mark()=0;<br>failed_CHW_S1->Mark()=1;<br>failed_CHW_S2->Mark()=1; |

| **Output Gate:** | **OG_FHW** |
|---|---|
| **Function** | Working_S1->Mark()=0;<br>Working_S2->Mark()=0;<br>failed_FHW_S1->Mark()=1;<br>failed_FHW_S2->Mark()=1; |

| **Output Gate:** | **OG_FHWt** |
|---|---|
| **Function** | Working_S1->Mark()=0;<br>Working_S2->Mark()=0;<br>failed_FHWt_S1->Mark()=1;<br>failed_FHWt_S2->Mark()=1; |

| **Output Gate:** | **OG_GR** |
|---|---|
| **Function** | Working_S1->Mark()=1;<br>Working_S2->Mark()=1; |

| **Output Gate:** | **OG_MR** |
|---|---|
| **Function** | Working_S1->Mark()=1;<br>Working_S2->Mark()=1; |

| **Output Gate:** | **OG_SW** |
|---|---|
| **Function** | Working_S1->Mark()=0;<br>Working_S2->Mark()=0;<br>failed_SW_S1->Mark()=1;<br>failed_SW_S2->Mark()=1; |

## Model: rrl

**Place Attributes**:

| Place Names | Initial Markings |
|---|---|
| Failed_L | 0 |
| GEO | 0 |
| Working_L | 1 |
| Working_S1 | 1 |
| Working_S2 | 1 |
| failed_CHW_S1 | 0 |
| failed_CHW_S2 | 0 |
| failed_FHW_S1 | 0 |
| failed_FHW_S2 | 0 |
| failed_FHWt_S1 | 0 |
| failed_FHWt_S2 | 0 |
| failed_MAN_S1 | 0 |
| failed_MAN_S2 | 0 |
| failed_SW_S1 | 0 |
| failed_SW_S2 | 0 |
| spare_CHW_S1 | 0 |
| spare_CHW_S2 | 0 |
| sys_down_S1 | 0 |
| sys_down_S2 | 0 |

| **Timed Activity:** | **CHW_F2_S1** |
|---|---|
| **Distribution Parameters** | Rate<br><br>chw_fail_rate |
| **Activation Predicate** | (none) |
| **Reactivation Predicate** | (none) |
| **Case Distributions** | case 1<br><br>if (Working_S2->Mark() == 1)<br>    return(heq_cvg); |

```
else
        return(1);
case 2

if (Working_S2->Mark() == 1)
        return(1-heq_cvg);
else
        return(0);
```

| Timed Activity: | CHW_F2_S2 |
|---|---|
| **Distribution Parameters** | **Rate**<br><br>chw_fail_rate |
| **Activation Predicate** | (none) |
| **Reactivation Predicate** | (none) |
| **Case Distributions** | **case 1**<br><br>`if (Working_S1->Mark() == 1)`<br>`        return(heq_cvg);`<br>`else`<br>`        return(1);`<br>**case 2**<br><br>`if (Working_S1->Mark() == 1)`<br>`        return(1-heq_cvg);`<br>`else`<br>`        return(0);` |

| Timed Activity: | CHW_F_S1 |
|---|---|
| **Distribution Parameters** | **Rate**<br><br>2 * chw_fail_rate |
| **Activation Predicate** | (none) |
| **Reactivation Predicate** | (none) |
| **Case Distributions** | **case 1**<br><br>1-chw_cvg<br>**case 2**<br><br>chw_cvg |

| Timed Activity: | CHW_F_S2 |
|---|---|
| **Distribution Parameters** | **Rate**<br><br>2 * chw_fail_rate |
| **Activation Predicate** | (none) |
| **Reactivation Predicate** | (none) |
| **Case Distributions** | **case 1**<br><br>1-chw_cvg<br>**case 2**<br><br>chw_cvg |

| Timed Activity: | CHW_R2_S1 |
|---|---|
| **Distribution Parameters** | **Rate**<br><br>chw_rcv_rate |
| **Activation Predicate** | (none) |
| **Reactivation Predicate** | (none) |

| Timed Activity: | CHW_R2_S2 |
|---|---|
| **Distribution Parameters** | **Rate**<br><br>chw_rcv_rate |
| **Activation Predicate** | (none) |
| **Reactivation Predicate** | (none) |

| Timed Activity: | CHW_R_S1 |
|---|---|
| **Distribution Parameters** | **Rate**<br><br>chw_fail_rate |
| **Activation Predicate** | (none) |

| Reactivation Predicate | (none) |
|---|---|

| Timed Activity: | **CHW_R_S2** |
|---|---|
| **Distribution Parameters** | **Rate**<br><br>chw_fail_rate |
| **Activation Predicate** | (none) |
| **Reactivation Predicate** | (none) |

| Timed Activity: | **FHW_F_S1** |
|---|---|
| **Distribution Parameters** | **Rate**<br><br>fhw_fail_rate |
| **Activation Predicate** | (none) |
| **Reactivation Predicate** | (none) |
| **Case Distributions** | **case 1**<br><br>```if(Working_S1->Mark()==1 && Working_S2->Mark()==1)```<br>```        return(1-heq_cvg);```<br>```else```<br>```        return(0);```<br>**case 2**<br><br>```if(Working_S1->Mark()==1 && Working_S2->Mark()==1)```<br>```        return(heq_cvg);```<br>```else```<br>```        return(1);``` |

| Timed Activity: | **FHW_F_S2** |
|---|---|
| **Distribution Parameters** | **Rate**<br><br>fhw_fail_rate |
| **Activation Predicate** | (none) |
| **Reactivation Predicate** | (none) |
| **Case Distributions** | **case 1**<br><br>```if(Working_S1->Mark()==1 && Working_S2->Mark()==1)```<br>```        return(1-heq_cvg);```<br>```else```<br>```        return(0);```<br>**case 2**<br><br>```if(Working_S1->Mark()==1 && Working_S2->Mark()==1)```<br>```        return(heq_cvg);```<br>```else```<br>```        return(1);``` |

| Timed Activity: | **FHW_R_S1** |
|---|---|
| **Distribution Parameters** | **Rate**<br><br>fhw_rcv_rate |
| **Activation Predicate** | (none) |
| **Reactivation Predicate** | (none) |

| Timed Activity: | **FHW_R_S2** |
|---|---|
| **Distribution Parameters** | **Rate**<br><br>fhw_rcv_rate |
| **Activation Predicate** | (none) |
| **Reactivation Predicate** | (none) |

| Timed Activity: | **FHWt_F_S1** |
|---|---|
| **Distribution Parameters** | **Rate**<br><br>fhwt_fail_rate |
| **Activation Predicate** | (none) |
| **Reactivation Predicate** | (none) |
| **Case Distributions** | **case 1**<br><br>```if(Working_S1->Mark()==1 && Working_S2->Mark()==1)```<br>```        return(1-heq_cvg);```<br>```else```<br>```        return(0);```<br>**case 2** |

```
if(Working_S1->Mark()==1 && Working_S2->Mark()==1)
        return(heq_cvg);
else
        return(1);
```

| Timed Activity: | FHWt_F_S2 |
|---|---|
| Distribution Parameters | **Rate**<br><br>fhwt_fail_rate |
| Activation Predicate | (none) |
| Reactivation Predicate | (none) |
| Case Distributions | **case 1**<br><br>if(Working_S1->Mark()==1 && Working_S2->Mark()==1)<br>        return(1-heq_cvg);<br>else<br>        return(0);<br>**case 2**<br><br>if(Working_S1->Mark()==1 && Working_S2->Mark()==1)<br>        return(heq_cvg);<br>else<br>        return(1); |

| Timed Activity: | FHWt_R_S1 |
|---|---|
| Distribution Parameters | **Rate**<br><br>fhwt_rcv_rate |
| Activation Predicate | (none) |
| Reactivation Predicate | (none) |

| Timed Activity: | FHWt_R_S2 |
|---|---|
| Distribution Parameters | **Rate**<br><br>fhwt_rcv_rate |
| Activation Predicate | (none) |
| Reactivation Predicate | (none) |

| Timed Activity: | GEO_F |
|---|---|
| Distribution Parameters | **Rate**<br><br>geo_fail_rate |
| Activation Predicate | (none) |
| Reactivation Predicate | (none) |

| Timed Activity: | GEO_R |
|---|---|
| Distribution Parameters | **Rate**<br><br>geo_rcv_rate |
| Activation Predicate | (none) |
| Reactivation Predicate | (none) |

| Timed Activity: | L_F |
|---|---|
| Distribution Parameters | **Rate**<br><br>link_fail_rate |
| Activation Predicate | (none) |
| Reactivation Predicate | (none) |

| Timed Activity: | L_R |
|---|---|
| Distribution Parameters | **Rate**<br><br>link_rcv_rate |
| Activation Predicate | (none) |
| Reactivation Predicate | (none) |

| Timed Activity: | MAN_F_S1 |
|---|---|
| Distribution Parameters | **Rate**<br><br>man_fail_rate |
| Activation Predicate | (none) |
| Reactivation Predicate | (none) |

| Case Distributions | case 1 |
|---|---|

```
if(Working_S1->Mark()==1 && Working_S2->Mark()==1)
        return(1-heq_cvg);
else
        return(0);
```

**case 2**

```
if(Working_S1->Mark()==1 && Working_S2->Mark()==1)
        return(heq_cvg);
else
        return(1);
```

| Timed Activity: | MAN_F_S2 |
|---|---|
| **Distribution Parameters** | **Rate**<br><br>man_fail_rate |
| **Activation Predicate** | (none) |
| **Reactivation Predicate** | (none) |
| **Case Distributions** | **case 1**<br><br>`if(Working_S1->Mark()==1 && Working_S2->Mark()==1)`<br>`        return(1-heq_cvg);`<br>`else`<br>`        return(0);`<br>**case 2**<br><br>`if(Working_S1->Mark()==1 && Working_S2->Mark()==1)`<br>`        return(heq_cvg);`<br>`else`<br>`        return(1);` |

| Timed Activity: | MAN_R_S1 |
|---|---|
| **Distribution Parameters** | **Rate**<br><br>man_rcv_rate |
| **Activation Predicate** | (none) |
| **Reactivation Predicate** | (none) |

| Timed Activity: | MAN_R_S2 |
|---|---|
| **Distribution Parameters** | **Rate**<br><br>man_rcv_rate |
| **Activation Predicate** | (none) |
| **Reactivation Predicate** | (none) |

| Timed Activity: | SW_F_S1 |
|---|---|
| **Distribution Parameters** | **Rate**<br><br>sw_fail_rate |
| **Activation Predicate** | (none) |
| **Reactivation Predicate** | (none) |
| **Case Distributions** | **case 1**<br><br>`if(Working_S1->Mark()==1 && Working_S2->Mark()==1)`<br>`        return(1-heq_cvg);`<br>`else`<br>`        return(0);`<br>**case 2**<br><br>`if(Working_S1->Mark()==1 && Working_S2->Mark()==1)`<br>`        return(heq_cvg);`<br>`else`<br>`        return(1);` |

| Timed Activity: | SW_F_S2 |
|---|---|
| **Distribution Parameters** | **Rate**<br><br>sw_fail_rate |
| **Activation Predicate** | (none) |
| **Reactivation Predicate** | (none) |
| **Case Distributions** | **case 1**<br><br>`if(Working_S1->Mark()==1 && Working_S2->Mark()==1)`<br>`        return(1-heq_cvg);` |

```
else
        return(0);

case 2

if(Working_S1->Mark()==1 && Working_S2->Mark()==1)
        return(heq_cvg);
else
        return(1);
```

| Timed Activity: | SW_R_S1 |
|---|---|
| Distribution Parameters | **Rate**<br><br>sw_rcv_rate |
| Activation Predicate | (none) |
| Reactivation Predicate | (none) |

| Timed Activity: | SW_R_S2 |
|---|---|
| Distribution Parameters | **Rate**<br><br>sw_rcv_rate |
| Activation Predicate | (none) |
| Reactivation Predicate | (none) |

| Timed Activity: | UCHW_R_S1 |
|---|---|
| Distribution Parameters | **Rate**<br><br>uchw_rcv_rate |
| Activation Predicate | (none) |
| Reactivation Predicate | (none) |

| Timed Activity: | UCHW_R_S2 |
|---|---|
| Distribution Parameters | **Rate**<br><br>uchw_rcv_rate |
| Activation Predicate | (none) |
| Reactivation Predicate | (none) |

| Input Gate: | IG_GF |
|---|---|
| Predicate | (Working_L->Mark()==1 && Working_S2->Mark()==1) |
| Function | Working_L->Mark()=0;<br>Working_S2->Mark()=0; |

| Output Gate: | OG_CHW |
|---|---|
| Function | Working_S1->Mark()=0;<br>Working_S2->Mark()=0;<br>failed_CHW_S1->Mark()=1;<br>failed_CHW_S2->Mark()=1; |

| Output Gate: | OG_FHW |
|---|---|
| Function | Working_S1->Mark()=0;<br>Working_S2->Mark()=0;<br>failed_FHW_S1->Mark()=1;<br>failed_FHW_S2->Mark()=1; |

| Output Gate: | OG_FHWt |
|---|---|
| Function | Working_S1->Mark()=0;<br>Working_S2->Mark()=0;<br>failed_FHWt_S1->Mark()=1;<br>failed_FHWt_S2->Mark()=1; |

| Output Gate: | OG_GR |
|---|---|
| Function | Working_L->Mark()=1;<br>Working_S2->Mark()=1; |

| Output Gate: | OG_MAN |
|---|---|
| Function | Working_S1->Mark()=0;<br>Working_S2->Mark()=0;<br>failed_SW_S1->Mark()=1;<br>failed_SW_S2->Mark()=1; |

| Output Gate: | OG_SW |
|---|---|
| Function | Working_S1->Mark()=0; |

```
Working_S2->Mark()=0;
failed_SW_S1->Mark()=1;
failed_SW_S2->Mark()=1;
```

## Model: rrr

**Place Attributes:**

| Place Names | Initial Markings |
|---|---|
| Working_S1 | 1 |
| Working_S2 | 1 |
| Working_S3 | 1 |
| failed_CHW_S1 | 0 |
| failed_CHW_S2 | 0 |
| failed_CHW_S3 | 0 |
| failed_FHW_S1 | 0 |
| failed_FHW_S2 | 0 |
| failed_FHW_S3 | 0 |
| failed_FHWt_S1 | 0 |
| failed_FHWt_S2 | 0 |
| failed_FHWt_S3 | 0 |
| failed_MAN_S1 | 0 |
| failed_MAN_S2 | 0 |
| failed_MAN_S3 | 0 |
| failed_SW_S1 | 0 |
| failed_SW_S2 | 0 |
| failed_SW_S3 | 0 |
| spare_CHW_S1 | 0 |
| spare_CHW_S2 | 0 |
| spare_CHW_S3 | 0 |
| sys_down_S1 | 0 |
| sys_down_S2 | 0 |
| sys_down_S3 | 0 |

| Timed Activity: | CHW_F2_S1 |
|---|---|
| **Distribution Parameters** | **Rate**<br><br>chw_fail_rate |
| **Activation Predicate** | (none) |
| **Reactivation Predicate** | (none) |
| **Case Distributions** | **case 1**<br><br>```if (Working_S2->Mark() == 1 && Working_S3->Mark()==1)```<br>    ```return(heq_cvg);```<br>```else```<br>    ```return(1);```<br>**case 2**<br><br>```if (Working_S2->Mark() == 1 && Working_S3->Mark()==1)```<br>    ```return(1-heq_cvg);```<br>```else```<br>    ```return(0);``` |

| Timed Activity: | CHW_F2_S2 |
|---|---|
| **Distribution Parameters** | **Rate**<br><br>chw_fail_rate |
| **Activation Predicate** | (none) |
| **Reactivation Predicate** | (none) |
| **Case Distributions** | **case 1**<br><br>```if (Working_S1->Mark() == 1 && Working_S3->Mark()==1)```<br>    ```return(heq_cvg);```<br>```else```<br>    ```return(1);```<br>**case 2**<br><br>```if (Working_S1->Mark() == 1 && Working_S3->Mark()==1)```<br>    ```return(1-heq_cvg);```<br>```else```<br>    ```return(0);``` |

| Timed Activity: | CHW_F2_S3 |
|---|---|
| | **Rate** |

| Distribution Parameters | chw_fail_rate |
|---|---|
| **Activation Predicate** | (none) |
| **Reactivation Predicate** | (none) |
| **Case Distributions** | case 1<br><br>if (Working_S1->Mark()==1 && Working_S2->Mark() == 1)<br>    return(heq_cvg);<br>else<br>    return(1);<br>case 2<br><br>if (Working_S1->Mark()==1 && Working_S2->Mark() == 1)<br>    return(1-heq_cvg);<br>else<br>    return(0); |

| **Timed Activity:** | **CHW_F_S1** |
|---|---|
| **Distribution Parameters** | **Rate**<br><br>2 * chw_fail_rate |
| **Activation Predicate** | (none) |
| **Reactivation Predicate** | (none) |
| **Case Distributions** | **case 1**<br><br>1-chw_cvg<br>**case 2**<br><br>chw_cvg |

| **Timed Activity:** | **CHW_F_S2** |
|---|---|
| **Distribution Parameters** | **Rate**<br><br>2 * chw_fail_rate |
| **Activation Predicate** | (none) |
| **Reactivation Predicate** | (none) |
| **Case Distributions** | **case 1**<br><br>1-chw_cvg<br>**case 2**<br><br>chw_cvg |

| **Timed Activity:** | **CHW_F_S3** |
|---|---|
| **Distribution Parameters** | **Rate**<br><br>2 * chw_fail_rate |
| **Activation Predicate** | (none) |
| **Reactivation Predicate** | (none) |
| **Case Distributions** | **case 1**<br><br>1-chw_cvg<br>**case 2**<br><br>chw_cvg |

| **Timed Activity:** | **CHW_R2_S1** |
|---|---|
| **Distribution Parameters** | **Rate**<br><br>chw_rcv_rate |
| **Activation Predicate** | (none) |
| **Reactivation Predicate** | (none) |

| **Timed Activity:** | **CHW_R2_S2** |
|---|---|
| **Distribution Parameters** | **Rate**<br><br>chw_rcv_rate |
| **Activation Predicate** | (none) |
| **Reactivation Predicate** | (none) |

| **Timed Activity:** | **CHW_R2_S3** |
|---|---|

| Distribution Parameters | Rate<br><br>chw_rcv_rate |
|---|---|
| **Activation Predicate** | (none) |
| **Reactivation Predicate** | (none) |

| Timed Activity: | **CHW_R_S1** |
|---|---|
| **Distribution Parameters** | Rate<br><br>chw_fail_rate |
| **Activation Predicate** | (none) |
| **Reactivation Predicate** | (none) |

| Timed Activity: | **CHW_R_S2** |
|---|---|
| **Distribution Parameters** | Rate<br><br>chw_fail_rate |
| **Activation Predicate** | (none) |
| **Reactivation Predicate** | (none) |

| Timed Activity: | **CHW_R_S3** |
|---|---|
| **Distribution Parameters** | Rate<br><br>chw_fail_rate |
| **Activation Predicate** | (none) |
| **Reactivation Predicate** | (none) |

| Timed Activity: | **FHW_F_S1** |
|---|---|
| **Distribution Parameters** | Rate<br><br>fhw_fail_rate |
| **Activation Predicate** | (none) |
| **Reactivation Predicate** | (none) |
| **Case Distributions** | case 1<br><br>```<br>if(Working_S1->Mark()==1 && Working_S2->Mark()==1 && Working_S3->Mark()==1)<br>        return(1-heq_cvg);<br>else<br>        return(0);<br>```<br>case 2<br><br>```<br>if(Working_S1->Mark()==1 && Working_S2->Mark()==1 && Working_S3->Mark()==1)<br>        return(heq_cvg);<br>else<br>        return(1);<br>``` |

| Timed Activity: | **FHW_F_S2** |
|---|---|
| **Distribution Parameters** | Rate<br><br>fhw_fail_rate |
| **Activation Predicate** | (none) |
| **Reactivation Predicate** | (none) |
| **Case Distributions** | case 1<br><br>```<br>if(Working_S1->Mark()==1 && Working_S2->Mark()==1&& Working_S3->Mark()==1)<br>        return(1-heq_cvg);<br>else<br>        return(0);<br>```<br>case 2<br><br>```<br>if(Working_S1->Mark()==1 && Working_S2->Mark()==1&& Working_S3->Mark()==1)<br>        return(heq_cvg);<br>else<br>        return(1);<br>``` |

| Timed Activity: | **FHW_F_S3** |
|---|---|
| **Distribution Parameters** | Rate<br><br>fhw_fail_rate |
| **Activation Predicate** | (none) |
| **Reactivation Predicate** | (none) |
| **Case Distributions** | case 1<br><br>```<br>if(Working_S1->Mark()==1 && Working_S2->Mark()==1 && Working_S3->Mark()==1)<br>``` |

```
                                return(1-heq_cvg);
                else
                                return(0);
        case 2

        if(Working_S1->Mark()==1 && Working_S2->Mark()==1 && Working_S3->Mark()==1)
                                return(heq_cvg);
        else
                                return(1);
```

| Timed Activity: | FHW_R_S1 |
|---|---|
| Distribution Parameters | **Rate**<br><br>fhw_rcv_rate |
| Activation Predicate | (none) |
| Reactivation Predicate | (none) |

| Timed Activity: | FHW_R_S2 |
|---|---|
| Distribution Parameters | **Rate**<br><br>fhw_rcv_rate |
| Activation Predicate | (none) |
| Reactivation Predicate | (none) |

| Timed Activity: | FHW_R_S3 |
|---|---|
| Distribution Parameters | **Rate**<br><br>fhw_rcv_rate |
| Activation Predicate | (none) |
| Reactivation Predicate | (none) |

| Timed Activity: | FHWt_F_S1 |
|---|---|
| Distribution Parameters | **Rate**<br><br>fhwt_fail_rate |
| Activation Predicate | (none) |
| Reactivation Predicate | (none) |
| Case Distributions | **case 1**<br><br>`if(Working_S1->Mark()==1 && Working_S2->Mark()==1 && Working_S3->Mark()==1)`<br>`        return(1-heq_cvg);`<br>`else`<br>`        return(0);`<br>**case 2**<br><br>`if(Working_S1->Mark()==1 && Working_S2->Mark()==1 && Working_S3->Mark()==1)`<br>`        return(heq_cvg);`<br>`else`<br>`        return(1);` |

| Timed Activity: | FHWt_F_S2 |
|---|---|
| Distribution Parameters | **Rate**<br><br>fhwt_fail_rate |
| Activation Predicate | (none) |
| Reactivation Predicate | (none) |
| Case Distributions | **case 1**<br><br>`if(Working_S1->Mark()==1 && Working_S2->Mark()==1 && Working_S3->Mark()==1)`<br>`        return(1-heq_cvg);`<br>`else`<br>`        return(0);`<br>**case 2**<br><br>`if(Working_S1->Mark()==1 && Working_S2->Mark()==1 && Working_S3->Mark()==1)`<br>`        return(heq_cvg);`<br>`else`<br>`        return(1);` |

| Timed Activity: | FHWt_F_S3 |
|---|---|
| Distribution Parameters | **Rate**<br><br>fhwt_fail_rate |
| Activation Predicate | (none) |
| Reactivation Predicate | (none) |

| Case Distributions | case 1 |
|---|---|
| | `if(Working_S1->Mark()==1 && Working_S2->Mark()==1 && Working_S3->Mark()==1)`<br>`        return(1-heq_cvg);`<br>`else`<br>`        return(0);` |
| | case 2 |
| | `if(Working_S1->Mark()==1 && Working_S2->Mark()==1 && Working_S3->Mark()==1)`<br>`        return(heq_cvg);`<br>`else`<br>`        return(1);` |

| Timed Activity: | FHWt_R_S1 |
|---|---|
| Distribution Parameters | **Rate**<br>`fhwt_rcv_rate` |
| Activation Predicate | (none) |
| Reactivation Predicate | (none) |

| Timed Activity: | FHWt_R_S2 |
|---|---|
| Distribution Parameters | **Rate**<br>`fhwt_rcv_rate` |
| Activation Predicate | (none) |
| Reactivation Predicate | (none) |

| Timed Activity: | FHWt_R_S3 |
|---|---|
| Distribution Parameters | **Rate**<br>`fhwt_rcv_rate` |
| Activation Predicate | (none) |
| Reactivation Predicate | (none) |

| Timed Activity: | MAN_F_S1 |
|---|---|
| Distribution Parameters | **Rate**<br>`man_fail_rate` |
| Activation Predicate | (none) |
| Reactivation Predicate | (none) |
| Case Distributions | case 1<br><br>`if(Working_S1->Mark()==1 && Working_S2->Mark()==1&& Working_S3->Mark()==1)`<br>`        return(1-heq_cvg);`<br>`else`<br>`        return(0);`<br>case 2<br><br>`if(Working_S1->Mark()==1 && Working_S2->Mark()==1&& Working_S3->Mark()==1)`<br>`        return(heq_cvg);`<br>`else`<br>`        return(1);` |

| Timed Activity: | MAN_F_S2 |
|---|---|
| Distribution Parameters | **Rate**<br>`man_fail_rate` |
| Activation Predicate | (none) |
| Reactivation Predicate | (none) |
| Case Distributions | case 1<br><br>`if(Working_S1->Mark()==1 && Working_S2->Mark()==1&& Working_S3->Mark()==1)`<br>`        return(1-heq_cvg);`<br>`else`<br>`        return(0);`<br>case 2<br><br>`if(Working_S1->Mark()==1 && Working_S2->Mark()==1&& Working_S3->Mark()==1)`<br>`        return(heq_cvg);`<br>`else`<br>`        return(1);` |

| Timed Activity: | MAN_F_S3 |
|---|---|
| Distribution Parameters | **Rate**<br>`man_fail_rate` |
| **Activation Predicate** | (none) |
| **Reactivation Predicate** | (none) |
| **Case Distributions** | **case 1**<br><br>```<br>if(Working_S1->Mark()==1 && Working_S2->Mark()==1 && Working_S3->Mark()==1)<br>        return(1-heq_cvg);<br>else<br>        return(0);<br>```<br>**case 2**<br><br>```<br>if(Working_S1->Mark()==1 && Working_S2->Mark()==1 && Working_S3->Mark()==1)<br>        return(heq_cvg);<br>else<br>        return(1);<br>``` |

| **Timed Activity:** | **MAN_R_S1** |
|---|---|
| **Distribution Parameters** | **Rate**<br><br>man_rcv_rate |
| **Activation Predicate** | (none) |
| **Reactivation Predicate** | (none) |

| **Timed Activity:** | **MAN_R_S2** |
|---|---|
| **Distribution Parameters** | **Rate**<br><br>man_rcv_rate |
| **Activation Predicate** | (none) |
| **Reactivation Predicate** | (none) |

| **Timed Activity:** | **MAN_R_S3** |
|---|---|
| **Distribution Parameters** | **Rate**<br><br>man_rcv_rate |
| **Activation Predicate** | (none) |
| **Reactivation Predicate** | (none) |

| **Timed Activity:** | **SW_F_S1** |
|---|---|
| **Distribution Parameters** | **Rate**<br><br>sw_fail_rate |
| **Activation Predicate** | (none) |
| **Reactivation Predicate** | (none) |
| **Case Distributions** | **case 1**<br><br>```<br>if(Working_S1->Mark()==1 && Working_S2->Mark()==1&& Working_S3->Mark()==1)<br>        return(1-heq_cvg);<br>else<br>        return(0);<br>```<br>**case 2**<br><br>```<br>if(Working_S1->Mark()==1 && Working_S2->Mark()==1&& Working_S3->Mark()==1)<br>        return(heq_cvg);<br>else<br>        return(1);<br>``` |

| **Timed Activity:** | **SW_F_S2** |
|---|---|
| **Distribution Parameters** | **Rate**<br><br>sw_fail_rate |
| **Activation Predicate** | (none) |
| **Reactivation Predicate** | (none) |
| **Case Distributions** | **case 1**<br><br>```<br>if(Working_S1->Mark()==1 && Working_S2->Mark()==1)<br>        return(1-heq_cvg);<br>else<br>        return(0);<br>```<br>**case 2**<br><br>```<br>if(Working_S1->Mark()==1 && Working_S2->Mark()==1)<br>        return(heq_cvg);<br>else<br>        return(1);<br>``` |

| **Timed Activity:** | **SW_F_S3** |
|---|---|

| Distribution Parameters | **Rate**<br><br>sw_fail_rate |
|---|---|
| **Activation Predicate** | (none) |
| **Reactivation Predicate** | (none) |

| Case Distributions | **case 1**<br><br>if(Working_S1->Mark()==1 && Working_S2->Mark()==1&& Working_S3->Mark()==1)<br>     return(1-heq_cvg);<br>else<br>     return(0);<br>**case 2**<br><br>if(Working_S1->Mark()==1 && Working_S2->Mark()==1&& Working_S3->Mark()==1)<br>     return(heq_cvg);<br>else<br>     return(1); |
|---|---|

| **Timed Activity:** | **SW_R_S1** |
|---|---|
| **Distribution Parameters** | **Rate**<br><br>sw_rcv_rate |
| **Activation Predicate** | (none) |
| **Reactivation Predicate** | (none) |

| **Timed Activity:** | **SW_R_S2** |
|---|---|
| **Distribution Parameters** | **Rate**<br><br>sw_rcv_rate |
| **Activation Predicate** | (none) |
| **Reactivation Predicate** | (none) |

| **Timed Activity:** | **SW_R_S3** |
|---|---|
| **Distribution Parameters** | **Rate**<br><br>sw_rcv_rate |
| **Activation Predicate** | (none) |
| **Reactivation Predicate** | (none) |

| **Timed Activity:** | **UCHW_R_S1** |
|---|---|
| **Distribution Parameters** | **Rate**<br><br>uchw_rcv_rate |
| **Activation Predicate** | (none) |
| **Reactivation Predicate** | (none) |

| **Timed Activity:** | **UCHW_R_S2** |
|---|---|
| **Distribution Parameters** | **Rate**<br><br>uchw_rcv_rate |
| **Activation Predicate** | (none) |
| **Reactivation Predicate** | (none) |

| **Timed Activity:** | **UCHW_R_S3** |
|---|---|
| **Distribution Parameters** | **Rate**<br><br>uchw_rcv_rate |
| **Activation Predicate** | (none) |
| **Reactivation Predicate** | (none) |

| **Output Gate:** | **OG_CHW** |
|---|---|
| **Function** | Working_S1->Mark()=0;<br>Working_S2->Mark()=0;<br>Working_S3->Mark()=0;<br>failed_CHW_S1->Mark()=1;<br>failed_CHW_S2->Mark()=1;<br>failed_CHW_S3->Mark()=1; |

| **Output Gate:** | **OG_FHW** |
|---|---|
| **Function** | Working_S1->Mark()=0;<br>Working_S2->Mark()=0;<br>Working_S3->Mark()=0;<br>failed_FHW_S1->Mark()=1; |

```
failed_FHW_S2->Mark()=1;
failed_FHW_S3->Mark()=1;
```

| Output Gate: | OG_FHWt |
|---|---|
| Function | `Working_S1->Mark()=0;`<br>`Working_S2->Mark()=0;`<br>`Working_S3->Mark()=0;`<br>`failed_FHWt_S1->Mark()=1;`<br>`failed_FHWt_S2->Mark()=1;`<br>`failed_FHWt_S3->Mark()=1;` |

| Output Gate: | OG_MAN |
|---|---|
| Function | `Working_S1->Mark()=0;`<br>`Working_S2->Mark()=0;`<br>`Working_S3->Mark()=0;`<br>`failed_SW_S1->Mark()=1;`<br>`failed_SW_S2->Mark()=1;`<br>`failed_SW_S3->Mark()=1;` |

| Output Gate: | OG_SW |
|---|---|
| Function | `Working_S1->Mark()=0;`<br>`Working_S2->Mark()=0;`<br>`Working_S3->Mark()=0;`<br>`failed_SW_S1->Mark()=1;`<br>`failed_SW_S2->Mark()=1;`<br>`failed_SW_S3->Mark()=1;` |

## Model: sll

Place Attributes:

| Place Names | Initial Markings |
|---|---|
| Failed_L1 | 0 |
| Failed_L2 | 0 |
| GEO | 0 |
| PHY | 0 |
| Working_L1 | 1 |
| Working_L2 | 1 |
| Working_S | 1 |
| failed_FHW | 0 |
| failed_FHWt | 0 |
| failed_SW | 0 |

| Timed Activity: | FHW_F |
|---|---|
| Distribution Parameters | Rate<br><br>`fhw_fail_rate` |
| Activation Predicate | (none) |
| Reactivation Predicate | (none) |

| Timed Activity: | FHW_R |
|---|---|
| Distribution Parameters | Rate<br><br>`fhw_rcv_rate` |
| Activation Predicate | (none) |
| Reactivation Predicate | (none) |

| Timed Activity: | FHWt_F |
|---|---|
| Distribution Parameters | Rate<br><br>`fhwt_fail_rate` |
| Activation Predicate | (none) |
| Reactivation Predicate | (none) |

| Timed Activity: | FHWt_R |
|---|---|
| Distribution Parameters | Rate<br><br>`fhwt_rcv_rate` |
| Activation Predicate | (none) |
| Reactivation Predicate | (none) |

| Timed Activity: | GEO_F |
|---|---|
| Distribution Parameters | Rate |
| | geo_fail_rate |
| **Activation Predicate** | (none) |
| **Reactivation Predicate** | (none) |

| Timed Activity: | **GEO_R** |
|---|---|
| **Distribution Parameters** | **Rate**<br><br>geo_rcv_rate |
| **Activation Predicate** | (none) |
| **Reactivation Predicate** | (none) |

| Timed Activity: | **L_F1** |
|---|---|
| **Distribution Parameters** | **Rate**<br><br>link_fail_rate |
| **Activation Predicate** | (none) |
| **Reactivation Predicate** | (none) |

| Timed Activity: | **L_F2** |
|---|---|
| **Distribution Parameters** | **Rate**<br><br>link_fail_rate |
| **Activation Predicate** | (none) |
| **Reactivation Predicate** | (none) |

| Timed Activity: | **L_R1** |
|---|---|
| **Distribution Parameters** | **Rate**<br><br>link_rcv_rate |
| **Activation Predicate** | (none) |
| **Reactivation Predicate** | (none) |

| Timed Activity: | **L_R2** |
|---|---|
| **Distribution Parameters** | **Rate**<br><br>link_rcv_rate |
| **Activation Predicate** | (none) |
| **Reactivation Predicate** | (none) |

| Timed Activity: | **PHY_F** |
|---|---|
| **Distribution Parameters** | **Rate**<br><br>phy_fail_rate |
| **Activation Predicate** | (none) |
| **Reactivation Predicate** | (none) |

| Timed Activity: | **PHY_R** |
|---|---|
| **Distribution Parameters** | **Rate**<br><br>phy_rcv_rate |
| **Activation Predicate** | (none) |
| **Reactivation Predicate** | (none) |

| Timed Activity: | **SW_F** |
|---|---|
| **Distribution Parameters** | **Rate**<br><br>sw_fail_rate |
| **Activation Predicate** | (none) |
| **Reactivation Predicate** | (none) |

| Timed Activity: | **SW_R** |
|---|---|
| **Distribution Parameters** | **Rate**<br><br>sw_rcv_rate |
| **Activation Predicate** | (none) |
| **Reactivation Predicate** | (none) |

| Input Gate: | IG_GF |
|---|---|
| Predicate | `(Working_L1->Mark()==1 && Working_L2->Mark()==1 && Working_S->Mark()==1)` |
| Function | `Working_L1->Mark()=0;`<br>`Working_L2->Mark()=0;`<br>`Working_S->Mark()=0;` |

| Input Gate: | IG_PF |
|---|---|
| Predicate | `(Working_L1->Mark()==1 && Working_L2->Mark()==1)` |
| Function | `Working_L1->Mark()=0;`<br>`Working_L2->Mark()=0;` |

| Output Gate: | OG_GR |
|---|---|
| Function | `Working_L1->Mark()=1;`<br>`Working_L2->Mark()=1;`<br>`Working_S->Mark()=1;` |

| Output Gate: | OG_PR |
|---|---|
| Function | `Working_L1->Mark()=1;`<br>`Working_L2->Mark()=1;` |

## Model: ss

**Place Attributes**:

| Place Names | Initial Markings |
|---|---|
| GEO | 0 |
| MIS | 0 |
| Working_S1 | 1 |
| Working_S2 | 1 |
| failed_FHW_S1 | 0 |
| failed_FHW_S2 | 0 |
| failed_FHWt_S1 | 0 |
| failed_FHWt_S2 | 0 |
| failed_SW_S1 | 0 |
| failed_SW_S2 | 0 |

| Timed Activity: | FHW_F_S1 |
|---|---|
| Distribution Parameters | **Rate**<br>`fhw_fail_rate` |
| Activation Predicate | (none) |
| Reactivation Predicate | (none) |
| Case Distributions | **case 1**<br><br>`if(Working_S1->Mark()==1 && Working_S2->Mark()==1)`<br>`        return(1-tmi_cvg);`<br>`else`<br>`        return(0);`<br>**case 2**<br><br>`if(Working_S1->Mark()==1 && Working_S2->Mark()==1)`<br>`        return(tmi_cvg);`<br>`else`<br>`        return(1);` |

| Timed Activity: | FHW_F_S2 |
|---|---|
| Distribution Parameters | **Rate**<br>`fhw_fail_rate` |
| Activation Predicate | (none) |
| Reactivation Predicate | (none) |
| Case Distributions | **case 1**<br><br>`if(Working_S1->Mark()==1 && Working_S2->Mark()==1)`<br>`        return(1-tmi_cvg);`<br>`else`<br>`        return(0);`<br>**case 2**<br><br>`if(Working_S1->Mark()==1 && Working_S2->Mark()==1)`<br>`        return(tmi_cvg);`<br>`else`<br>`        return(1);` |

| Timed Activity: | FHW_R_S1 |
|---|---|
| **Distribution Parameters** | **Rate**<br><br>fhw_rcv_rate |
| **Activation Predicate** | (none) |
| **Reactivation Predicate** | (none) |

| Timed Activity: | FHW_R_S2 |
|---|---|
| **Distribution Parameters** | **Rate**<br><br>fhw_rcv_rate |
| **Activation Predicate** | (none) |
| **Reactivation Predicate** | (none) |

| Timed Activity: | FHWt_F_S1 |
|---|---|
| **Distribution Parameters** | **Rate**<br><br>fhwt_fail_rate |
| **Activation Predicate** | (none) |
| **Reactivation Predicate** | (none) |
| **Case Distributions** | **case 1**<br><br>if(Working_S1->Mark()==1 && Working_S2->Mark()==1)<br>    return(1-tmi_cvg);<br>else<br>    return(0);<br>**case 2**<br><br>if(Working_S1->Mark()==1 && Working_S2->Mark()==1)<br>    return(tmi_cvg);<br>else<br>    return(1); |

| Timed Activity: | FHWt_F_S2 |
|---|---|
| **Distribution Parameters** | **Rate**<br><br>fhwt_fail_rate |
| **Activation Predicate** | (none) |
| **Reactivation Predicate** | (none) |
| **Case Distributions** | **case 1**<br><br>if(Working_S1->Mark()==1 && Working_S2->Mark()==1)<br>    return(1-tmi_cvg);<br>else<br>    return(0);<br>**case 2**<br><br>if(Working_S1->Mark()==1 && Working_S2->Mark()==1)<br>    return(tmi_cvg);<br>else<br>    return(1); |

| Timed Activity: | FHWt_R_S1 |
|---|---|
| **Distribution Parameters** | **Rate**<br><br>fhwt_rcv_rate |
| **Activation Predicate** | (none) |
| **Reactivation Predicate** | (none) |

| Timed Activity: | FHWt_R_S2 |
|---|---|
| **Distribution Parameters** | **Rate**<br><br>fhwt_rcv_rate |
| **Activation Predicate** | (none) |
| **Reactivation Predicate** | (none) |

| Timed Activity: | GEO_F |
|---|---|
| **Distribution Parameters** | **Rate**<br><br>geo_fail_rate |
| **Activation Predicate** | (none) |
| **Reactivation Predicate** | (none) |

| Timed Activity: | GEO_R |
|---|---|
| **Distribution Parameters** | **Rate**<br>geo_rcv_rate |
| **Activation Predicate** | (none) |
| **Reactivation Predicate** | (none) |

| Timed Activity: | MIS_F |
|---|---|
| **Distribution Parameters** | **Rate**<br>mis_fail_rate |
| **Activation Predicate** | (none) |
| **Reactivation Predicate** | (none) |

| Timed Activity: | MIS_R |
|---|---|
| **Distribution Parameters** | **Rate**<br>mis_rcv_rate |
| **Activation Predicate** | (none) |
| **Reactivation Predicate** | (none) |

| Timed Activity: | SW_F_S1 |
|---|---|
| **Distribution Parameters** | **Rate**<br>sw_fail_rate |
| **Activation Predicate** | (none) |
| **Reactivation Predicate** | (none) |
| **Case Distributions** | **case 1**<br><br>```if(Working_S1->Mark()==1 && Working_S2->Mark()==1)```<br>```        return(1-tmi_cvg);```<br>```else```<br>```        return(0);```<br>**case 2**<br><br>```if(Working_S1->Mark()==1 && Working_S2->Mark()==1)```<br>```        return(tmi_cvg);```<br>```else```<br>```        return(1);``` |

| Timed Activity: | SW_F_S2 |
|---|---|
| **Distribution Parameters** | **Rate**<br>sw_fail_rate |
| **Activation Predicate** | (none) |
| **Reactivation Predicate** | (none) |
| **Case Distributions** | **case 1**<br><br>```if(Working_S1->Mark()==1 && Working_S2->Mark()==1)```<br>```        return(1-tmi_cvg);```<br>```else```<br>```        return(0);```<br>**case 2**<br><br>```if(Working_S1->Mark()==1 && Working_S2->Mark()==1)```<br>```        return(tmi_cvg);```<br>```else```<br>```        return(1);``` |

| Timed Activity: | SW_R_S1 |
|---|---|
| **Distribution Parameters** | **Rate**<br>sw_rcv_rate |
| **Activation Predicate** | (none) |
| **Reactivation Predicate** | (none) |

| Timed Activity: | SW_R_S2 |
|---|---|
| **Distribution Parameters** | **Rate**<br>sw_rcv_rate |
| **Activation Predicate** | (none) |
| **Reactivation Predicate** | (none) |

| Input Gate: | IG_GF |
|---|---|
| Predicate | (Working_S1->Mark()==1 && Working_S2->Mark()==1) |
| Function | Working_S1->Mark()=0;<br>Working_S2->Mark()=0; |

| Input Gate: | IG_MF |
|---|---|
| Predicate | (Working_S1->Mark()==1 && Working_S2->Mark()==1) |
| Function | Working_S1->Mark()=0;<br>Working_S2->Mark()=0; |

| Output Gate: | OG_FHW |
|---|---|
| Function | Working_S1->Mark()=0;<br>Working_S2->Mark()=0;<br>failed_FHW_S1->Mark()=1;<br>failed_FHW_S2->Mark()=1; |

| Output Gate: | OG_FHWt |
|---|---|
| Function | Working_S1->Mark()=0;<br>Working_S2->Mark()=0;<br>failed_FHWt_S1->Mark()=1;<br>failed_FHWt_S2->Mark()=1; |

| Output Gate: | OG_GR |
|---|---|
| Function | Working_S1->Mark()=1;<br>Working_S2->Mark()=1; |

| Output Gate: | OG_MR |
|---|---|
| Function | Working_S1->Mark()=1;<br>Working_S2->Mark()=1; |

| Output Gate: | OG_SW |
|---|---|
| Function | Working_S1->Mark()=0;<br>Working_S2->Mark()=0;<br>failed_SW_S1->Mark()=1;<br>failed_SW_S2->Mark()=1; |

## Model: ssl

**Place Attributes**:

| Place Names | Initial Markings |
|---|---|
| Failed_L | 0 |
| GEO | 0 |
| Working_L | 1 |
| Working_S1 | 1 |
| Working_S2 | 1 |
| failed_FHW_S1 | 0 |
| failed_FHW_S2 | 0 |
| failed_FHWt_S1 | 0 |
| failed_FHWt_S2 | 0 |
| failed_SW_S1 | 0 |
| failed_SW_S2 | 0 |

| Timed Activity: | FHW_F_S1 |
|---|---|
| Distribution Parameters | Rate<br><br>fhw_fail_rate |
| Activation Predicate | (none) |
| Reactivation Predicate | (none) |
| Case Distributions | case 1<br><br>if(Working_S1->Mark()==1 && Working_S2->Mark()==1)<br>    return(1-heq_cvg);<br>else<br>    return(0);<br>case 2<br><br>if(Working_S1->Mark()==1 && Working_S2->Mark()==1)<br>    return(heq_cvg);<br>else<br>    return(1); |

| Timed Activity: | FHW_F_S2 |
|---|---|
| Distribution Parameters | **Rate**<br><br>fhw_fail_rate |
| Activation Predicate | (none) |
| Reactivation Predicate | (none) |
| Case Distributions | **case 1**<br><br>if(Working_S1->Mark()==1 && Working_S2->Mark()==1)<br>    return(1-heq_cvg);<br>else<br>    return(0);<br>**case 2**<br><br>if(Working_S1->Mark()==1 && Working_S2->Mark()==1)<br>    return(heq_cvg);<br>else<br>    return(1); |

| Timed Activity: | FHW_R_S1 |
|---|---|
| Distribution Parameters | **Rate**<br><br>fhw_rcv_rate |
| Activation Predicate | (none) |
| Reactivation Predicate | (none) |

| Timed Activity: | FHW_R_S2 |
|---|---|
| Distribution Parameters | **Rate**<br><br>fhw_rcv_rate |
| Activation Predicate | (none) |
| Reactivation Predicate | (none) |

| Timed Activity: | FHWt_F_S1 |
|---|---|
| Distribution Parameters | **Rate**<br><br>fhwt_fail_rate |
| Activation Predicate | (none) |
| Reactivation Predicate | (none) |
| Case Distributions | **case 1**<br><br>if(Working_S1->Mark()==1 && Working_S2->Mark()==1)<br>    return(1-heq_cvg);<br>else<br>    return(0);<br>**case 2**<br><br>if(Working_S1->Mark()==1 && Working_S2->Mark()==1)<br>    return(heq_cvg);<br>else<br>    return(1); |

| Timed Activity: | FHWt_F_S2 |
|---|---|
| Distribution Parameters | **Rate**<br><br>fhwt_fail_rate |
| Activation Predicate | (none) |
| Reactivation Predicate | (none) |
| Case Distributions | **case 1**<br><br>if(Working_S1->Mark()==1 && Working_S2->Mark()==1)<br>    return(1-heq_cvg);<br>else<br>    return(0);<br>**case 2**<br><br>if(Working_S1->Mark()==1 && Working_S2->Mark()==1)<br>    return(heq_cvg);<br>else<br>    return(1); |

| Timed Activity: | FHWt_R_S1 |
|---|---|
| Distribution Parameters | **Rate**<br><br>fhwt_rcv_rate |
| Activation Predicate | (none) |

| Reactivation Predicate | (none) |
|---|---|

| Timed Activity: | **FHWt_R_S2** |
|---|---|
| **Distribution Parameters** | **Rate**<br><br>fhwt_rcv_rate |
| **Activation Predicate** | (none) |
| **Reactivation Predicate** | (none) |

| Timed Activity: | **GEO_F** |
|---|---|
| **Distribution Parameters** | **Rate**<br><br>geo_fail_rate |
| **Activation Predicate** | (none) |
| **Reactivation Predicate** | (none) |

| Timed Activity: | **GEO_R** |
|---|---|
| **Distribution Parameters** | **Rate**<br><br>geo_rcv_rate |
| **Activation Predicate** | (none) |
| **Reactivation Predicate** | (none) |

| Timed Activity: | **L_F** |
|---|---|
| **Distribution Parameters** | **Rate**<br><br>link_fail_rate |
| **Activation Predicate** | (none) |
| **Reactivation Predicate** | (none) |

| Timed Activity: | **L_R** |
|---|---|
| **Distribution Parameters** | **Rate**<br><br>link_rcv_rate |
| **Activation Predicate** | (none) |
| **Reactivation Predicate** | (none) |

| Timed Activity: | **SW_F_S1** |
|---|---|
| **Distribution Parameters** | **Rate**<br><br>sw_fail_rate |
| **Activation Predicate** | (none) |
| **Reactivation Predicate** | (none) |
| **Case Distributions** | **case 1**<br><br>if(Working_S1->Mark()==1 && Working_S2->Mark()==1)<br>        return(1-heq_cvg);<br>else<br>        return(0);<br>**case 2**<br><br>if(Working_S1->Mark()==1 && Working_S2->Mark()==1)<br>        return(heq_cvg);<br>else<br>        return(1); |

| Timed Activity: | **SW_F_S2** |
|---|---|
| **Distribution Parameters** | **Rate**<br><br>sw_fail_rate |
| **Activation Predicate** | (none) |
| **Reactivation Predicate** | (none) |
| **Case Distributions** | **case 1**<br><br>if(Working_S1->Mark()==1 && Working_S2->Mark()==1)<br>        return(1-heq_cvg);<br>else<br>        return(0);<br>**case 2**<br><br>if(Working_S1->Mark()==1 && Working_S2->Mark()==1) |

```
                              return(heq_cvg);
                     else
                              return(1);
```

| Timed Activity: | SW_R_S1 |
|---|---|
| Distribution Parameters | **Rate**<br><br>sw_rcv_rate |
| Activation Predicate | (none) |
| Reactivation Predicate | (none) |

| Timed Activity: | SW_R_S2 |
|---|---|
| Distribution Parameters | **Rate**<br><br>sw_rcv_rate |
| Activation Predicate | (none) |
| Reactivation Predicate | (none) |

| Input Gate: | IG_GF |
|---|---|
| Predicate | `(Working_L->Mark()==1 && Working_S2->Mark()==1)` |
| Function | `Working_L->Mark()=0;`<br>`Working_S2->Mark()=0;` |

| Output Gate: | OG_FHW |
|---|---|
| Function | `Working_S1->Mark()=0;`<br>`Working_S2->Mark()=0;`<br>`failed_FHW_S1->Mark()=1;`<br>`failed_FHW_S2->Mark()=1;` |

| Output Gate: | OG_FHWt |
|---|---|
| Function | `Working_S1->Mark()=0;`<br>`Working_S2->Mark()=0;`<br>`failed_FHWt_S1->Mark()=1;`<br>`failed_FHWt_S2->Mark()=1;` |

| Output Gate: | OG_GR |
|---|---|
| Function | `Working_L->Mark()=1;`<br>`Working_S2->Mark()=1;` |

| Output Gate: | OG_SW |
|---|---|
| Function | `Working_S1->Mark()=0;`<br>`Working_S2->Mark()=0;`<br>`failed_SW_S1->Mark()=1;`<br>`failed_SW_S2->Mark()=1;` |

## Model: sss

**Place Attributes**:

| Place Names | Initial Markings |
|---|---|
| Working_S1 | 1 |
| Working_S2 | 1 |
| Working_S3 | 1 |
| failed_FHW_S1 | 0 |
| failed_FHW_S2 | 0 |
| failed_FHW_S3 | 0 |
| failed_FHWt_S1 | 0 |
| failed_FHWt_S2 | 0 |
| failed_FHWt_S3 | 0 |
| failed_SW_S1 | 0 |
| failed_SW_S2 | 0 |
| failed_SW_S3 | 0 |

| Timed Activity: | FHW_F_S1 |
|---|---|
| Distribution Parameters | **Rate**<br><br>fhw_fail_rate |
| Activation Predicate | (none) |
| Reactivation Predicate | (none) |
| Case Distributions | **case 1**<br><br>`if (Working_S1->Mark() == 1 && Working_S2->Mark() == 1 && Working_S3->Mark() ==1)` |

```
                    return(1-heq_cvg);
else
                    return(0);
case 2

if (Working_S1->Mark() == 1 && Working_S2->Mark() == 1 && Working_S3->Mark() ==1)
              return(heq_cvg);
else
              return(1);
```

| Timed Activity: | FHW_F_S2 |
|---|---|
| Distribution Parameters | **Rate**<br><br>fhw_fail_rate |
| Activation Predicate | (none) |
| Reactivation Predicate | (none) |
| Case Distributions | **case 1**<br><br>`if (Working_S1->Mark() == 1 && Working_S2->Mark() == 1 && Working_S3->Mark() ==1)`<br>`        return(1-heq_cvg);`<br>`else`<br>`        return(0);`<br>**case 2**<br><br>`if (Working_S1->Mark() == 1 && Working_S2->Mark() == 1 && Working_S3->Mark() ==1)`<br>`        return(heq_cvg);`<br>`else`<br>`        return(1);` |

| Timed Activity: | FHW_F_S3 |
|---|---|
| Distribution Parameters | **Rate**<br><br>fhw_fail_rate |
| Activation Predicate | (none) |
| Reactivation Predicate | (none) |
| Case Distributions | **case 1**<br><br>`if (Working_S1->Mark() == 1 && Working_S2->Mark() == 1 && Working_S3->Mark() ==1)`<br>`        return(1-heq_cvg);`<br>`else`<br>`        return(0);`<br>**case 2**<br><br>`if (Working_S1->Mark() == 1 && Working_S2->Mark() == 1 && Working_S3->Mark() ==1)`<br>`        return(heq_cvg);`<br>`else`<br>`        return(1);` |

| Timed Activity: | FHW_R_S1 |
|---|---|
| Distribution Parameters | **Rate**<br><br>fhw_rcv_rate |
| Activation Predicate | (none) |
| Reactivation Predicate | (none) |

| Timed Activity: | FHW_R_S2 |
|---|---|
| Distribution Parameters | **Rate**<br><br>fhw_rcv_rate |
| Activation Predicate | (none) |
| Reactivation Predicate | (none) |

| Timed Activity: | FHW_R_S3 |
|---|---|
| Distribution Parameters | **Rate**<br><br>fhw_rcv_rate |
| Activation Predicate | (none) |
| Reactivation Predicate | (none) |

| Timed Activity: | FHWt_F_S1 |
|---|---|
| Distribution Parameters | **Rate**<br><br>fhwt_fail_rate |
| Activation Predicate | (none) |
| Reactivation Predicate | (none) |

| Case Distributions | case 1 |
|---|---|

```
if (Working_S1->Mark() == 1 && Working_S2->Mark() == 1 && Working_S3->Mark() ==1)
        return(1-heq_cvg);
else
        return(0);
```

**case 2**

```
if (Working_S1->Mark() == 1 && Working_S2->Mark() == 1 && Working_S3->Mark() ==1)
        return(heq_cvg);
else
        return(1);
```

| Timed Activity: | **FHWt_F_S2** |
|---|---|
| **Distribution Parameters** | **Rate**<br><br>fhwt_fail_rate |
| **Activation Predicate** | (none) |
| **Reactivation Predicate** | (none) |
| **Case Distributions** | **case 1**<br><br>`if (Working_S1->Mark() == 1 && Working_S2->Mark() == 1 && Working_S3->Mark() ==1)`<br>`        return(1-heq_cvg);`<br>`else`<br>`        return(0);`<br>**case 2**<br><br>`if (Working_S1->Mark() == 1 && Working_S2->Mark() == 1 && Working_S3->Mark() ==1)`<br>`        return(heq_cvg);`<br>`else`<br>`        return(1);` |

| Timed Activity: | **FHWt_F_S3** |
|---|---|
| **Distribution Parameters** | **Rate**<br><br>fhwt_fail_rate |
| **Activation Predicate** | (none) |
| **Reactivation Predicate** | (none) |
| **Case Distributions** | **case 1**<br><br>`if (Working_S1->Mark() == 1 && Working_S2->Mark() == 1 && Working_S3->Mark() ==1)`<br>`        return(1-heq_cvg);`<br>`else`<br>`        return(0);`<br>**case 2**<br><br>`if (Working_S1->Mark() == 1 && Working_S2->Mark() == 1 && Working_S3->Mark() ==1)`<br>`        return(heq_cvg);`<br>`else`<br>`        return(1);` |

| Timed Activity: | **FHWt_R_S1** |
|---|---|
| **Distribution Parameters** | **Rate**<br><br>fhwt_rcv_rate |
| **Activation Predicate** | (none) |
| **Reactivation Predicate** | (none) |

| Timed Activity: | **FHWt_R_S2** |
|---|---|
| **Distribution Parameters** | **Rate**<br><br>fhwt_rcv_rate |
| **Activation Predicate** | (none) |
| **Reactivation Predicate** | (none) |

| Timed Activity: | **FHWt_R_S3** |
|---|---|
| **Distribution Parameters** | **Rate**<br><br>fhwt_rcv_rate |
| **Activation Predicate** | (none) |
| **Reactivation Predicate** | (none) |

| Timed Activity: | **SW_F_S1** |
|---|---|
| **Distribution Parameters** | **Rate**<br><br>sw_fail_rate |

| Activation Predicate | (none) |
|---|---|
| Reactivation Predicate | (none) |

| Case Distributions | **case 1**<br><br>`if (Working_S1->Mark() == 1 && Working_S2->Mark() == 1 && Working_S3->Mark() ==1)`<br>`        return(1-heq_cvg);`<br>`else`<br>`        return(0);`<br>**case 2**<br><br>`if (Working_S1->Mark() == 1 && Working_S2->Mark() == 1 && Working_S3->Mark() ==1)`<br>`        return(heq_cvg);`<br>`else`<br>`        return(1);` |
|---|---|

| Timed Activity: | **SW_F_S2** |
|---|---|
| Distribution Parameters | **Rate**<br><br>`sw_fail_rate` |
| Activation Predicate | (none) |
| Reactivation Predicate | (none) |
| Case Distributions | **case 1**<br><br>`if (Working_S1->Mark() == 1 && Working_S2->Mark() == 1 && Working_S3->Mark() ==1)`<br>`        return(1-heq_cvg);`<br>`else`<br>`        return(0);`<br>**case 2**<br><br>`if (Working_S1->Mark() == 1 && Working_S2->Mark() == 1 && Working_S3->Mark() ==1)`<br>`        return(heq_cvg);`<br>`else`<br>`        return(1);` |

| Timed Activity: | **SW_F_S3** |
|---|---|
| Distribution Parameters | **Rate**<br><br>`sw_fail_rate` |
| Activation Predicate | (none) |
| Reactivation Predicate | (none) |
| Case Distributions | **case 1**<br><br>`if (Working_S1->Mark() == 1 && Working_S2->Mark() == 1 && Working_S3->Mark() ==1)`<br>`        return(1-heq_cvg);`<br>`else`<br>`        return(0);`<br>**case 2**<br><br>`if (Working_S1->Mark() == 1 && Working_S2->Mark() == 1 && Working_S3->Mark() ==1)`<br>`        return(heq_cvg);`<br>`else`<br>`        return(1);` |

| Timed Activity: | **SW_R_S1** |
|---|---|
| Distribution Parameters | **Rate**<br><br>`sw_rcv_rate` |
| Activation Predicate | (none) |
| Reactivation Predicate | (none) |

| Timed Activity: | **SW_R_S2** |
|---|---|
| Distribution Parameters | **Rate**<br><br>`sw_rcv_rate` |
| Activation Predicate | (none) |
| Reactivation Predicate | (none) |

| Timed Activity: | **SW_R_S3** |
|---|---|
| Distribution Parameters | **Rate**<br><br>`sw_rcv_rate` |
| Activation Predicate | (none) |
| Reactivation Predicate | (none) |

| Output Gate: | OG_FHW |
|---|---|
| Function | `Working_S1->Mark()=0;`<br>`Working_S2->Mark()=0;`<br>`Working_S3->Mark()=0;`<br>`failed_FHW_S1->Mark()=1;`<br>`failed_FHW_S2->Mark()=1;`<br>`failed_FHW_S3->Mark()=1;` |

| Output Gate: | OG_FHWt |
|---|---|
| Function | `Working_S1->Mark()=0;`<br>`Working_S2->Mark()=0;`<br>`Working_S3->Mark()=0;`<br>`failed_FHWt_S1->Mark()=1;`<br>`failed_FHWt_S2->Mark()=1;`<br>`failed_FHWt_S3->Mark()=1;` |

| Output Gate: | OG_SW |
|---|---|
| Function | `Working_S1->Mark()=0;`<br>`Working_S2->Mark()=0;`<br>`Working_S3->Mark()=0;`<br>`failed_SW_S1->Mark()=1;`<br>`failed_SW_S2->Mark()=1;`<br>`failed_SW_S3->Mark()=1;` |

**Range Study Variable Assignments for Study *CC_study* in Project *SDNbackbone* :**

| Variable | Type | Range Type | Range | Increment | Increment Type | Function | n |
|----------|------|-----------|-------|-----------|----------------|----------|---|
| K_th | int | Fixed | 8 | - | - | - | - |
| N_proc | int | Fixed | 10 | - | - | - | - |
| hw_cvg | double | Fixed | 0.97 | - | - | - | - |
| hw_fail_rate | double | Fixed | 1.0E-8 | - | - | - | - |
| hw_rcv_rate | double | Fixed | 2.0E-5 | - | - | - | - |
| man_fail_rate | double | Fixed | 1.0E-6 | - | - | - | - |
| man_rcv_rate | double | Fixed | 9.0E-5 | - | - | - | - |
| mis_fail_rate | double | Manual | [5.0E-6, 5.0E-7, 5.0E-8, 5.0E-9, 5.0E-10] | - | - | - | - |
| mis_rcv_rate | double | Fixed | 9.0E-5 | - | - | - | - |
| sw_cvg | double | Fixed | 0.9 | - | - | - | - |
| sw_fail_rate | double | Fixed | 2.0E-5 | - | - | - | - |
| sw_rcv_rate | double | Fixed | 0.006 | - | - | - | - |
| tmi_cvg | double | Manual | [0.9, 0.93, 0.95, 0.97, 1.0] | - | - | - | - |
| uhw_rcv_rate | double | Fixed | 6.0E-4 | - | - | - | - |
| usw_rcv_rate | double | Fixed | 6.0E-4 | - | - | - | - |

| Performance Variable Model: CC_unavailability | | |
|---|---|---|
| Top Level Model Information | Child Model Name | cc |
| | Model Type | SAN Model |

| Performance Variable : U_cc | |
|---|---|
| Affecting Models | cc |
| Impulse Functions | |
| Reward Function | *(Reward is over all Available Models)*<br><br>```<br>if (( (cc->Active_proc_C1->Mark()<K_th \|\| cc->failed_MAN_C1->Mark()==1 \|\| cc->sys_down_C1->Mark()==1 \|\| cc->sw_sys_down_C1->Mark()==1)<br>         && (cc->Active_proc_C2->Mark()<K_th \|\| cc->failed_MAN_C2->Mark()==1 \|\| cc->sys_down_C2->Mark()==1 \|\| cc->sw_sys_down_C2->Mark()==1))<br>    \|\| cc->MIS->Mark()==1){<br>         return(1);<br>}<br>else{<br>         return(0);<br>}<br>``` |

| Simulator Statistics | Type | Time Averaged Interval of Time | |
|---|---|---|---|
| | Options | Estimate Mean | |
| | | Include Lower Bound on Interval Estimate | |
| | | Include Upper Bound on Interval Estimate | |
| | | Estimate out of Range Probabilities | |
| | | Confidence Level is Relative | |
| | Parameters | Start Time | 0.0, |
| | | Stop Time | 10000000, |
| | Confidence | Confidence Level | 0.95 |
| | | Confidence Interval | 0.1 |

**Range Study Variable Assignments for Study *CSL_study* in Project *SDNbackbone* :**

| Variable | Type | Range Type | Range | Increment | Increment Type | Function | n |
|----------|------|-----------|-------|-----------|----------------|----------|---|
| K_th | int | Fixed | 8 | - | - | - | - |
| N_proc | int | Fixed | 10 | - | - | - | - |
| cis_fail_rate | double | Manual | [2.0E-4, 2.0E-5, 2.0E-6, 2.0E-7, 2.0E-8] | - | - | - | - |
| cis_rcv_rate | double | Fixed | 0.002 | - | - | - | - |
| csw_fail_rate | double | Fixed | 2.0E-5 | - | - | | - |
| csw_rcv_rate | double | Fixed | 0.006 | - | - | - | - |
| fhw_fail_rate | double | Fixed | 9.0E-9 | - | - | - | - |
| fhw_rcv_rate | double | Fixed | 2.0E-5 | - | - | - | - |
| fhwt_fail_rate | double | Fixed | 2.0E-6 | - | - | - | - |
| fhwt_rcv_rate | double | Fixed | 0.006 | - | - | - | - |
| geo_fail_rate | double | Manual | [9.0E-8, 9.0E-9, 9.0E-10, 9.0E-11, 9.0E-12] | - | - | - | - |
| geo_rcv_rate | double | Fixed | 7.0E-6 | - | - | - | - |
| hw_cvg | double | Fixed | 0.97 | - | - | - | - |
| hw_fail_rate | double | Fixed | 1.0E-8 | - | - | - | - |
| hw_rcv_rate | double | Fixed | 2.0E-5 | - | - | - | - |
| link_fail_rate | double | Fixed | 1.0E-6 | - | - | - | - |
| link_rcv_rate | double | Fixed | 0.01 | - | - | - | - |
| man_fail_rate | double | Fixed | 1.0E-6 | - | - | - | - |
| man_rcv_rate | double | Fixed | 9.0E-5 | - | - | - | - |
| sw_cvg | double | Fixed | 0.9 | - | - | - | - |
| sw_fail_rate | double | Fixed | 2.0E-20 | - | - | - | - |
| sw_rcv_rate | double | Fixed | 0.006 | - | - | - | - |
| uhw_rcv_rate | double | Fixed | 6.0E-4 | - | - | - | - |
| usw_rcv_rate | double | Fixed | 6.0E-4 | - | - | - | - |

| Performance Variable Model: CSL_unavailability | | |
|---|---|---|
| Top Level Model Information | Child Model Name | csl |
| | Model Type | SAN Model |

| Performance Variable : U_csl | |
|---|---|
| Affecting Models | csl |
| Impulse Functions | |
| Reward Function | *(Reward is over all Available Models)*<br><br>```<br>if (csl->Working_S->Mark()==0 && csl->Working_L->Mark()==0 &&<br>    (csl->Active_proc->Mark()<K_th || csl->failed_MAN->Mark()==1 || csl->sys_down->Mark()==1 || csl->sw_sys_down-<br>>Mark()==1<br>    || csl->CIS->Mark()==1)){<br>        return(1);<br>}<br>else{<br>        return(0);<br>}<br>``` |
| Simulator Statistics | Type | Time Averaged Interval of Time |
| | Options | Estimate Mean |
| | | Include Lower Bound on Interval Estimate |
| | | Include Upper Bound on Interval Estimate |
| | | Estimate out of Range Probabilities |
| | | Confidence Level is Relative |
| | Parameters | Start Time | 0.0, |
| | | Stop Time | 10000000, |
| | Confidence | Confidence Level | 0.95 |
| | | Confidence Interval | 0.1 |

**Range Study Variable Assignments for Study *CSS_study* in Project *SDNbackbone*** :

| Variable | Type | Range Type | Range | Increment | Increment Type | Function | n |
|---|---|---|---|---|---|---|---|
| K_th | int | Fixed | 8 | - | - | - | - |
| N_proc | int | Fixed | 10 | - | - | - | - |
| cis_fail_rate | double | Manual | [2.0E-4, 2.0E-5, 2.0E-6, 2.0E-7, 2.0E-8] | - | - | - | - |

| cis_rcv_rate | double | Fixed | 0.002 | - | - | - | - |
| csw_fail_rate | double | Fixed | 2.0E-5 | - | - | - | - |
| csw_rcv_rate | double | Fixed | 0.006 | - | - | - | - |
| fhw_fail_rate | double | Fixed | 9.0E-9 | - | - | - | - |
| fhw_rcv_rate | double | Fixed | 2.0E-5 | - | - | - | - |
| fhwt_fail_rate | double | Fixed | 2.0E-6 | - | - | - | - |
| fhwt_rcv_rate | double | Fixed | 0.006 | - | - | - | - |
| geo_fail_rate | double | Manual | [9.0E-8, 9.0E-9, 9.0E-10, 9.0E-11, 9.0E-12] | - | - | - | - |
| geo_rcv_rate | double | Fixed | 7.0E-6 | - | - | - | - |
| hw_cvg | double | Fixed | 0.97 | - | - | - | - |
| hw_fail_rate | double | Fixed | 1.0E-8 | - | - | - | - |
| hw_rcv_rate | double | Fixed | 2.0E-5 | - | - | - | - |
| man_fail_rate | double | Fixed | 1.0E-6 | - | - | - | - |
| man_rcv_rate | double | Fixed | 9.0E-5 | - | - | - | - |
| sw_cvg | double | Fixed | 0.9 | - | - | - | - |
| sw_fail_rate | double | Fixed | 2.0E-20 | - | - | - | - |
| sw_rcv_rate | double | Fixed | 0.006 | - | - | - | - |
| uhw_rcv_rate | double | Fixed | 6.0E-4 | - | - | - | - |
| usw_rcv_rate | double | Fixed | 6.0E-4 | - | - | - | - |

| Performance Variable Model: CSS_unavailability | | |
|---|---|---|
| Top Level Model Information | Child Model Name | css |
| | Model Type | SAN Model |

| Performance Variable : U_css | |
|---|---|
| Affecting Models | css |
| Impulse Functions | |
| Reward Function | *(Reward is over all Available Models)*<br><br>`if (css->Working_S1->Mark()==0 && css->Working_S2->Mark()==0 &&`<br>`    (css->Active_proc->Mark()<K_th || css->failed_MAN->Mark()==1 || css->sys_down->Mark()==1 || css->sw_sys_down->Mark()==1`<br>`    || css->CIS->Mark()==1|| css->CIS_S1->Mark()==1|| css->CIS_S2->Mark()==1)){`<br>`        return(1);`<br>`}:q`<br>`else{`<br>`        return(0);`<br>`}` |
| Simulator Statistics | Type | Time Averaged Interval of Time | |
| | Options | Estimate Mean | |
| | | Include Lower Bound on Interval Estimate | |
| | | Include Upper Bound on Interval Estimate | |
| | | Estimate out of Range Probabilities | |
| | | Confidence Level is Relative | |
| | Parameters | Start Time | 0.0, |
| | | Stop Time | 10000000, |
| | Confidence | Confidence Level | 0.95 |
| | | Confidence Interval | 0.1 |

**Range Study Variable Assignments for Study *C_study* in Project *SDNbackbone* :**

| Variable | Type | Range Type | Range | Increment | Increment Type | Function | n |
|---|---|---|---|---|---|---|---|
| K_th | int | Fixed | 8 | - | - | - | - |
| N_proc | int | Fixed | 10 | - | - | - | - |
| hw_cvg | double | Fixed | 0.97 | - | - | - | - |
| hw_fail_rate | double | Fixed | 1.0E-8 | - | - | - | - |

| hw_rcv_rate | double | Fixed | 2.0E-5 | - | - | - | - |
|---|---|---|---|---|---|---|---|
| man_fail_rate | double | Fixed | 1.0E-6 | - | - | - | - |
| man_rcv_rate | double | Fixed | 9.0E-5 | - | - | - | - |
| sw_cvg | double | Fixed | 0.9 | - | - | - | - |
| sw_fail_rate | double | Fixed | 2.0E-5 | - | - | - | - |
| sw_rcv_rate | double | Fixed | 0.006 | - | - | - | - |
| uhw_rcv_rate | double | Fixed | 6.0E-4 | - | - | - | - |
| usw_rcv_rate | double | Fixed | 6.0E-4 | - | - | - | - |

| Performance Variable Model: C_unavailability | | |
|---|---|---|
| Top Level Model Information | Child Model Name | SDNcontroller |
| | Model Type | SAN Model |

| Performance Variable : U_c | |
|---|---|
| Affecting Models | SDNcontroller |
| Impulse Functions | |
| Reward Function | *(Reward is over all Available Models)*<br><br>`if (SDNcontroller->Active_proc->Mark()<K_th || SDNcontroller->failed_MAN->Mark()==1 || SDNcontroller->sys_down->Mark()==1`<br>`|| SDNcontroller->sw_sys_down->Mark()==1){`<br>`        return(1);`<br>`}`<br>`else{`<br>`        return(0);`<br>`}` |
| Simulator Statistics | Type | Time Averaged Interval of Time | |
| | Options | Estimate Mean | |
| | | Include Lower Bound on Interval Estimate | |
| | | Include Upper Bound on Interval Estimate | |
| | | Estimate out of Range Probabilities | |
| | | Confidence Level is Relative | |
| | Parameters | Start Time | 0.0, |
| | | Stop Time | 10000000, |
| | Confidence | Confidence Level | 0.95 |
| | | Confidence Interval | 0.1 |

**Range Study Variable Assignments for Study** *LL_study* **in Project** *SDNbackbone* :

| Variable | Type | Range Type | Range | Increment | Increment Type | Function | n |
|---|---|---|---|---|---|---|---|
| geo_fail_rate | double | Manual | [1.0E-5, 1.0E-6, 1.0E-7, 1.0E-8, 1.0E-9] | - | - | - | - |
| geo_rcv_rate | double | Fixed | 0.01 | - | - | - | - |
| link_fail_rate | double | Fixed | 1.0E-6 | - | - | - | - |
| link_rcv_rate | double | Fixed | 0.01 | - | - | - | - |
| phy_fail_rate | double | Manual | [1.0E-5, 1.0E-6, 1.0E-7, 1.0E-8, 1.0E-9] | - | - | - | - |
| phy_rcv_rate | double | Fixed | 0.003 | - | - | - | - |

| Performance Variable Model: LL_unavailability | | |
|---|---|---|
| Top Level Model Information | Child Model Name | ll |
| | Model Type | SAN Model |

| Performance Variable : U_ll | |
|---|---|
| Affecting Models | ll |
| Impulse Functions | |
| Reward Function | *(Reward is over all Available Models)* |

```
if (ll->Working_L1->Mark()==0 && ll->Working_L2->Mark()==0){
        return(1);
}
else{
        return(0);
}
```

| Simulator Statistics | Type | Time Averaged Interval of Time | |
|---|---|---|---|
| | Options | Estimate Mean | |
| | | Include Lower Bound on Interval Estimate | |
| | | Include Upper Bound on Interval Estimate | |
| | | Estimate out of Range Probabilities | |
| | | Confidence Level is Relative | |
| | Parameters | Start Time | 0.0, |
| | | Stop Time | 10000000, |
| | Confidence | Confidence Level | 0.95 |
| | | Confidence Interval | 0.1 |

**Range Study Variable Assignments for Study *RLL_study* in Project *SDNbackbone* :**

| Variable | Type | Range Type | Range | Increment | Increment Type | Function | n |
|---|---|---|---|---|---|---|---|
| chw_cvg | double | Fixed | 0.97 | - | - | - | - |
| chw_fail_rate | double | Fixed | 9.0E-9 | - | - | - | - |
| chw_rcv_rate | double | Fixed | 2.0E-5 | - | - | - | - |
| fhw_fail_rate | double | Fixed | 9.0E-9 | - | - | - | - |
| fhw_rcv_rate | double | Fixed | 2.0E-5 | - | - | - | - |
| fhwt_fail_rate | double | Fixed | 2.0E-6 | - | - | - | - |
| fhwt_rcv_rate | double | Fixed | 0.006 | - | - | - | - |
| geo_fail_rate | double | Manual | [9.0E-8, 9.0E-9, 9.0E-10, 9.0E-11, 9.0E-12] | - | - | - | - |
| geo_rcv_rate | double | Fixed | 7.0E-6 | - | - | - | - |
| link_fail_rate | double | Fixed | 1.0E-6 | - | - | - | - |
| link_rcv_rate | double | Fixed | 0.01 | - | - | - | - |
| man_fail_rate | double | Fixed | 5.0E-7 | - | - | - | - |
| man_rcv_rate | double | Fixed | 9.0E-5 | - | - | - | - |
| phy_fail_rate | double | Manual | [1.0E-5, 1.0E-6, 1.0E-7, 1.0E-8, 1.0E-9] | - | - | - | - |
| phy_rcv_rate | double | Fixed | 0.003 | - | - | - | - |
| sw_fail_rate | double | Fixed | 2.0E-6 | - | - | - | - |
| sw_rcv_rate | double | Fixed | 0.006 | - | - | - | - |
| uchw_rcv_rate | double | Fixed | 3.0E-5 | - | - | - | - |

| Performance Variable Model: RLL_unavailability | | |
|---|---|---|
| Top Level Model Information | Child Model Name | rll |
| | Model Type | SAN Model |

| Performance Variable : U_rll | |
|---|---|
| Affecting Models | rll |
| Impulse Functions | |
| Reward Function | *(Reward is over all Available Models)* <br><br> ```if (rll->Working_L1->Mark()==0 && rll->Working_L2->Mark()==0 && rll->Working_R->Mark()==0 && rll->spare_CHW->Mark()==0){\n        return(1);\n}\nelse{\n        return(0);\n}``` |
| Simulator Statistics | Type | Time Averaged Interval of Time |
| | Options | Estimate Mean |
| | | Include Lower Bound on Interval Estimate |

| | | Include Upper Bound on Interval Estimate | |
|---|---|---|---|
| | | Estimate out of Range Probabilities | |
| | | Confidence Level is Relative | |
| | Parameters | Start Time | 0.0, |
| | | Stop Time | 10000000, |
| | Confidence | Confidence Level | 0.95 |
| | | Confidence Interval | 0.1 |

**Range Study Variable Assignments for Study *RRL_study* in Project *SDNbackbone*** :

| Variable | Type | Range Type | Range | Increment | Increment Type | Function | n |
|---|---|---|---|---|---|---|---|
| chw_cvg | double | Fixed | 0.97 | - | - | - | - |
| chw_fail_rate | double | Fixed | 9.0E-9 | - | - | - | - |
| chw_rcv_rate | double | Fixed | 2.0E-5 | - | - | - | - |
| fhw_fail_rate | double | Fixed | 9.0E-9 | - | - | - | - |
| fhw_rcv_rate | double | Fixed | 2.0E-5 | - | - | - | - |
| fhwt_fail_rate | double | Fixed | 2.0E-6 | - | - | - | - |
| fhwt_rcv_rate | double | Fixed | 0.006 | - | - | - | - |
| geo_fail_rate | double | Manual | [9.0E-8, 9.0E-9, 9.0E-10, 9.0E-11, 9.0E-12] | - | - | - | - |
| geo_rcv_rate | double | Fixed | 7.0E-6 | - | - | - | - |
| heq_cvg | double | Manual | [0.98, 0.99, 1.0] | - | - | - | - |
| link_fail_rate | double | Fixed | 1.0E-6 | - | - | - | - |
| link_rcv_rate | double | Fixed | 0.01 | - | - | - | - |
| man_fail_rate | double | Fixed | 5.0E-7 | - | - | - | - |
| man_rcv_rate | double | Fixed | 9.0E-5 | - | - | - | - |
| sw_fail_rate | double | Fixed | 2.0E-6 | - | - | - | - |
| sw_rcv_rate | double | Fixed | 0.006 | - | - | - | - |
| uchw_rcv_rate | double | Fixed | 3.0E-5 | - | - | - | - |

| **Performance Variable Model: RRL_unavailability** | | |
|---|---|---|
| Top Level Model Information | Child Model Name | rrl |
| | Model Type | SAN Model |

| **Performance Variable : U_rrl** | |
|---|---|
| Affecting Models | rrl |
| Impulse Functions | |
| Reward Function | *(Reward is over all Available Models)*<br><br>`if (rrl->Working_S1->Mark()==0 && rrl->Working_S2->Mark()==0 && rrl->Working_L->Mark()==0 &&`<br>`    rrl->spare_CHW_S1->Mark()==0 && rrl->spare_CHW_S2->Mark()==0){`<br>`        return(1);`<br>`}`<br>`else{`<br>`        return(0);`<br>`}` |
| Simulator Statistics | Type | Time Averaged Interval of Time | |
| | Options | Estimate Mean | |
| | | Include Lower Bound on Interval Estimate | |
| | | Include Upper Bound on Interval Estimate | |
| | | Estimate out of Range Probabilities | |
| | | Confidence Level is Relative | |
| | Parameters | Start Time | 0.0, |
| | | Stop Time | 10000000, |
| | Confidence | Confidence Level | 0.95 |
| | | Confidence Interval | 0.1 |

**Range Study Variable Assignments for Study *RRR_study* in Project *SDNbackbone*** :

| Variable | Type | Range Type | Range | Increment | Increment Type | Function | n |
|----------|------|-----------|-------|-----------|----------------|----------|---|
| chw_cvg | double | Fixed | 0.97 | - | - | - | - |
| chw_fail_rate | double | Fixed | 9.0E-9 | - | - | - | - |
| chw_rcv_rate | double | Fixed | 2.0E-5 | - | - | - | - |
| fhw_fail_rate | double | Fixed | 9.0E-9 | - | - | - | - |
| fhw_rcv_rate | double | Fixed | 2.0E-5 | - | - | - | - |
| fhwt_fail_rate | double | Fixed | 2.0E-6 | - | - | - | - |
| fhwt_rcv_rate | double | Fixed | 0.006 | - | - | - | - |
| heq_cvg | double | Manual | [0.98, 0.99, 1.0] | - | - | - | - |
| man_fail_rate | double | Fixed | 5.0E-7 | - | - | - | - |
| man_rcv_rate | double | Fixed | 9.0E-5 | - | - | - | - |
| sw_fail_rate | double | Fixed | 2.0E-6 | - | - | - | - |
| sw_rcv_rate | double | Fixed | 0.006 | - | - | - | - |
| uchw_rcv_rate | double | Fixed | 3.0E-5 | - | - | - | - |

| Performance Variable Model: RRR_unavailability | | |
|---|---|---|
| Top Level Model Information | Child Model Name | rrr |
| | Model Type | SAN Model |

| Performance Variable : U_rrr | |
|---|---|
| Affecting Models | rrr |
| Impulse Functions | |
| Reward Function | *(Reward is over all Available Models)*<br><br>`if (rrr->Working_S1->Mark()==0 && rrr->Working_S2->Mark()==0 && rrr->Working_S3->Mark()==0 &&`<br>`    rrr->spare_CHW_S1->Mark()==0 && rrr->spare_CHW_S2->Mark()==0 && rrr->spare_CHW_S3->Mark()==0){`<br>`    return(1);`<br>`}`<br>`else{`<br>`    return(0);`<br>`}` |
| Simulator Statistics | |

| Simulator Statistics | Type | Time Averaged Interval of Time | |
|---|---|---|---|
| | Options | Estimate Mean | |
| | | Include Lower Bound on Interval Estimate | |
| | | Include Upper Bound on Interval Estimate | |
| | | Estimate out of Range Probabilities | |
| | | Confidence Level is Relative | |
| | Parameters | Start Time | 0.0, |
| | | Stop Time | 10000000, |
| | Confidence | Confidence Level | 0.95 |
| | | Confidence Interval | 0.1 |

**Range Study Variable Assignments for Study *RR_study* in Project *SDNbackbone*** :

| Variable | Type | Range Type | Range | Increment | Increment Type | Function | n |
|----------|------|-----------|-------|-----------|----------------|----------|---|
| chw_cvg | double | Fixed | 0.97 | | - | - | - |
| chw_fail_rate | double | Fixed | 9.0E-9 | - | - | - | - |
| chw_rcv_rate | double | Fixed | 2.0E-5 | - | - | - | - |
| fhw_fail_rate | double | Fixed | 9.0E-9 | - | - | - | - |
| fhw_rcv_rate | double | Fixed | 2.0E-5 | - | - | - | - |
| fhwt_fail_rate | double | Fixed | 2.0E-6 | - | - | - | - |
| fhwt_rcv_rate | double | Fixed | 0.006 | - | - | - | - |
| geo_fail_rate | double | Manual | [9.0E-8, 9.0E-9, 9.0E-10, 9.0E-11, 9.0E-12] | - | - | - | - |
| geo_rcv_rate | double | Fixed | 7.0E-6 | - | - | - | - |
| man_fail_rate | double | Manual | [5.0E-5, 5.0E-6, 5.0E-7, 5.0E-8, | | | | |

| | | | [5.0E-9] | | | | |
|---|---|---|---|---|---|---|---|
| man_rcv_rate | double | Fixed | 9.0E-5 | - | - | - | - |
| sw_fail_rate | double | Fixed | 2.0E-6 | - | - | - | - |
| sw_rcv_rate | double | Fixed | 0.006 | - | - | - | - |
| tmi_cvg | double | Manual | [0.9, 0.93, 0.95, 0.97, 1.0] | - | - | - | - |
| uchw_rcv_rate | double | Fixed | 3.0E-5 | - | - | - | - |

| Performance Variable Model: RR_unavailability | | |
|---|---|---|
| Top Level Model Information | Child Model Name | rr |
| | Model Type | SAN Model |

| Performance Variable : U_rr | | |
|---|---|---|
| Affecting Models | rr | |
| Impulse Functions | | |
| Reward Function | *(Reward is over all Available Models)*<br><br>`if (rr->Working_S1->Mark()==0 && rr->Working_S2->Mark()==0 &&`<br>`    rr->spare_CHW_S1->Mark()==0 && rr->spare_CHW_S2->Mark()==0){`<br>`      return(1);`<br>`}`<br>`else{`<br>`      return(0);`<br>`}` | |
| Simulator Statistics | Type | Time Averaged Interval of Time |
| | Options | Estimate Mean |
| | | Include Lower Bound on Interval Estimate |
| | | Include Upper Bound on Interval Estimate |
| | | Estimate out of Range Probabilities |
| | | Confidence Level is Relative |
| | Parameters | Start Time | 0.0, |
| | | Stop Time | 10000000, |
| | Confidence | Confidence Level | 0.95 |
| | | Confidence Interval | 0.1 |

## Model: SDNcontroller

**Place Attributes**:

| Place Names | Initial Markings |
|---|---|
| Active_proc | N_proc |
| failed_HW | 0 |
| failed_MAN | 0 |
| failed_SW | 0 |
| sw_sys_down | 0 |
| sys_down | 0 |

| Timed Activity: | HW_F1 |
|---|---|
| Distribution Parameters | Rate<br><br>`Active_proc->Mark() * hw_fail_rate` |
| Activation Predicate | (none) |
| Reactivation Predicate | (none) |
| Case Distributions | **case 1**<br><br>`if (sys_down->Mark() == 0 && sw_sys_down->Mark() == 0 && failed_MAN->Mark() == 0)`<br>`        return(1-hw_cvg);`<br>`else`<br>`        return(0);`<br><br>**case 2** |

```
if (sys_down->Mark() == 0 && sw_sys_down->Mark() == 0 && failed_MAN->Mark() == 0)
        return(hw_cvg);
else
        return(1);
```

| Timed Activity: | HW_F2 |
|---|---|
| Distribution Parameters | **Rate**<br><br>hw_fail_rate * failed_SW->Mark() |
| Activation Predicate | (none) |
| Reactivation Predicate | (none) |

| Timed Activity: | HW_R |
|---|---|
| Distribution Parameters | **Rate**<br><br>hw_rcv_rate |
| Activation Predicate | (none) |
| Reactivation Predicate | (none) |

| Timed Activity: | MAN_F |
|---|---|
| Distribution Parameters | **Rate**<br><br>man_fail_rate |
| Activation Predicate | (none) |
| Reactivation Predicate | (none) |

| Timed Activity: | MAN_R |
|---|---|
| Distribution Parameters | **Rate**<br><br>man_rcv_rate |
| Activation Predicate | (none) |
| Reactivation Predicate | (none) |

| Timed Activity: | SW_F |
|---|---|
| Distribution Parameters | **Rate**<br><br>if(Active_proc->Mark() >= K_th)<br>        return(sw_fail_rate);<br>else<br>        return(sw_fail_rate * Active_proc->Mark()); |
| Activation Predicate | (none) |
| Reactivation Predicate | (none) |
| Case Distributions | **case 1**<br><br>1-sw_cvg<br>**case 2**<br><br>sw_cvg |

| Timed Activity: | SW_R |
|---|---|
| Distribution Parameters | **Rate**<br><br>sw_rcv_rate |
| Activation Predicate | (none) |
| Reactivation Predicate | (none) |

| Timed Activity: | UHW_R |
|---|---|
| | **Rate** |

| Distribution Parameters | uhw_rcv_rate |
|---|---|
| **Activation Predicate** | (none) |
| **Reactivation Predicate** | (none) |

| **Timed Activity:** | **USW_R** |
|---|---|
| **Distribution Parameters** | Rate<br><br>usw_rcv_rate |
| **Activation Predicate** | (none) |
| **Reactivation Predicate** | (none) |

| **Input Gate:** | **IG_MAN** |
|---|---|
| **Predicate** | (failed_MAN->Mark() == 0 && sys_down->Mark() == 0 && sw_sys_down->Mark() == 0) |
| **Function** | ; |

| **Input Gate:** | **IG_SW** |
|---|---|
| **Predicate** | (failed_MAN->Mark() ==0 && sys_down->Mark() ==0 && sw_sys_down->Mark() == 0 && Active_proc->Mark() > 0) |
| **Function** | Active_proc->Mark()--; |

| **Output Gate:** | **OG_MAN** |
|---|---|
| **Function** | Active_proc->Mark() = N_proc - failed_HW->Mark();<br>failed_SW->Mark()=0; |

| **Output Gate:** | **OG_SD** |
|---|---|
| **Function** | failed_HW->Mark()++;<br>Active_proc->Mark() = N_proc - failed_HW->Mark();<br>failed_SW->Mark()=0; |

| **Output Gate:** | **OG_SSD** |
|---|---|
| **Function** | Active_proc->Mark() = N_proc - failed_HW->Mark();<br>failed_SW->Mark()=0; |

**Range Study Variable Assignments for Study** *SLL_study* **in Project** *SDNbackbone* :

| Variable | Type | Range Type | Range | Increment | Increment Type | Function | n |
|---|---|---|---|---|---|---|---|
| fhw_fail_rate | double | Fixed | 9.0E-9 | - | - | - | - |
| fhw_rcv_rate | double | Fixed | 2.0E-5 | - | - | - | - |
| fhwt_fail_rate | double | Fixed | 2.0E-6 | - | - | - | - |
| fhwt_rcv_rate | double | Fixed | 0.006 | - | - | - | - |
| geo_fail_rate | double | Manual | [9.0E-8, 9.0E-9, 9.0E-10, 9.0E-11, 9.0E-12] | - | - | - | - |
| geo_rcv_rate | double | Fixed | 7.0E-6 | - | - | - | - |
| link_fail_rate | double | Fixed | 1.0E-6 | - | - | - | - |
| link_rcv_rate | double | Fixed | 0.01 | - | - | - | - |
| phy_fail_rate | double | Manual | [1.0E-5, 1.0E-6, 1.0E-7, 1.0E-8, 1.0E-9] | - | - | - | - |
| phy_rcv_rate | double | Fixed | 0.003 | - | - | - | - |
| sw_fail_rate | double | Fixed | 2.0E-20 | - | - | - | - |
| sw_rcv_rate | double | Fixed | 0.006 | - | - | - | - |

| **Performance Variable Model: SLL_unavailability** | | |
|---|---|---|
| Top Level Model Information | Child Model Name | sll |
| | Model Type | SAN Model |

| Performance Variable : U_sll | |
|---|---|
| Affecting Models | sll |
| Impulse Functions | |

| Reward Function | *(Reward is over all Available Models)*<br><br>```<br>if (sll->Working_L1->Mark()==0 && sll->Working_L2->Mark()==0 && sll->Working_S->Mark()==0){<br>        return(1);<br>}<br>else{<br>        return(0);<br>}<br>``` |
|---|---|

| Simulator Statistics | Type | Time Averaged Interval of Time | |
|---|---|---|---|
| | Options | Estimate Mean | |
| | | Include Lower Bound on Interval Estimate | |
| | | Include Upper Bound on Interval Estimate | |
| | | Estimate out of Range Probabilities | |
| | | Confidence Level is Relative | |
| | Parameters | Start Time | 0.0, |
| | | Stop Time | 10000000, |
| | Confidence | Confidence Level | 0.95 |
| | | Confidence Interval | 0.1 |

**Range Study Variable Assignments for Study *SSL_study* in Project *SDNbackbone* :**

| Variable | Type | Range Type | Range | Increment | Increment Type | Function | n |
|---|---|---|---|---|---|---|---|
| fhw_fail_rate | double | Fixed | 9.0E-9 | - | - | - | - |
| fhw_rcv_rate | double | Fixed | 2.0E-5 | - | - | - | - |
| fhwt_fail_rate | double | Fixed | 2.0E-6 | - | - | - | - |
| fhwt_rcv_rate | double | Fixed | 0.006 | - | - | - | - |
| geo_fail_rate | double | Manual | [9.0E-8, 9.0E-9, 9.0E-10, 9.0E-11, 9.0E-12] | - | - | - | - |
| geo_rcv_rate | double | Fixed | 7.0E-6 | - | - | - | - |
| heq_cvg | double | Manual | [0.98, 0.99, 1.0] | - | - | - | - |
| link_fail_rate | double | Fixed | 1.0E-6 | - | - | - | - |
| link_rcv_rate | double | Fixed | 0.01 | - | - | - | - |
| sw_fail_rate | double | Fixed | 2.0E-20 | - | - | - | - |
| sw_rcv_rate | double | Fixed | 0.006 | - | - | - | - |

| Performance Variable Model: SSL_unavailability | | |
|---|---|---|
| Top Level Model Information | Child Model Name | ssl |
| | Model Type | SAN Model |

| Performance Variable : U_ssl | |
|---|---|
| Affecting Models | ssl |
| Impulse Functions | |

| Reward Function | *(Reward is over all Available Models)*<br><br>```<br>if (ssl->Working_S1->Mark()==0 && ssl->Working_S2->Mark()==0 && ssl->Working_L->Mark()==0){<br>        return(1);<br>}<br>else{<br>        return(0);<br>}<br>``` |
|---|---|

| Simulator Statistics | Type | Time Averaged Interval of Time | |
|---|---|---|---|
| | Options | Estimate Mean | |
| | | Include Lower Bound on Interval Estimate | |
| | | Include Upper Bound on Interval Estimate | |
| | | Estimate out of Range Probabilities | |
| | | Confidence Level is Relative | |
| | Parameters | Start Time | 0.0, |

| | | Stop Time | 10000000, |
|---|---|---|---|
| | Confidence | Confidence Level | 0.95 |
| | | Confidence Interval | 0.1 |

**Range Study Variable Assignments for Study** *SSS_study* **in Project** *SDNbackbone* :

| Variable | Type | Range Type | Range | Increment | Increment Type | Function | n |
|---|---|---|---|---|---|---|---|
| fhw_fail_rate | double | Fixed | 9.0E-9 | - | - | - | - |
| fhw_rcv_rate | double | Fixed | 2.0E-5 | - | - | - | - |
| fhwt_fail_rate | double | Fixed | 2.0E-6 | - | - | - | - |
| fhwt_rcv_rate | double | Fixed | 0.006 | - | - | - | - |
| heq_cvg | double | Manual | [0.98, 0.99, 1.0] | - | - | - | - |
| sw_fail_rate | double | Fixed | 2.0E-20 | - | - | - | - |
| sw_rcv_rate | double | Fixed | 0.006 | - | - | - | - |

| Performance Variable Model: SSS_unavailability | | |
|---|---|---|
| Top Level Model Information | Child Model Name | sss |
| | Model Type | SAN Model |

| Performance Variable : U_sss | |
|---|---|
| Affecting Models | sss |
| Impulse Functions | |
| Reward Function | *(Reward is over all Available Models)*<br><br>if (sss->Working_S1->Mark()==0 && sss->Working_S2->Mark()==0 && sss->Working_S3->Mark()==0){<br>    return(1);<br>}<br>else{<br>    return(0);<br>} |
| Simulator Statistics | Type | Time Averaged Interval of Time |
| | Options | Estimate Mean |
| | | Include Lower Bound on Interval Estimate |
| | | Include Upper Bound on Interval Estimate |
| | | Estimate out of Range Probabilities |
| | | Confidence Level is Relative |
| | Parameters | Start Time | 0.0, |
| | | Stop Time | 10000000, |
| | Confidence | Confidence Level | 0.95 |
| | | Confidence Interval | 0.1 |

**Range Study Variable Assignments for Study** *SS_study* **in Project** *SDNbackbone* :

| Variable | Type | Range Type | Range | Increment | Increment Type | Function | n |
|---|---|---|---|---|---|---|---|
| fhw_fail_rate | double | Fixed | 9.0E-9 | - | - | - | - |
| fhw_rcv_rate | double | Fixed | 2.0E-5 | - | - | - | - |
| fhwt_fail_rate | double | Fixed | 2.0E-6 | - | - | - | - |
| fhwt_rcv_rate | double | Fixed | 0.006 | - | - | - | - |
| geo_fail_rate | double | Manual | [9.0E-8, 9.0E-9, 9.0E-10, 9.0E-11, 9.0E-12] | - | - | - | - |
| geo_rcv_rate | double | Fixed | 7.0E-6 | - | - | - | - |
| mis_fail_rate | double | Manual | [5.0E-6, 5.0E-7, 5.0E-8, 5.0E-9, 5.0E-10] | - | - | - | - |
| mis_rcv_rate | double | Fixed | 9.0E-5 | - | - | - | - |
| sw_fail_rate | double | Fixed | 2.0E-20 | - | - | - | - |
| sw_rcv_rate | double | Fixed | 0.006 | - | - | - | - |
| tmi_cvg | double | Manual | [0.9, 0.93, 0.95, 0.97, 1.0] | - | - | - | - |

| Performance Variable Model: SS_unavailability | | |
|---|---|---|
| Top Level Model Information | Child Model Name | ss |
| | Model Type | SAN Model |

| Performance Variable : U_ss | | | |
|---|---|---|---|
| Affecting Models | ss | | |
| Impulse Functions | | | |
| Reward Function | *(Reward is over all Available Models)*<br><br>```if (ss->Working_S1->Mark()==0 && ss->Working_S2->Mark()==0){`<br>`        return(1);`<br>`}`<br>`else{`<br>`        return(0);`<br>`}``` | | |
| Simulator Statistics | Type | Time Averaged Interval of Time | |
| | Options | Estimate Mean | |
| | | Include Lower Bound on Interval Estimate | |
| | | Include Upper Bound on Interval Estimate | |
| | | Estimate out of Range Probabilities | |
| | | Confidence Level is Relative | |
| | Parameters | Start Time | 0.0, |
| | | Stop Time | 10000000, |
| | Confidence | Confidence Level | 0.95 |
| | | Confidence Interval | 0.1 |